\documentclass[aps,superscriptaddress,eqsecnum,nofootinbib,showpacs,preprintnumbers]{revtex4-2}
\usepackage{amsmath,mathtools}
\usepackage{amssymb}
\usepackage{amsfonts,amsthm,bm}
\usepackage{color,xcolor}
\usepackage[greek,english]{babel}
\usepackage{slashed}
\usepackage{titlesec}
\usepackage{comment}
\usepackage{appendix}
\usepackage{graphicx,epsfig}
\graphicspath{{figures/}{./}}
\usepackage{float}
\usepackage{subfigure}
\usepackage{tikz}
\newcommand{\abs}[1]{\left| #1 \right|}

\newcommand{\be}{\begin{eqnarray}}
	\newcommand{\ee}{\end{eqnarray}}
\newcommand{\bea}{\begin{eqnarray}}
	
	\newcommand{\eea}{\end{eqnarray}}

\newcommand{\beq}{\begin{equation}}
	\newcommand{\eeq}{\end{equation}}
\newcommand{\bseq}{\begin{subequations}}
	\newcommand{\eseq}{\end{subequations}}
\begin{document}

\preprint{\leftline{KCL-PH-TH/2024-{\bf 56}}}
 
	\title{Quantum-Ordering Ambiguities in Weak Chern - Simons 4D Gravity and Metastability of the Condensate-Induced Inflation}
	
	\author{Panagiotis  Dorlis }
	\email{psdorlis0@gmail.com} 
	\affiliation{Physics Division, School of Applied Mathematical and Physical Sciences,
		National Technical University of Athens,  Zografou Campus,
		Athens 15780, Greece.}
	\author{Nick E. Mavromatos}
	\affiliation{Physics Division, School of Applied Mathematical and Physical Sciences,
		National Technical University of Athens,  Zografou Campus,
		Athens 15780, Greece.}
	\affiliation{Theoretical Particle Physics and Cosmology Group, Physics Department, King's College London, Strand, London WC2R 2LS, UK.}

	\author{Sotirios-Neilos Vlachos }
	\email{sovlacho@gmail.com}
	\affiliation{Physics Division, School of Applied Mathematical and Physical Sciences,
		National Technical University of Athens,  Zografou Campus,
		Athens 15780, Greece.}

	\vspace{17.5cm}
	\begin{abstract}
In this work, we elaborate further on a (3+1)-dimensional cosmological Running-Vacuum-type Model (RVM) of inflation that characterises string-inspired Chern-Simons (CS) gravity.  These low-energy field theories involve axion fields coupled to gravitational CS (gCS) anomalous terms. It has been shown that inflation in such models is caused by a condensation of the gCS terms induced by primordial gravitational waves (GW), which leads to a linear-axion potential, thus breaking the shift symmetry, and lifting its periodicity (monodromy). We demonstrate here, rather rigorously, that this inflationary phase  may  be metastable, due to the existence of non-trivial imaginary parts of the gCS condensate.
 These imaginary parts are quantum effects, proportional to appropriate commutators of GW perturbations, and therefore vanish in the classical theory.  As we stress, their existence is quantum-ordering-scheme dependent.
 We argue here in favor of a physical importance of such imaginary parts, which we compute to second order in the GW (tensor) perturbations in the framework of a specific gauge-fixed effective Lagrangian, within a (mean field) weak-quantum-gravity path integral approach. We pay specific attention
 to the various space-time boundary terms. We thus provide an estimate of the life time of inflation. On matching our results with the relevant inflationary phenomenology, we fix the quantum-ordering ambiguities, and obtain an order-of-magnitude constraint on the ratio of the string energy scale $M_s$ in this model to the (four-spacetime-dimensional) reduced Planck mass $M_{\rm Pl}$, specifically, $M_s/M_{\rm Pl} = \mathcal O(10^{-1})$. This is consistent with the corresponding estimate obtained in previous analyses by the authors in this framework, based on a dynamical-system approach to linear-axion RVM inflation. Finally, we examine the r\^ole of periodic modulations in the axion potential induced by non-perturbative stringy effects on the slow-roll inflationary parameters, and find  compatibility with the cosmological data, upon appropriately selecting the relevant scales in the effective (multi)axion potential. 
   
	\end{abstract}
	\vspace{3.5cm}

	\maketitle
	
	\flushbottom
	
	\tableofcontents

\section{Introduction}

  In a series of previous works~\cite{bms,bms2,ms1,ms2,Mavromatos:2022xdo,Dorlis:2024yqw,Gomez-Valent:2024tdb},
a rather novel approach to string-inspired cosmology, 
the so-called Stringy Running Vacuum Model (StRVM), has been developed. 
The StRVM embeds, in a non-trivial way, the Running-Vacuum-Model (RVM) approach to modern cosmology~\cite{Shapiro:2000dz,rvm1,rvm2,rvmdata,SolaPeracaula:2024iil} into string theory~\cite{str1,str2}.
The RVM is a phenomenologically consistent and innovative cosmological framework, with important predictions on observable deviations from $\Lambda$CDM~\cite{rvmqft1,rvmqft2,rvmqft3,rvmqft4,Gomez-Valent:2024tdb}, especially in the modern era, where it may also provide alleviations of the cosmological tensions~\cite{rvmtens,Gomez-Valent:2023hov}, features which are fully shared by (variants of) the StRVM~\cite{Gomez-Valent:2024tdb}. Moreover, the RVM approach describes a smooth evolution of the Universe from the inflationary to the modern eras, explaining also the thermodynamic properties of the Universe~\cite{Lima:2013dmf,Lima:2015mca,rvm3}.  

In the StRVM
there are highly non-trivial and diverse r\^oles played by the antisymmetric tensor field of the massless multiplet of the underlying string model~\cite{str1,str2,pol1,pol2,kaloper},  
from driving inflation to the 
generation of matter-antimatter asymmetry in this Universe, during the early postinflationary epochs.\footnote{  The StRVM is based on massless fields of the closed  sector of the underlying microscopic string theories~\cite{str1,str2}, which is characterized by an Abelian gauge invariance of the spin-1 antisymmetric tensor (Kalb-Ramond (KR)) field $B_{\mu\nu}(x) =  - B_{\nu\mu}(x) \to B_{\mu\nu}(x) + \partial_{[\mu}\theta_{\nu]}(x)$ (with $[\,,\,]$ denoting antisymmetry in the respective spacetime indices). Hence, the corresponding gravitational action of the StRVM depends only on the field strength $H_{\mu\nu\rho}$ of $B_{\mu\nu}$. In the framework of brane models~\cite{pol1,pol2,Polchinski:1996na}, on the other hand, this symmetry might be broken, in which case there are other types of gravitational theories that depend explicitly on the $B$-field as well, which may also acquire a mass. Such models lead to more general cosmologies, with a rich and quite interesting structure, complementary in several aspects to that of StRVM, which have only recently started to be exploited~\cite{Capanelli:2023uwv}. } 
An important feature, which by the way is shared between the StRVM and the works in Refs.~\cite{Antoniadis:1988aa,Antoniadis:1988vi,Antoniadis:1990uu} 
on string-dilaton cosmologies, is the linear time dependence of the (3+1)-dimensional dual of the antisymmetric-tensor field strength after string compactification. This dual field is a pseudoscalar, and is commonly known as a KR, or gravitational, or string-model-independent, axion, $b(x)$, to be distinguished from the string-compactification-induced axions, $a^{(i)}(x)$, $i=1, \dots N$ (axion species), which depend on the specific string model~\cite{kaloper,svrcek}.  In \cite{Antoniadis:1988aa,Antoniadis:1988vi,Antoniadis:1990uu}, such linearly dependent KR axions characterize static Einstein Universes with positive constant curvature (perfect-fluid de Sitter solutions). 
Although our approach and situation in this article will be different from that discussed in ~\cite{Antoniadis:1988aa,Antoniadis:1988vi,Antoniadis:1990uu}, nonetheless in our previous works~\cite{bms,bms2,ms1,ms2} we have 
demonstrated, as already mentioned, the importance of such a linear dependence in cosmic time of the KR axion for the evolution of the string-inspired Universe in a Friedman-Lemaitre-Robertson-Walker (FLRW) expanding-Universe spacetime background, specifically the existence of an inflationary phase, 
as well as the post-inflationary dominance of matter over antimatter in the Cosmos, 
via the scenario advocated in~\cite{deCesare:2014dga,Bossingham:2017gtm,Bossingham:2018ivs,Mavromatos:2020dkh}.   

As regards the former property, inflation in this model is induced~\cite{bms,bms2,ms1,ms2,Mavromatos:2022xdo,Dorlis:2024yqw} by the condensation of primordial Gravitational-Wave (GW) tensor perturbations, which in turn lead to  a non trivial condensate of the anomalous gravitational Chern-Simons (gCS) terms.
As for the generation of matter-antimatter asymmetry in the (early) post-inflationary epoch of this Universe, 
as discussed in \cite{bms,bms2,ms1,ms2,Mavromatos:2022xdo,Dorlis:2024yqw}, the presence of a constant $\dot b$ (where the dot denotes cosmic time derivative) 
during inflation, implies that the cosmic KR axion background remains undiluted at the end of inflation. At that epoch, chiral fermionic matter is generated through the decay of the metastable (as we shall show here) vacuum. Such fermions are characterized by their own gravitational and chiral anomalies in the gauge sector. In the scenario of \cite{bms,ms1,ms2}, the fermion-induced chiral anomalies cancel the primordial ones, the latter assumed to exist in the early-eras of the compactified string Universe due to the Green-Schwarz mechanism~\cite{GS}, introduced for anomaly cancellation in (the higher-dimensional) string theories~\cite{str1,str2,pol1,pol2}. This cancellation implies that during the early radiation era, that succeeds inflation, the KR background axion field $b$ scales with the scale factor of the universe as $a^{-3} \sim T^3$, where $T$ is the Universe temperature. Under such scaling, it is possible to show, then, that in theories with right-handed neutrinos one can have the generation of a Lepton asymmetry (Leptogenesis) according to the mechanism of \cite{Bossingham:2017gtm,Bossingham:2018ivs}, which for the short time scales of the process resembles the situation advocated in \cite{deCesare:2014dga}, for a Leptogenesis induced by an approximately constant $\dot b$ background. This Leptogenesis can then be communicated to the baryon sector~\cite{Mavromatos:2020dkh} via, say, Baryon-minus-Lepton-number preserving sphaleron processes~\cite{Manton:1983nd,Klinkhamer:1984di, Arnold:1987mh,Cline:1993bd,Manton:2019qka} in the standard model sector (Baryogenesis~\cite{Kuzmin:1985mm,Gavela:1994ds,Gavela:1994dt,Rubakov:1996vz}), thus leading to an explanation of the observable matter-antimatter asymmetry in the Universe~\cite{bms}, in the framework of this string-inspired CS gravity model. 

The presence of anomalous gCS terms is therefore an important feature of our approach, which has not been considered in \cite{Antoniadis:1988aa,Antoniadis:1988vi,Antoniadis:1990uu}. Such terms couple to the KR axion, as a consequence of a Bianchi identity of a modified 
field strength of the antisymmetric tensor field of the massless gravitational string multiplet~\cite{kaloper,svrcek}. The presence of the pertinent modifications is dictated by the appropriate Green-Schwarz counterterms that are added to the effective action in order to potentially cancel gauge and gravitational anomalies~\cite{GS} in string theory~\cite{str1,str2,pol1,pol2}. 
In the approach of \cite{bms,bms2,ms1,ms2,Mavromatos:2022xdo},  (3+1)-dimensional gravitational anomalies are assumed {\it not} to be canceling. Moreover, only fields from the massless gravitational string multiplet are assumed to appear as external lines in the pertinent Feynman graphs of the effective theory that describes the primordial Universe. Supergravity is assumed to have been broken dynamically in this scenario at the pre-inflationary epoch, with the partners having obtained masses of order of the reduced Planck mass $M_{\rm Pl}= 2.4 \times 10^{18}$~GeV and thus decoupled during the inflationary epoch of interest to us here~\cite{ms1,ms2}. Therefore, for our purposes, we deal with effective low-energy gravitational theories emerging from string theory, after compactification to (3+1)-dimensions, involving graviton, dilaton and antisymmetric tensor fields, which are at most quadratic in derivatives, or, equivalently, ${\mathcal O}(\alpha^\prime)$ effective actions, where $\alpha^\prime = M_s^{-2}$ is the Regge slope of the string, and $M_s$ the string mass scale. The latter is treated as a phenomenological parameter in our approach, which is in general different from the four-dimensional reduced Planck scale $\kappa^{-1}=M_{\rm Pl}$, where $\kappa = \sqrt{8\pi\, \rm G}$ denotes the (3+1)-dimensional gravitational constant, with $\rm G$ Newton's constant. For our purposes, following \cite{bms,bms2}, we shall assume that the dilaton field is constant, which can be self-consistently arranged in string theory by an appropriate choice of the dilaton potential (possibly generated by string loops). Under this assumption, the $\mathcal O(\alpha^\prime)$ effective action we shall be dealing with is a Chern-Simons gravitational theory of the form discussed in \cite{Jackiw,Alexander:2009tp}.\footnote{In the absence of a non-trivial dilaton configuration, the Gauss-Bonnet quadratic curvature term of the Bosonic sector of the $\mathcal O(\alpha^\prime)$ string-inspired gravitational effective theory~\cite{Gross:1986mw,Metsaev:1987zx} 
is a total derivative in (3+1)-dimensions, and plays no r\^ole in our approach.}

In the presence of a sufficient amount of {\it chiral} primordial gravitational waves (GW), the gravitational CS terms condense, leading to an effective potential which is linear in the KR axion, thus driving inflation, as in the so-called axion-monodromy situation encountered in strings~\cite{silver}.
However, there is an important difference between conventional-string axion-monodromy scenarios for inflation, and our condensate-induced inflation. In our case, the CS condensate depends~\cite{bms,ms1,ms2,Dorlis:2024yqw} on the fourth power of the Hubble parameter $H_I$ during inflation, which varies very mildly with the cosmic time. This has, as a consequence, the non-linear dependence of the vacuum energy density on $H_I^4$, which dominates and leads to a running-vacuum-model (RVM) type of inflation~\cite{rvm1,rvm2,rvm3,Lima:2015mca,Lima:2013dmf}. In fact, the RVM form of the pressure and energy densities of the corresponding cosmic fluid, that is their dependence on even powers of the Hubble parameter, of which terms of order $H^2$ and $H^4$ are the dominant ones, persist during the post-inflationary period as well, until the current epoch~\cite{bms}. It is for this reason that the abovedescribed cosmological model is termed Stringy Running Vacuum Model (StRVM)~\cite{bms,ms1,ms2}.\footnote{  It should be mentioned that RVM-type cosmologies arise quite naturally in quantum field theories, after integrating out massive matter modes in the path-integral, and appropriately subtracting the corresponding Ultra-Violet (UV) divergent vacuum contributions~\cite{rvmqft1,rvmqft2,rvmqft3,rvmqft4}. In such situations, the corresponding vacuum energy densities contain $H^2$ and $H^6$ (and higher order) terms, in contrast to our StRVM, where the highest power of the vacuum energy density is $H^4$~\cite{bms,ms1,Mavromatos:2022xdo,Dorlis:2024yqw}. Nonetheless the metastable nature of the RVM vacuum energy is manifest in both approaches.  } 

Phenomenology requires that the slow-roll parameters of this inflationary model fall into the range of values inferred from the plethora of the currently available cosmological data~\cite{Planck:2018jri,Planck:2018vyg}. A linear axion alone cannot yield optimal fits to the data, as we shall discuss in this article. Nonetheless, one may have periodic modulations of the axion potentials in string theory due, e.g., to non-perturbative instanton effects of appropriate gauge groups, that exist during the RVM inflation. Such modulations do yield consistent results with the data, provided one fixes appropriately the (target-space) energy scales of the  instantons, which in our approach are treated as phenomenological, string-model independent parameters. 
  In addition to target-space gauge-group instantons in string theories, there are also world-sheet instantons which couple to the compactification axions~\cite{svrcek,silver}, but not to the KR action. These world-sheet instantons also contribute appropriate periodic modulations in the multiaxion potential, which contribute to the slow-roll inflationary parameters. 
In the current work we shall treat all such modulations as being characterized by phenomenological scales. 
We stress, however, that in principle one should start from a consistent microscopic string theory model and study in detail the world-sheet or target-space gauge instanton effects. This is a highly non-trivial task, and at present, as far as we know, the situation is incomplete. For our purposes, therefore, we shall follow the aforementioned bottom-up approach, by means of which we shall determine the energy scales of the non-perturbatively-induced modulations of the axion potential phenomenologically, by requiring that they yield agreement 
with the cosmological (inflationary) data~\cite{Planck:2018jri,Planck:2018vyg,Martin:2013tda}.   

Another important novel aspect of the model, which we shall discuss here, concerns the microscopic origin of the imaginary parts of the primordial-GW-induced CS condensate. As we shall demonstrate, the latter are associated with {\it quantum}  effects of the effective Hamiltonian of the GW, that is, they vanish classically. Specifically, we shall show that the pertinent imaginary parts are proportional to quantum commutators of {\it chiral} GW creation and annihilation operators. 
Hence, it appears that it is the quantum aspects of the chiral GW that lead to an effective violation of unitarity, and the metastability of the condensate-induced inflationary vacuum. 

 It is important to stress that the presence of imaginary parts in the gravitational CS condensate appears to be quantum-operator-ordering dependent. Par contrast, 
the real part of the condensate is independent of such orderings. 
In a symmetric ordering, such parts vanish. However, we argue in the current article, that choosing ordering schemes in which they are present has physical significance, in that such schemes imply instabilities of the inflationary vacuum and a finite life-time, as indicated in the classical dynamical-system approach~\cite{Dorlis:2024yqw}. Matching the resulting finite life-time with the phenomenologically acceptable 50-60 e-foldings~\cite{Planck:2018jri,Martin:2013tda}, will lead, as we shall see, in 
adopting a specific ordering scheme, and also 
a constraint on the magnitude of the string scale $M_s$ as compared to $M_{\rm Pl}$, in agreement with the findings of the classical analysis of \cite{Dorlis:2024yqw}. 

\color{black} Before concluding the introduction with the outline of the article, we feel placing our work in perspective of the more general effective framework of Chern-Simons gravity and its applications~\cite{Jackiw,Alexander:2009tp}, in the presence of contorted geometries~\cite{Hehl:1976kj,Shapiro:2001rz}. As is well known, in the context of string theories~\cite{str1,str2,pol1,pol2}, the  field strength of the antisymmetric tensor spin-one field of the string gravitational multiplet, $\mathcal H_{\mu\nu\rho}$, plays the r\^ole of a totally antisymmetric torsion in the low energy effective field theory, up to and including fourth-order derivative terms~\cite{Gross:1986mw, Metsaev:1987zx,kaloper}. Although in our present work the torsion interpretation will not play a r\^ole, nonetheless for completeness we shall make some comments in this direction, which will hopefully constitute  future research avenues in this context.

First, of all, as discussed in some detail in \cite{Mavromatos:2021hai}, where we refer the interested reader, the KR axion in string theory, which is dual in (3+1)-dimensions to the antisymemtric-tensor field-strength three form, $\mathcal H_{\mu\nu\rho}$,  can be thought of as completely analogous to the Barbero-Immirzi parameter~\cite{BarberoG:1994eia,Immirzi:1996di} of loop quantum gravity~\cite{Ashtekar:1986yd,Ashtekar:1987gu,Ashtekar:2004eh}, when the latter is promoted to a (pseudoscalar) field~\cite{Taveras:2008yf,Calcagni:2009xz,Mercuri:2009zt,Lattanzi:2009mg}, in analogy to the case of the instanton angle in Quantum Chromodynamics (QCD)~\cite{Mercuri:2010yj}.
According to \cite{Mavromatos:2021hai}, the string effective action of the StRVM, in the presence of CS gravitational anomalous terms, 
is linked to the 
so-called Nieh-Yan invariant~\cite{nieh,nieh2,nieh3,nieh4,nieh5}, which is a topological torsional invariant, in contrast to the Holst term~\cite{Holst:1995pc} 
$\int d^4 x \, \varepsilon^{\mu\nu\rho\sigma} \, \widehat R_{\mu\nu\rho\sigma} (\overline \Gamma)$ ($\overline \Gamma$ denotes the generalised torsionful Christoffel connection),  which  alone is not a topological invariant quantity in the presence of torsion~\cite{mercuri,mercuri2}. The Nieh-Yan term is defined as
\begin{align}\label{nieh}
S_{\rm Nieh-Yan} \propto \int d^4 x  \, \Big(\varepsilon^{\mu\nu\rho\sigma} \, T^\lambda_{\,\,\,\,\mu\rho} \, T_{\lambda\nu\sigma} - \varepsilon^{\mu\nu\rho\sigma} \, \widehat R_{\mu\nu\rho\sigma} (\overline \Gamma) \Big)=  \int d^4 x  \, \partial_\mu \Big(\varepsilon^{\mu\nu\rho\sigma} \, T_{\nu\rho\sigma} \Big),  
\end{align}
where $T^\mu_{\,\,\,\,\nu\rho}$ is the torsion tensor, which in the context of the Einstein-Cartan theory is a non-propagating field, since the corresponding action contains non-derivative terms of the torsion tensor. Thus, the Nieh-Yan term is indeed a total derivative even in  the presence of torsion and replaces~\cite{mercuri,mercuri2} the Holst term~\cite{Holst:1995pc}.
  In our string-theory case, the torsion, as already mentioned, has a single totally antisymmetric component, proportional to the antisymemtric-tensor field strength $T_{\mu\nu\rho} \propto \mathcal H_{\mu\nu\rho}$. The BI can be considered as a constant coefficient $\beta$ of the torsional invariant term \eqref{nieh}.

The promotion of the BI parameter to a field would naively be  equivalent to considering adding to the effective gravitational action 
in the presence of torsion
a term \eqref{nieh} but with a coordinate-dependent coefficient $\beta (x)$, which is viewed as a field variable, integrated over in the path-integral (with a measure $\mathcal D \beta(x)$):
\begin{align}\label{nieh2}
S_{\rm Nieh-Yan}^{\rm BI-field} &=  \int d^4 x\,  \beta(x) \, \Big(\varepsilon^{\mu\nu\rho\sigma} \, T^\lambda_{\,\,\,\,\mu\rho} \, T_{\lambda\nu\sigma} - \varepsilon^{\mu\nu\rho\sigma} \, \widehat R_{\mu\nu\rho\sigma} (\overline \Gamma) \Big)
\nonumber \\
&=  \int d^4 x  \, \beta (x)\, \partial_\mu \Big(\varepsilon^{\mu\nu\rho\sigma} \, T_{\nu\rho\sigma} \Big) \propto  \int d^4 x  \, \beta (x) \,  \partial_\mu \Big(\varepsilon^{\mu\nu\rho\sigma} \, \mathcal H_{\nu\rho\sigma} \Big)
\end{align}
where in the last equality the proportionality factor takes into account numerical coefficients that appear in the precise relation between the torsion and the KR field strength~\cite{Mavromatos:2021hai}. 
As discussed in previous works~\cite{kaloper,bms,Mavromatos:2021hai}, path-integrating over (the non-propagating field) $\mathcal H$ in the low-energy string-inspired effective action,  would produce a dynamical propagating field $\beta (x)$, with canonically normalised kinetic terms, which would lead~\cite{Mavromatos:2021hai}to the result of \cite{Taveras:2008yf,Calcagni:2009xz,Mercuri:2009zt}. 

However, the above procedure would imply a Bianchi identity constraint $\varepsilon^{\mu\nu\rho\sigma} \, \partial_\mu\,\mathcal H_{\nu\rho\sigma} =0$, which does not take into account 
the Green-Schwarz (GS) counteterms 
required in string theory for 
cancellation of gauge and gravitational anomalies in the extra-dimensional target spacetime of strings~\cite{GS}. The GS counterterms lead to
appropriate modifications in the definition of $\mathcal H_{\mu\nu\rho}$, by appropriate gravitational and (non-Abelian) gauge CS three forms~\cite{kaloper}, the gauge terms pertaining to the appropriate gauge groups that characterise the underlying string theory model. The modified $\mathcal H_{\mu\nu\rho}$ obeys a Bianchi-identity constraint~\cite{bms}, which, when implemented in the path integral via a 
pseudoscalar Lagrange multiplier field
~\cite{kaloper,svrcek,bms,Mavromatos:2021hai}, leads to the emergence of 
a dynamical 
KR axion field, $b(x)$, dual 
to  the modified field strength $\mathcal H_{\mu\nu\rho}$, which is analogous to the BI field.  Thus, in the string case, the promotion of the BI parameter accompanying the Nieh-Yan invariant \eqref{nieh} to a dynamical pseudoscalar field cannot be done solely via \eqref{nieh2}, as in \cite{Taveras:2008yf,Calcagni:2009xz,Mercuri:2009zt,Lattanzi:2009mg}, but by adding instead to the low-energy, string-inspired, effective gravitational field-theory action  an {\it appropriate combination} of topological invariants, the Nieh-Yan invariant and the gravitational (without torsion) Chern-Simons term, with the BI field appearing as its coefficient~\cite{Mavromatos:2021hai}:
\begin{align}\label{nieh3}
S_{\rm Nieh-Yan}^{\rm BI-field} + S_{\rm Grav.~Chern~Simons} &=  
 \int d^4 x  \, b(x) \, \Big(\varepsilon^{\mu\nu\rho\sigma} \, \partial_\mu \, \mathcal H_{\nu\rho\sigma}  -  \frac{\alpha^\prime}{32\, \kappa} \, \sqrt{-g}\, R_{\mu\nu\rho\sigma}\, \widetilde R^{\mu\nu\rho\sigma} \Big)\,,
\end{align} 
where $\widetilde{(...)}$ denotes the dual Riemann tensor, defined in the next section \ref{sec:cond} ({\it cf.} \eqref{dualriem}), $\kappa$ is the gravitational constant in (3+1)-dimensions, and $\alpha^\prime$ the Regge slope of the string. 
The pseudoscalar (axion-like) field $b(x)$ is the so-called string-model independent axion~\cite{svrcek}. The reader should notice that the fields $b(x)$ \eqref{nieh3} and BI $\beta(x)$ \eqref{nieh2} share the standard axionic shift symmetry, that is, the corresponding action terms are invariant under the shift of the fields by constants.

Another aspect of the CS theories is the coupling of the axion fields $b(x)$ (or, in general the axions arising from string compactification $a(x)$) to mixed anomalies, that is gravitational but also gauge CS terms, the latter being of the form 
\begin{align}\label{axions}
\propto \int d^4 x \, \sqrt{-g}\, (b(x)~{\rm or}~a(x))\, {\mathbf F}_{\mu\nu} \, \widetilde{{\mathbf F}}^{\mu\nu}\,,
\end{align}
where the dual of the gauge field strength is defined in \eqref{gaugedual}, in section \ref{sec:cond} below. 

In our current paper, such gauge anomalous terms shall not be considered, the reason being that we
are interested in the inflationary epoch of the StRVM, which is not characterised by gauge fields. As discussed in \cite{bms,ms1} in the StRVM, the latter are generated at the end of the inflationary period, as a result of the decay of the unstable running vacuum. In the primordial eras one could at most have non-perturbative instanton-type configurations of appropriate non-Abelian gauge fields, which are integrated out in a path integral to yield periodic modulation terms in the axion potentials, as we shall discuss in section \ref{sec:gauginst}. The latter can contribute positively to the correct phenomenology of the inflationary slow-roll parameters, leading to agreement with the data~\cite{Planck:2018jri,Planck:2018vyg}, but lead to no further physical effects.  
Par contrast, in the post-inflationary period, such chiral gauge anomalies can play a crucial r\^ole in several respects~\cite{bms,bms2,ms1,ms2}. For instance, during the QCD epoch, color-group instanton effects can lead, via chiral anomalies that survive the post-inflationary eras of the StRVM, to non-perturbative generated masses of the KR (or other) axions, which can thus play the r\^ole of dark matter with mass of the order of the QCD axion~\cite{bms2}. In general, in string models, instantons of the appropriate non-Abelian gauge groups can lead to massive compactification axions, which can play the r\^ole of even ultralight DM~\cite{Marsh:2015xka}. In addition, Abelian electromagnetic U(1) chiral-anomaly terms \eqref{axions} coupled to axions can exist at late epochs of the post-inflationary StRVM Universe~\cite{bms}, and can lead to interesting effects, via axion-photon couplings. At such epochs, the effects of the KR field torsion could also be significant in leading to the generation of cosmic magnetic fields from galactic dynamo instabilities in the Universe, in similar spirit to what happens~\cite{GarciadeAndrade:2012zz,GarciadeAndrade:2018acd, GarciadeAndrade:2022fik,GarciadeAndrade:2022vyy} in generic torsional Einstein-Cartan theories~\cite{Cartan:1923zea}. Moreover, on the related front of the BI parameter of the loop quantum gravity~\cite{Ashtekar:1986yd,Ashtekar:1987gu,Ashtekar:2004eh}, which, as we conjectured above~\cite{Mavromatos:2021hai}, appears to be linked to the StRVM,
one should make the important reamark that the BI parameter may be constrained via photon-axion conversions and axion mixing in the presence of (generic) torsion, since such phenomena are found to depend on the BI parameter~\cite{Gao:2024czf}. In our case, there are additional constraints on the magnitude of the KR axion background (which plays the r\^ole of the BI field, as discussed above) at late eras of the Universe evolution, which arise from the characteristic cosmic evolution of the KR field and its r\^ole in Leptogenesis~\cite{bms,deCesare:2014dga,Bossingham:2017gtm,Bossingham:2018ivs,Mavromatos:2024szb}. 
All these avenues of research are interesting ones to pursue, and we hope to come back to such issues in future publications. \color{black}

The structure of this article is the following: in the next section \ref{sec:cond}, we shall review briefly the emergence of a linear axion potential in the context of our string-inspired effective theory, as a result of the (real part of the) gCS condensate induced by {\it chiral} primordial GW. In the subsequent section \ref{sec:reclass},  as part of the original material of this work, we shall demonstrate - by means of formal path integral methods- the r\^ole of the presence of $N$ sources of GW as a multiplicative factor in front of the real part of the CS condensate induced by the GW due to a single source. This has been conjectured in \cite{Mavromatos:2022xdo}, using hand-waving arguments, but here we shall prove it rather rigorously, within the assumptions made in our perturbative (weak) quantum gravity approach.
In section \ref{sec:imagin}, we shall 
demonstrate in detail the presence of imaginary parts of the gCS condensate, 
 in generic quantum-ordering schemes,  
and show that such parts are linked to quantum commutators of the pertinent chiral GW. These imaginary parts of the effective gravitational action lead to a finite lifetime of the inflationary period.   
The latter is independent of the number of sources of GW, $N$, as it is determined by the imaginary parts of the GW Hamiltonian due to a single source. We shall then estimate the magnitude of these imaginary parts,   and thus the life time of the inflationary vacuum.
Compatibility with the inflationary phenomenology~\cite{Planck:2018jri,Planck:2018vyg,Martin:2013tda} will then constrain the ratio of the energy scales $M_s/M_{\rm Pl}$ in this string-inspired cosmological model.
  Section \ref{sec:slowrollparam} deals with the derivation of the slow-roll parameters of a variant of the StRVM, in which periodic modulations in the axion potential are included, which are induced by non-perturbative stringy instanton effects. The inclusion of such modulations yields better fits of the StRVM to the cosmological inflationary data.
We shall first consider in subsection \ref{sec:gauginst} target-space-gauge-group-instanton-induced periodic modulations of the KR axion potential, which can lead to the correct slow-parameter inflationary phenomenology, upon appropriate fixing of the relevant energy scales.   
In subsection \ref{sec:wsinst},
we shall discuss briefly the r\^ole of world-sheet instantons that characterize other axions that arise from string compactification, which are in general present together with the KR axion in string theories (the KR axion does not couple to world-sheet instantons). 
Finally, conclusions and outlook will be given in section \ref{sec:concl}.   
Some conjectures on the microscopic origin of the observed acceleration of the Universe in the current-era, 
within the context of GW condensation, are also given briefly at the end of the section.
Technical aspects of our approach are discussed in several Appendices. Appendix \ref{sec:bound} deals for completeness with a technical discussion    concerning the necessary conditions for obtaining a consistent variational principle of the CS gravity, by means of the addition of appropriate boundary terms. This discussion also justifies our use of the GW as proper solutions of the gravitational equations.  In Appendix \ref{sec:AppB} we construct the perturbed CS action to second order in GW perturbations around an arbitrary spacetime background, which will give the full graviton propagator to this order. It is then shown that, in the context of a FLRW background, the latter is equivalent to the propagation of two independent scalar modes. Our study contains a careful inclusion of the boundary terms.

\section{Linear Axion Potential from Gravitational-anomaly Condensates: A brief review }\label{sec:cond}

The CS gravitational theory~\cite{Jackiw,Alexander:2009tp,kaloper} introduces a linear coupling of the (pseudo)scalar field $b$, with the so-called mixed anomaly term, that is a specific linear combination of the gauge CS term $\mathbf F \widetilde{\mathbf F}$ and 
the gCS term, $R_{CS}$ (throughout this article Greek indices denote (3+1)-dimensional spacetime indices):\footnote{Our conventions and definitions used throughout this work are: signature of metric $(-, +,+,+ )$, Riemann Curvature tensor 
$R^\lambda_{\,\,\,\,\mu \nu \sigma} = \partial_\nu \, \Gamma^\lambda_{\,\,\mu\sigma} + \Gamma^\rho_{\,\, \mu\sigma} \, \Gamma^\lambda_{\,\, \rho\nu} - (\nu \leftrightarrow \sigma)$, Ricci tensor $R_{\mu\nu} = R^\lambda_{\,\,\,\,\mu \lambda \nu}$, and Ricci scalar $R_{\mu\nu}g^{\mu\nu}$. We also work in units $\hbar=c=1$.} 
\begin{equation}
	S=\int d^4x \,\sqrt{-g}\,  \left[\frac{R}{2\kappa^2}-\frac{1}{2}(\partial_\mu b)(\partial^\mu b) - A\, b\,\Big(R_{CS} + \mathbf F_{\mu\nu} \widetilde{\mathbf F}^{\mu\nu} \Big) + \dots \right]\,,
\label{eq:Action}
\end{equation}
where  
 $A$ is a coupling constant, to be discussed further below. The $\dots$ denote the ``kinetic'' terms of the gauge fields, which will not be important for our discussion until section \ref{sec:slowrollparam}. The quantity $\mathbf F_{\mu\nu}$ is a (non-Abelian, in general) gauge field strength, with 
\begin{align}\label{gaugedual}
\widetilde{\mathbf F}_{\mu\nu} =  \frac{1}{2} \, \varepsilon_{\mu\nu\alpha\beta} \, \mathbf F^{\alpha\beta}
\end{align}
its dual, where $\varepsilon_{\mu\nu\alpha\beta} = \sqrt{-g(x)} \, \hat \epsilon_{\mu\nu\alpha\beta} $ is the covariant Levi-Civita tensor, and $\hat \epsilon_{\mu\nu\rho\sigma}$ denotes the Minkoswki space-time Levi-Civita totally antisymmetric symbol, with the convention 
$\hat \epsilon_{0123}=1$, {\it etc}.

The gCS term $R_{CS}$ in \eqref{eq:Action} is given by:
\begin{equation}
	\label{RCS}
	R_{CS}= \frac{1}{2}R^{\mu}_{\,\,\,\nu\rho\sigma}\widetilde{R}^{\nu\,\,\,\,\rho\sigma}_{\,\,\,\mu},
	\end{equation}
with the symbol $\widetilde{(\dots)}$ denoting the dual of the Riemann tensor, defined as
\begin{equation}
\label{dualriem}
\widetilde{R}_{\alpha\beta\gamma\delta}=\frac{1}{2}R_{\alpha\beta}^{\,\,\,\,\,\,\,\,\rho\sigma}\varepsilon_{\rho\sigma\gamma\delta}\,,
\end{equation}
to be contrasted with the Hodge-dual 
\begin{equation}
\label{hodgedualriem}
^\star{R}_{\alpha\beta\gamma\delta}=\frac{1}{2}R_{\alpha\beta}^{\,\,\,\,\,\,\,\,\rho\sigma}\hat \epsilon_{\rho\sigma\gamma\delta}\,,
\end{equation}
which uses the Minkowskian Levi-Civita symbol $\hat \epsilon_{\mu\nu\rho\sigma}$.
In what follows we shall alternatively make use of both duals, exploring the identity 
\begin{align}\label{identity}
\sqrt{-g} \, R_{\mu\nu\rho\sigma} \, \widetilde R^{\mu\nu\rho\sigma} = R_{\mu\nu\rho\sigma} \, ^\star R^{\mu\nu\rho\sigma} \,.
\end{align}

The mixed anomaly $R_{CS} + \mathbf F \widetilde{\mathbf F} $ is a total derivative~\cite{Alvarez-Gaume:1983ihn,Eguchi:1980jx}. Consequently, the theory \eqref{eq:Action} is shift symmetric, i.e. invariant under global transformations of the pseudoscalar field  
\begin{align}\label{shiftsym}
b\rightarrow b+ {\rm constant}\,. 
\end{align}

The gauge anomaly term $\mathbf F^{\mu\nu} \, \widetilde{\mathbf F}_{\mu\nu}$ is topological, that is independent of variations of the metric tensor, so it does not contribute to the gravitational (Einstein) equations of motion. Par contrast, the gCS term $b\,R_{CS}$ is characterized by a non trivial variation with respect to the metric tensor, namely the Cotton tensor $C_{\mu\nu}$~\cite{Jackiw}: 
\begin{equation}\label{cottdef}
	C_{\mu\nu}=-\frac{1}{2}\nabla^{\alpha}\left[(\nabla^{\beta} b) \widetilde{R}_{\alpha\mu\beta\nu}+(\nabla^{\beta} b) \widetilde{R}_{\alpha\nu\beta\mu}\right]~.
\end{equation}
Variation of the action with respect to the metric  and the axion field~\footnote{For a rather detailed discussion on the well-definiteness of the variational principle in CS gravity we refer the reader to Appendix \ref{sec:bound}.} yields the following equations of motion~\cite{kaloper,Jackiw}
\begin{align}
	\label{grav}
	&G_{\mu\nu}=\kappa^2 T^{(b)}_{\mu\nu}+4 \kappa^2 A C_{\mu\nu}~,\\
	\label{Axion}
	&\square\,b=A \, R_{CS}~,
\end{align}
where $T^{(b)}_{\mu\nu}$ is the  stress energy-momentum tensor associated with the kinetic term of a matter field,
\begin{equation}\label{stressb}
	T^{(b)}_{\mu\nu}=\nabla_\mu b\nabla_\nu b-\frac{1}{2}g_{\mu\nu}(\nabla b)^2~.
\end{equation}
The property of the Cotton tensor~\cite{Jackiw} $C_{\mu\nu}^{\quad ;\mu} = -\frac{1}{4} \, (\partial_\nu b)\, R_{CS}$, where $;$ denotes gravitational covariant derivative (with respect to a torsion-free connection),  implies that the naive covariant conservation of the axion-matter stress tensor fails, 
\begin{align}\label{nonconsstress}
T^{(b)\,\,;\mu}_{\mu\nu} = A (\partial_\nu b) \, R_{CS}\,.
\end{align}
This implies  the non conservation of the naive axion stress tensor, which is physically interpreted as indicating a non trivial exchange of energy between the axion matter and the gravitational field. The $R_{CS}$ term is CP violating,\footnote{The reader should note that, if the field $b$ is a scalar, then the coupled axion-mixed-anomaly interaction term in the action \eqref{eq:Action} violates CP. Par contrast, the latter symmetry is preserved in the case of a pseudoscalar $b$ field, which will be the focus of our interest here.} and, as such, it vanishes for spherically symmetric or isotropic and homogeneous spacetime backgrounds. This is the case when a Friedman-Lemaitre-Robertson-Walker (FLRW) spacetime background is considered, which means that in such a case the modified Lagrangian is equivalent to the Einstein - Hilbert of General Relativity. Consequently, in that case, the scalar field is minimally coupled to gravity and the cosmological evolution is governed by a stiff equation of state $w^b_{\rm stiff}=+1$.

However, when {\it chiral} GW are produced, through non-spherically symmetric coalescence of primordial black holes or collisions of domain walls, the gCS term becomes non-trivial because different helicities of the tensorial perturbations propagate in a different way~\cite{Lue:1998mq,Alexander:2004wk,Alexander:2004us,Lyth:2005jf}. This difference to the wave equations for the left and right - handed polarizations arises because of the presence of the Cotton tensor \eqref{cottdef} in the gravitational equations \eqref{grav}, leading to gravitational-wave birefringence of cosmological origin.

In \cite{Dorlis:2024yqw} the GW were treated quantum-field theoretically (in a weak-gravity setting) by applying the process of second quantization~\cite{Lyth:2005jf}, {\it i.e.} by promoting the perturbations to operators through the definition of the corresponding creation and annihilation operators. In this sense, the gCS term becomes also an operator, $\widehat{R}_{CS}$, which we calculate up to second order in the tensorial perturbations. The gCS operator backreacts onto the effective Lagrangian through its vacuum expectation value (vev), $\langle  R_{CS}\rangle \equiv \langle 0|\widehat{R}_{CS}|0\rangle$, where $|0\rangle$ denotes the appropriate gravitational ground state of the system, and the symbol $\widehat{(...)}$ is used to denote quantum operators. 
If  $\langle R_{CS}\rangle$
acquires a constant value, thus having the form of a (translationally invariant) gravitational condensate, then in this case a linear potential for the KR axion arises, 
\begin{align}\label{effectivepot}
V_{\rm eff}^{\rm lin} = A\, \langle R_{CS}\rangle b\,,
\end{align}
which breaks the shift symmetry \eqref{shiftsym} of the effective gravitational theory \eqref{eq:Action}.
We argued in \cite{Dorlis:2024yqw}, 
by making use of a dynamical-system approach, that the linear-axion potential \eqref{effectivepot} leads to inflation. A linear-axion potential also characterizes axion monodromy in conventional string/brane theory~\cite{silver}, but in our case its origin its entirely different, given that it arises as a result of the formation of the GW-induced condensate of the gravitational anomaly term.

We now come to the string-inspired CS theories~\cite{kaloper}, which we shall restrict our attention upon for the remainder of this article. In particular, 
we shall concentrate on the stringy running vacuum model (StRVM) of cosmology~\cite{bms,ms1,ms2}, 
  which, as already mentioned, is based on the assumption that in the primordial Universe only fields from the massless bosonic gravitational multiplet of the string appear as external fields. In such models, the gauge fields $\mathbf F$, together with other matter excitations, are assumed absent, and generated only at the end of the inflationary period, due to the decay of the metastable ground state (vacuum) of the StRVM~\cite{bms,ms1}. 
The action in this model, which we shall restrict ourselves on in the next couple of sections, is described by \eqref{eq:Action} upon setting $\mathbf F=0$, that is: 
\begin{equation}
	S=\int d^4x \,\sqrt{-g}\,  \left[\frac{R}{2\kappa^2}-\frac{1}{2}(\partial_\mu b)(\partial^\mu b) - A\, b\,R_{CS}  \right]\,.
\label{eq:ActionAB}
\end{equation}

In this class of models the coupling $A$ 
is determined by string theory considerations~\cite{kaloper}, and is 
given by 
\begin{equation}\label{Aval}
    A=\sqrt{\frac{2}{3}}\frac{\alpha^\prime}{48\kappa} = 
  \sqrt{\frac{2}{3}}\frac{M_{\rm Pl}}{48\, M_s^2}   \,.
\end{equation}
Since the StRVM \eqref{eq:Action} is viewed as a low-energy string effective gravitational theory, it is natural to consider that the UV cutoff $\mu$ of the momenta of the pertinent low-energy excitations, including  the (quantum) graviton modes,  
    is provided by the string mass scale~\cite{Mavromatos:2022xdo}, 
\begin{align}\label{muMs}
\mu \approx M_s \le M_{\rm Pl}\,,
\end{align}
so that the transplanckian conjecture is satisfied~\cite{TCH1,TCH2}.

The detailed analysis of \cite{Dorlis:2024yqw}
has then demonstrated that the GW-induced gCS condensate is a complex quantity. 
Its real part 
(to leading order in  the assumed small quantity $|\kappa^2 \, \dot b | \ll 1$) has been estimated to be:
\begin{equation}
      {\rm Re} \langle R_{CS} \rangle^I =-\frac{ A }{\pi ^2}\frac{\dot{b}_I}{M_{\rm Pl}} \left(\frac{H_{I}}{M_{\rm Pl}}\right)^3 \mu ^4<0\,.
      \label{VEV_inflation}
\end{equation}
 Notably, the negative sign in front of the quantity on the right-hand side of the above equality guarantees a positive sign for the  effective cosmological constant. The quantity $H_I$ is the Hubble rate at inflation, whose upper bound has been imposed by the Planck Collaboration data \cite{Planck:2018jri} to have the order of magnitude 
 \begin{align}\label{HI}
 H_I \lesssim  2.5 \times 10^{-5}M_{\rm Pl}\,. 
\end{align}
 We next recall~\cite{Mavromatos:2022xdo,Dorlis:2024yqw} that 
for the actual value of the condensate in the inflationary epoch, we have to multiply the result \eqref{VEV_inflation}
with the number $\mathcal{N}_I$ of all kinds of sources of gravitational waves during inflation, thus obtaining as a final result:
\begin{equation}      
{\rm Re}\left<R_{CS}\right>^{total}_{I}=-\mathcal{N}_I\frac{ A \,  \kappa ^4 \, \mu ^4}{\pi ^2}\, \dot{b}_I \, H_{I}^3\,.
      \label{VEV_inflation_source}
\end{equation}
A detailed proof of this feature, within the path-integral approach of weak quantum gravity, will be given in the next section \ref{sec:reclass}.

In the analysis of \cite{Dorlis:2024yqw} an estimate of the order of magnitude of $\dot{b}$, which remains approximately constant during the entire inflation era, has also been obtained as a consequence of the imposition of appropriate initial conditions in the dynamical-system approach: 
\begin{equation}
    \dot{b}_I\sim 10^{-1} H_I M_{\rm Pl}\,,
    \label{bdotInflation}
\end{equation}
confirming in this way the assumptions made in \cite{bms,ms1,ms2,Mavromatos:2022xdo}, including the assumption on the smallness of $|\kappa^2 \dot b|$ that lead to \eqref{VEV_inflation}.

We stress at this stage that the approximate constancy of the rate of change of the KR axion background in our StRVM approach is due to 
the presence of a gCS condensate $\langle R_{CS}\rangle = {\rm const.} \ne 0$. Indeed, if we recall
that the gCS-Hirzebruch term in \eqref{eq:Action} is a total derivative~\cite{Eguchi:1980jx,kaloper}
\begin{align}\label{totalder}
 R_{CS}= \mathcal J^\mu_{\,\,\,\,\,;\mu}\, ,  
\end{align}
where $;$ denotes covariant derivative in the curved background, and make the following plausible approximation
in our homogeneous and isotropic FLRW inflationary cosmology with an (approximately) constant  Hubble parameter $H$:\cite{bms,ms1}  
\begin{align}\label{approx}
\langle R_{CS} \rangle = \langle 
{\mathcal J}^\mu_{\,\,\,\,\,;\mu}\rangle ~ \simeq ~
\partial_0 \langle {\mathcal J}^0 \rangle + 3 \, H\, \langle {\mathcal J}^0\rangle\,,   
\end{align}
then, from the equations of motion for the $b$-field stemming from the action \eqref{eq:Action}, we obtain (ignoring the gauge terms, as in the model of \cite{bms,ms1})
\begin{align}
\Box b &= A \langle R_{CS}\rangle \quad \Rightarrow \nonumber \\
{\rm for~FLRW~background}: \quad 
 \ddot b + 3 \, H\, \dot b &= A\, \langle R_{CS}\rangle = {\rm constant} \ne 0\,,
\label{eqinflb}
\end{align}
where $\Box$ denotes the covariant D'Alembertian. 
For $H={\rm constant}$, and $\langle R_{CS}\rangle ={\rm constant}$, one obtains from \eqref{eqinflb} that 
\begin{align}\label{bdotconst2}
\dot b = {\rm constant}\,. 
\end{align}
Equivalently, on using \eqref{approx}, we may also write the $b$ equation as:
\begin{align}\label{bdotconst3}
\frac{1}{\sqrt{-g}} \, \partial_0 \Big(\sqrt{-g} \, [\dot b -A\, \langle \mathcal J^0 \rangle ] \Big) =0 \,, 
\end{align}
which also accepts as a solution a constant
$\dot b$ axion background:\footnote{ We cannot resist the temptation of noting at this point that in the action \eqref{eq:Action}, the shift symmetry \eqref{shiftsym} implies that the corresponding Noether current is given by $J_N^\mu = \partial^\mu b - A\, \mathcal J^\mu$, which is thus covariantly conserved on account of the $b$ field Euler-Lagrange field equation \eqref{eqinflb}. The presence of a gCS $R_{CS}$ condensate, though, breaks this shift symmetry.
Thus, although this implies a constant rate $\dot b$ \eqref{bdotconst4}, nonetheless one does not encounter the case of Refs.~\cite{Bakopoulos:2023fmv,Bakopoulos:2023sdm}
in which the exact shift symmetry of the Horndeski-like scalar field, $\varphi$, which exhibits a linear time dependence, leads to a primary hair for the corresponding stationary black hole solution, provided by the $\dot \varphi$ constant rate.  
}
\begin{align}\label{bdotconst4}
 \dot b =  A\, \langle {\mathcal J}^0\rangle = {\rm constant} \ne 0\,.   
\end{align}

Thus, Eq.~\eqref{bdotInflation}, which in \cite{Dorlis:2024yqw} has been obtained as a consistent condition in the dynamical-system approach to linear-axion inflation, 
constitutes, in view of \eqref{eqinflb}, \eqref{bdotconst2}, \eqref{bdotconst3}, a highly non-trivial consistency check of the StRVM approach to inflation via primordial-gravitational-wave-induced CS condensates, which in this way is mapped into a dynamical evolution of a single-(axion) field system with a linear potential. The reader should therefore notice the appearance of linear in time KR axion backgrounds in this StRVM system, as happened in the string-inspired cosmological models of \cite{Antoniadis:1988aa,Antoniadis:1988vi,Antoniadis:1990uu}, but here the origin of this dependence is different. 

  Moreover, in view of \eqref{VEV_inflation_source}, \eqref{bdotInflation}, the gCS condensate is of $\mathcal O(H_I^4)$, which, as already mentioned in the introductory section of the article, makes the corresponding energy density having a form that is encountered in the RVM approach to cosmology~\cite{rvm1,rvm2,rvm3,Lima:2013dmf,Lima:2015mca,rvmdata,rvmqft1,rvmqft2,rvmqft3,rvmqft4,rvmtens,Gomez-Valent:2024tdb}, that is being a function of even powers of the Hubble parameter, as dictated by the underlying general covariance of the formalism. 

In the context of the dynamical-system approach to inflation, the analysis of \cite{Dorlis:2024yqw} also provides a restriction on the magnitude of the string scale $M_s$, compared to $M_{\rm Pl}$:
\begin{equation}
    M_s\sim 10^{-1}M_{\rm Pl}<M_{\rm Pl}\,,
    \label{cutoff}
\end{equation}
which is consistent with the transplanckian censorship hypothesis~\cite{TCH1,TCH2,ms2}.
 
From \eqref{bdotInflation}, \eqref{muMs}, \eqref{Aval},\eqref{HI} and \eqref{cutoff}, we therefore obtain for \eqref{VEV_inflation_source}:
\begin{align}\label{vevRCSfinal}
        {\rm Re}\left<R_{CS}\right>^{total}_{I} \simeq -\mathcal{N}_I\, \sqrt{\frac{2}{3}} \, \frac{1}{480\, \pi^2}
    \, \Big(\frac{M_s}{M_{\rm Pl}}\Big)^2 \, H_I^4 \, \gtrsim\, -6.7 \times 10^{-25} \, \mathcal N_I \, M_{\rm Pl}^4 \,.
\end{align}

Finally, an estimate of the number of sources of GW during the inflationary epoch, $\mathcal N_I$, in terms of the corresponding number $\mathcal N_S$ during the stiff-axion era, which, in the StRVM Cosmology, precedes the inflationary era~\cite{bms,ms1,ms2}, is also provided
as~\cite{Dorlis:2024yqw}
\begin{equation}\label{nosourcesI}
    \frac{\mathcal{N}_I}{\mathcal{N}_S}\sim 7\cdot 10^{16}\,,
\end{equation}
which lies in the range given in \cite{Mavromatos:2022xdo}, stemming from the assumption of the constancy of the CS condensate during inflation, which thus avoids exponential dilution~\cite{bms,ms1}. For our purposes we may take without loss of generality that $\mathcal N_S = \mathcal O(1)$.

  We  proceed in the next section to a detailed discussion on the emergence of the enhancement \eqref{vevRCSfinal} of the value of the real part of the gCS condensate 
by the number of GW sources, following a path integral approach in the framework of  (gauge fixed) weak quantum gravity of GW perturbations.

\section{Effective (mean-field) Theory from "re-classicalization" and  the effect of the number of GW sources}\label{sec:reclass}
We now turn our attention to the path integral formulation which, as we shall see, can indeed lead to a better physical insight   on the StRVM approach to Cosmology, albeit there are still several formal questions that remain unanswered, especially the ones related to the still wide-open issue of gauge invariance in quantum gravity~\cite{Batalin:1981jr,Batalin:1983ggl,Barnich:2000zw}. \par 
The full path integral of the StRVM theory reads, 
\begin{equation}
    \mathcal{Z}=\int \mathcal{D}b\;\mathcal{D}g \;\;e^{iS[b,g]}
\end{equation}
where $S[b,g]$ is given by \eqref{eq:ActionAB}. This, of course, is a formal expression, since both the measure and the action of quantum gravity are still not known. However, since in our case we are dealing with an effective low-energy gravitational theory obtained from strings~\cite{Gross:1986mw,Metsaev:1987zx,str1,str2,pol1,pol2}, the above quantum theory is valid only at its semiclassical approximation. In general, when assuming the semiclassical approximation, the partition function becomes a functional of the background fields, 
\begin{equation}
     \mathcal{Z}=e^{iW[g,b]}
\end{equation}
where $W[g,b]$ denotes the effective action of the background fields, which obeys:
\begin{equation}\label{station}
    -i\frac{\delta \ln \mathcal{Z}}{\delta b}=\frac{\delta W[g,b]}{\delta b}=0\,,\qquad -i\frac{\delta \ln \mathcal{Z}}{\delta g_{\mu\nu}}=\frac{\delta W[g,b]}{\delta g_{\mu\nu}}=0\,,
\end{equation}
at appropriate saddle points of the path integral. 

Upon ignoring any perturbation of the fields, the functional $W[g,b]$ can be easily identified with the effective action, $W[g,b]= S^{(0)}[g,b]$,
where $S^{(0)}[g,b]$ is the action in the background spacetime without perturbations. However, upon considering  variations of the metric about the stationary points \eqref{station}, as we did previously in the form of GW~\cite{Dorlis:2024yqw}, we obtain
\begin{equation}
    S[b,g+h]= S^{(0)}[b,g] + S^{(2)}[b,g,h] +\dots,
\end{equation}
where $S^{(0)}[b,g]$ is the unperturbed  background action \eqref{eq:ActionAB}, and 
$S^{(2)}[b,g,h]$ denotes the action up to second order in the metric variations, $h_{\mu\nu}$, with the latter treated as weak quantum variables (fluctuations).   This order in the expansion about the stationary point suffices for the purposes of this work.   
Then, the path integral has to be of the following form: 
\begin{equation}
    \mathcal{Z}= e^{i S^{(0)}[b,g]} \int \mathcal{D}h\;\; e^{iS^{(2)}[b,g,h]}=e^{i S^{(0)}[b,g]} \mathcal{Z}_h[b,g]
\end{equation}
where $b$ and $g$ act as background fields, while $S^{(2)}[b,g,h]$ is the action up to second order in the metric perturbations and $\mathcal{Z}_h$ denotes the path integral of the GW perturbations,
\begin{equation}
 \mathcal{Z}_h[b,g]= \int \mathcal{D}h\;\; e^{iS^{(2)}[b,g,h]}\,.
\end{equation}
From now on we focus on the FLRW spacetime background, in the presence of chiral GW perturbations. In such a case, 
$h$ is an abbreviation for the two independent modes, $h_{L}$ and $h_R$ of the GW ({\it cf.} section \ref{sec:imagin}). These perturbations correspond to a GW induced by some source. In this case,  $S^{(2)}[b,g,h]$ denotes the action of the GW, which has the form,  
\begin{equation}
    S^{(2)}[b,g,h] \equiv 
    S^{(2)}[b,g,h]\Big |_{A=0} - 
    \int d^4x \, A \, b \, \mathcal{O}_h \,.  
\end{equation}
In the above expression, $S^{(2)}[b,g,h]\Big |_{A=0}$ denotes the terms of the gravitational action \eqref{eq:Action}, 
perturbed to second order, 
in the {\it absence} of gCS and chiral gauge anomalies, which is derived in Appendix \ref{sec:AppB} ({\it cf.} Eq.~\eqref{GravandMatterWithoutGaugeFixing}). 
The quantity
$\mathcal{O}_h=R_{\mu\nu\rho\sigma}\!^*R^{\mu\nu\rho\sigma}$, with $R_{\mu\nu\rho\sigma}\,^*R^{\nu\mu\rho\sigma}$ expressed up to second order in the (weak) GW perturbations, and we used \eqref{identity}, \eqref{hodgedualriem}. Then, the path integral reads, 
\begin{equation}
    \mathcal{Z}= e^{iS^{(0)}[b,g]} \int \mathcal{D}h \; e^{iS^{(2)}[b,g,h]\Big|_{A=0}- i A\, \int d^4x \, b\, \mathcal{O}_h} \,,
\end{equation}
implying that the effective action, upon assuming metric perturbations, $W[b,g]$, differs from $S^{(0)}[b,g]$. In order to find $W[b,g]$, we take the functional derivative (assuming that the variations with respect to the background fields commute with the measure of the GW path-integration)
\begin{align}\label{N=1station}
 0 &= -i \frac{\delta \ln \mathcal{Z}}{\delta b}=
\frac{\delta S^{(0)}[b,g]}{\delta b} 
+ 
    \frac{1}{\mathcal Z}\, \int \mathcal{D}h  
    \Big(\frac{\delta S^{(2)}[b,g,h]\Big|_{A=0}}{\delta b} 
-A\;\mathcal{O}_h \Big)\; e^{iS^{(2)}[b,g,h]\Big|_{A=0}- i A\, \int d^4x \,\mathcal{O}_h \, b} 
\nonumber \\
  &= \frac{\delta S^{(0)}[b,g]}{\delta b} 
  +    \Big\langle \frac{\delta S^{(2)}[b,g,h]\Big|_{A=0}}{\delta b} \Big\rangle   
  -A\langle\mathcal{O}_h\rangle        
\end{align}
where 
\begin{equation}\label{Oaver}
  \langle\mathcal{\mathcal O}_h\rangle= \frac{1}{\mathcal{Z}}   \int \mathcal{D}h \;\mathcal{\mathcal O}_h\; e^{iS^{(2)}[b,g,h]\Big|_{A=0}-A i \int d^4x \, \mathcal{\mathcal O}_h \, b}\,,     
\end{equation}
is the vev of the composite operator, $\mathcal{\mathcal O}_h$, which  contains the metric perturbations. 
Such a vev  could depend only on the background metric and in the case of FLRW background only on even powers of  the Hubble parameter, $H$, as implied by  diffeomorphism invariance of the theory~\cite{rvm1}. 

  We now proceed to incorporate a number of GW sources $\mathcal N > 1$ in the path integral. This is introduced by treating the GW coming from various sources as ``an ideal gas of non - interacting GW perturbations"- in the sense that the partition function of the $\mathcal N$ sources of GW is constructed as a product of $\mathcal N$ partition functions
of the individual GW:
\begin{equation}\label{ZN}
\mathcal{Z}=e^{iS^{(0)}[b,g]} \int\prod_{i=1}^{\mathcal N} \mathcal{D}h_i\;\;e^{i\sum_iS^{(2)}[b,g,h_i]\Big|_{A=0}- i \sum_i\int d^4x \;A\;\mathcal{O}_{h_i} b} = e^{iS^{(0)}[b,g]} \left[   \int \mathcal{D}h \; e^{iS^{(2)}[b,g,h]\Big|_{A=0}-A i \int d^4x \mathcal{O}_h b}      \right]^{\mathcal N} \  .
\end{equation}
This explains,  
within the path integral formulation, the linear superposition of GW effects assumed in \cite{Mavromatos:2022xdo}.

Then, following the same procedure as in the single GW case above, we have, 
\begin{align}
    0 &= -i \frac{\delta \ln \mathcal{Z}}{\delta b}=\frac{\delta S^{(0)}}{\delta b} + 
    \mathcal{N}\, \Big\langle \frac{\delta S^{(2)}[b,g,h]\Big|_{A=0}}{\delta b} \Big\rangle 
     -A \, \mathcal{N}\, \langle\mathcal{O}_{h}\rangle
     \,,
\label{eqbeforeeffectiveaction}
\end{align}
where by the symbol $\langle \dots \rangle $ we have defined the following vev of a composite operator $\mathcal B [\dots, h_i]$ (for every GW perturbation $h_i$)
\begin{equation}
\langle\mathcal{B}[\dots, h_i]\rangle=  \frac{\int\mathcal{D}h_i \;\mathcal{B}[\dots, h_i]\, e^{i\sum_iS^{(2)}[b,g,h_i]\Big|_{A=0}-A i \sum_i\int d^4x \mathcal{O}_{h_i} b}}{\int\mathcal{D}h_i\;e^{iS^{(2)}[b,g,h_i]-A i \int d^4x \mathcal{O}_{h_i} b}} \,\,,
\end{equation}
which obviously depends only on the background fields, and as such we denote it simply as $\langle\mathcal{B} [\dots, h]\rangle$ or
$\langle\mathcal{B}_h\rangle$, i.e. we omit  the index referring to each GW perturbation, and instead we use a generic $h$. Therefore, Eq.~\eqref{eqbeforeeffectiveaction} reads, 
\begin{align}\label{stationb}
   0 = -i \frac{\delta \ln \mathcal{Z}}{\delta b}&=\frac{\delta S^{(0)}[b,g]}{\delta b} 
   + 
    \mathcal{N}\, \Big\langle \frac{\delta S^{(2)}[b,g,h]\Big|_{A=0}}{\delta b} \Big\rangle 
   -A \mathcal{N}\langle\mathcal{O}_{h}\rangle
   \,.
\end{align}

  Introducing the notion of the proper number density $n$ of GW sources in the spacetime background corresponding to the metric $g_{\mu\nu}$:
\begin{equation}
  n=\mathcal{N}/\sqrt{-g}\,,
    \label{properdensity}
\end{equation}
and using the identity \eqref{identity}, we define the following quantity,  
\begin{equation}\label{Ncond}
\langle R_{\mu\nu\rho\sigma}\;^*R^{\mu\nu\rho\sigma}\rangle_n \equiv n\langle R_{\mu\nu\rho\sigma}\;^*R^{\mu\nu\rho\sigma}\rangle\,,
\end{equation}

  We next make the assumption on the approximate constancy of the gCS condensate during inflation (which is confirmed by explicit calculations in our case~\cite{bms,ms1,ms2,Mavromatos:2022xdo,Dorlis:2024yqw})
\begin{align}\label{Ncondconst}
\langle R_{\mu\nu\rho\sigma}\;^*R^{\mu\nu\rho\sigma}\rangle_n=~{\rm constant}\,.
\end{align}

 As we have already shown in previous works~\cite{Mavromatos:2022xdo}, the introduction of the sources is crucial in order to 
satisfy \eqref{Ncondconst}
but also not to dilute the condensate during inflation~\cite{bms,Mavromatos:2022xdo}, while respecting \eqref{muMs}. Assuming that these conditions are valid (i.e. assuming a sufficiently large number of sources),  indeed an  effective linear potential for the axion is introduced. In this interpretation, such a linear potential comes from a ``re-classicalization" process of the quantum perturbations around the classical background.   The term ``re-classicalization" is used here to stress that, upon its formation, the condensate $\langle R_{\mu\nu\rho\sigma}\;^*R^{\mu\nu\rho\sigma}\rangle$ acts as a ``classical" source for the axion,    affecting in this way its already classical background, and producing an effective linear potential for the (pseudo-) scalar field.

The reader should then observe that any variation of the condensate with respect to the background fields vanishes.  
As a consequence, a linear potential for the axion 
arises leading to inflation, as discussed in detail in~\cite{Dorlis:2024yqw}.
We recall from our previous analysis in that work
that the number of GW sources is not constant throughout the evolution of the primordial universe, but it necessarily increases abruptly as we pass from the stiff to the RVM inflationary epoch \eqref{nosourcesI}. This is an essential behavior to guarantee the onset of inflation in our approach. The nature of the GW sources depends on the underlying microscopic model. There are various mechanisms for GW production during the pre-inflationary era, ranging from non-spherically-symmetric merging of primordial black holes to  annihilation/non-spherically-symmetric collapse of domain walls~\cite{ms1}. 

Using \eqref{Ncondconst} we now apply a mean-field Hartree-Fock (HF) approximation~\cite{HF1,HF2,HF3} for the evaluation of the source path integral \eqref{ZN}. According to the HF treatment, the gCS term is split into a condensate term and quantum fluctuations about it (denoted by the symbol $: \dots :$)~\cite{bms,ms1}:
\begin{align}\label{HF}
\int d^4x \, \mathcal{N}\, \mathcal{O}_h b =  \int d^4x \sqrt{-g} \, \langle R_{\mu\nu\rho\sigma}\;^*R^{\mu\nu\rho\sigma}\rangle_n \,  b \, + :\int d^4x \, \mathcal{N}\, \mathcal{O}_h \, b :\,,     
\end{align}
where we used the identity \eqref{identity}, and the definition \eqref{Ncond}, with the  property \eqref{Ncondconst} in mind. On account of \eqref{HF}, 
the corresponding variational equations with respect to the $b$-axion
and graviton fields, obtained from \eqref{ZN}, read:
\begin{align}
    b-{\rm axion~equation:} \quad 0 &=\frac{\delta S^{(0)}}{\delta b} 
    - A \, \langle R_{\mu\nu\rho\sigma}\;^*R^{\mu\nu\rho\sigma}\rangle_n
    + 
    \mathcal{N}\, \Big\langle \frac{\delta S^{(2)[b,g,h]\Big|_{A=0}}}{\delta b} \Big\rangle 
     -A \, \mathcal{N}\, \langle : \mathcal{O}_{h} :\rangle
     \,, \nonumber \\
     {\rm graviton~equation:} \quad 0 &= \frac{\delta S^{(0)}}{\delta g_{\mu\nu}} 
     + \frac{A}{2}g^{\mu\nu} \langle R_{\mu\nu\rho\sigma}\;^*R^{\mu\nu\rho\sigma}\rangle_n \, b 
    + 
    {\mathcal N}\Big\langle \frac{\delta S^{(2)}[b,g,h]\Big|_{A=0}}{\delta g_{\mu\nu}} \Big\rangle  
    - {\mathcal N}\, A\, \Big\langle \frac{\delta}{\delta g_{\mu\nu}} \int d^4 x\, : O_h \, b : \Big\rangle \,.
\label{HFstation}
\end{align} 
The normal-ordered ($: \dots :$) terms 
of the composite operators bring in new UV divergencies, which in general may lead to background-dependent terms in both equations \eqref{HFstation}. Such terms can be arranged to vanish by an appropriate choice of counterterms.
This lies at the heart of our mean-field HF treatment, which entails ignoring quantum fluctuations about the gCS condensate. 
Moreover, from our analysis in Appendix \ref{sec:AppB} ({\it cf.} Eq.~\eqref{mattercontributiontogravitonlagrangian}), we observe that 
in the Traceless-Transverse (TT) gauge of GW \eqref{TTgauge}, the third term on the right-hand-side of the $b$-axion equation \eqref{HFstation} vanishes, in a self consistent way to second order in the graviton perturbation, in which the corrections $\mathcal S^{(2)}$ are gauge invariant, as discussed in Appendix \ref{sec:AppB}.

On account of \eqref{Ncondconst}, it is evident that an (approximate) linear-axion 
contribution to the effective stress tensor arises during the inflationary period ({\it cf.} Eq.~\eqref{VEV_inflation_source}), which is equivalent to an approximate positive-cosmological-constant (de-Sitter type) 
contribution, since, for the duration of inflation, the $b$ field remains of the same order of magnitude,  
as explained in \cite{bms,ms1,ms2,Mavromatos:2022xdo,Dorlis:2024yqw}.

  At this stage we would like to make some brief but important remarks regarding the limitations of our results. The above analysis was somewhat formal and did not deal with the important issue of gauge invariance of the quantum gravitational path integral. 
The estimates \eqref{VEV_inflation_source}
of the gravitational CS condensate in our approach~\cite{bms,Dorlis:2024yqw} have been performed in a specific gauge (TT gauge \eqref{TTgauge} of GW) within the framework of weak quantum gravity. 
Unfortunately, due to the lack of knowledge of the full theory of quantum gravity, we cannot demonstrate gauge invariance of our condensate at a full non-perturbative level. Such a task would require the application of methods as in \cite{Batalin:1981jr,Batalin:1983ggl,Barnich:2000zw,Grassi:2024kif}, 
 which, in the current state of affairs of our model, lies well beyond the scope of the current article. Our aim here was to provide evidence for the existence of a complex non-zero condensate of the gravitational CS anomaly term in the effective action \eqref{eq:Action} during inflation, and estimate the life time of the associated unstable vacuum. Hopefully, when a fully UV completion of the model is at hand, which in our case implies a specific string model of phenomenological relevance to particle physics,  our estimates will not change much quantitatively.

The reader should then notice that in this HF form, the equations \eqref{HFstation}
include the (classical) background parts, associated with the respective variations of the $S^{(0)}$ part of the effective action (not involving GW corrections), plus non-trivial averages of the quadratic GW corrections, such as the third term on the right-hand-side of the graviton equation in \eqref{HFstation}, which expresses the quantum average of the contribution of the second order GW perturbations to the GW stress-energy tensor. The latter is {\it not} gauge invariant. From dimensional considerations, we expect such terms to yield corrections to the Ricci and scalar-curvature parts of the Einstein tensor, plus a potential vacuum-energy-like contribution. If we restrict ourselves to terms of second-order in derivatives, for consistency with our discussion so far, such vacuum energy contributions in a FLRW background are expected to be of order $\mathcal O(H^2)$, where $H$ is the Hubble parameter. 
Such terms would be subdominant at early eras of the Universe, like the RVM inflation we are interested in here, as compared to the $\mathcal O(H^4)$ terms coming from the gCS condensate $\frac{A}{2}g^{\mu\nu}\, \langle R_{\mu\nu\rho\sigma}\;^*R^{\mu\nu\rho\sigma}\rangle_\mathcal{N} $ appearing in the graviton equation \eqref{HFstation}. However, their presence would complete the description of the energy density of the vacuum in this StRVM approach, which would thus be in agreement with the considerations of \cite{bms,ms1}. In this work we shall not proceed further to evaluate such terms, which will be the subject of a forthcoming publication. We may conjecture though that, if the above terms are indeed of order $H^2$, then one might encounter similar contributions to the 
vacuum energy density in modern eras, 
from appropriately condensing sources of GW in the current epoch, in analogy to the inflationary era.
Such terms, of order 
$\mathcal O(H_0^2)$
(where $H_0$ denotes today's value of the Hubble parameter), would thus provide a contribution to today's dark energy, according to  generic RVM features~\cite{rvm1,rvm2,rvmdata,Gomez-Valent:2024tdb}, which the StRVM would share. 

In the next section we shall evaluate the imaginary parts of the gCS condensate, thereby proving the metastability of the inflationary vacuum, and provide an estimate of the associated duration of inflation. On comparing with cosmological data~\cite{Planck:2018jri,Planck:2018vyg,Martin:2013tda}, then, we shall constrain the string scale $M_s$ of the StRVM.

\section{Quantum-Ordering Ambiguities and Metastability of the gCS-Condensate-Induced Inflation}\label{sec:imagin}

In order to estimate the gravitational anomaly condensate, we use canonical quantization techniques. Applying such methods for GWs in an FLRW background, we are also able to investigate the quantum effects that show up regarding the behavior of the composite operator of the Hirzebruch signature, 
\begin{align}\label{Hirop} \widehat{R_{\mu\nu\rho\sigma}^{\ \ \ \ \ *}R^{\mu\nu\rho\sigma}}\,.
\end{align}
As we shall show, this operator {\it may} contain imaginary parts, which are absent in the classical theory. In our approach these imaginary parts play an important r\^ole in inducing metastability of the inflationary quantum vacuum, and thus a finite duration of  inflation
in the StRVM. Such a metastability is in agreement with the swampland criteria of de-Sitter-like inflation in string theory~\cite{swamp1,swamp2,swamp3,swamp4,swamp5,srvmswamp}.\footnote{It is worthy of mentioning at this 
stage that the complex nature of the composite Hirzebruch operator \eqref{Hirop} has been noted in passing in \cite{Lyth:2005jf}, but the authors attributed this behavior to the naive definition of the operator, implying that one should consider the hermitian conjugate so as to eliminate the imaginary parts. Par contrast, in our approach, as we shall discuss below, we attribute a non-trivial physical significance to the complex nature of \eqref{Hirop}, associated with the metastable nature of the quantum StRVM vacuum.}\par 
The method of quantizing GWs followed in this work lies on the fact that the tensorial metric perturbations have two degrees of freedom. Then, one can show that their evolution is equivalent with that of two independent scalar modes (see Appendix \ref{sec:AppB} for a detailed analysis in the case of a background with a minimally coupled scalar field). In this way, the gauge-fixing issues encountered in the quantization of gauge theories, such as the theory of the graviton field, can be bypassed by identifying the two scalar modes and quantize them independently using appropriate mode expansions \cite{Ford:1977dj}. The case of the 4D Chern-Simons gravity is similar. The graviton in this theory also contains two degrees of freedom~\cite{Jackiw}, and the quantization procedure, as we shall demonstrate below, can be applied in a similar fashion \cite{Lyth:2005jf, Alexander:2004us, Creque-Sarbinowski:2023wmb,Alexander:2004wk}. The main difference occurs because of the derivative couplings arising due to the Chern-Simons interaction and especially due to the presence of the Levi-Civita tensor. Due to the latter, such derivative couplings contain odd powers of spatial derivatives, thereby inducing a dependence on the direction of propagation, which explains the parity-violating nature of $R_{CS}$, but also a form of derivatives with respect to the conformal time, implying ordering ambiguities upon quantization. As we shall discuss below, this implies the presence of (quantum-operator-ordering dependent) imaginary parts.   \par 

For a spatially-flat FLRW spacetime,
\begin{equation}
    ds^2=-dt^2+\alpha^2(t)\delta_{ij}dx^idx^j
\end{equation}
where $\alpha(t)$ denotes the scale 
factor, the tensor perturbations (GWs) have the following form: 
\begin{equation}
    ds^2= -dt^2 + \alpha^2(t) (\delta_{ij}+h_{ij})dx^idx^j\; , \quad i,j =1,2,3\,.
    \label{FLRWmetricPerturbed}
\end{equation}
We can express $h_{ij}$ in the linear polarization basis \cite{Misner:1973prb}, expressed as:
\begin{equation}
    h_{ij}=h_+ \epsilon^{(+)}_{ij} + h_\times \epsilon^{(\times)}_{ij} \,,
    \label{plus_cross_polarizations}
\end{equation}
where the polarization tensors are defined through:
\begin{align}
  &  \epsilon_{ij}^{(+)}= [e_1(\vec{k})]_i[e_1(\vec{k})]_j-[e_2(\vec{k})]_i[e_2(\vec{k})]_j \,,\\
  &\epsilon_{ij}^{(\times)}= [e_1(\vec{k})]_i[e_2(\vec{k})]_j+[e_1(\vec{k})]_j[e_2(\vec{k})]_i \,,
\end{align}
where $(e_1(\vec{k}),\, e_2(\vec{k}), \, e_3(\vec{k}))$, with $e_3(\vec{k})=\vec{k}/\vert\vec{k}\vert$,  form a right-handed orthogonal triad of unit vectors. In the presence of the gCS coupling, it proves more appropriate to expand the perturbation \eqref{plus_cross_polarizations} with respect to the helicity basis tensors $\epsilon_{ij}^{L,R}$, as~\cite{Lyth:2005jf},
\begin{equation}
    h_{ij}(t,\Vec{x})=h_L \ \epsilon_{ij}^{(L)}+h_R \ \epsilon_{ij}^{(R)}=\sum_{\lambda=L,R} h_\lambda(t,\vec{x})\epsilon_{ij}^{(\lambda)}\,,
    \label{helicityexpansion}
\end{equation}
where $\epsilon_{ij}^{L,R}$ are defined as follows:
\begin{equation}
  \left [ \epsilon_{ij}^{(R)}\right]=\frac{1}{\sqrt{2}}\left( \left[\epsilon^{(+)}_{ij}\right]+ i \left[\epsilon^{(\times)}_{ij} \right] \right)=\left[ \epsilon_{ij}^{(L)}\right]^\dagger
\label{helicitybasis}\,.
\end{equation}  
The polarization tensors obey the following normalization:
\begin{equation}
\epsilon^{*(\lambda)}_{ij}\epsilon_{(\lambda^\prime)}^{ij}=2\delta_{\lambda\lambda^\prime}\,.
\end{equation}
where $\lambda=L,R$ denotes Left and Right polarization. We can then obtain the following property:
\begin{equation}
k_j\epsilon^{ijk}e^{(\lambda^\prime)}_{kq}=l_{\vec{k}}l_{(\lambda^\prime)}i\vert\vec{k}\vert e^{{(\lambda^\prime)}\;i}_{\;\;\;q}
\end{equation}
with $l_R=+1$, $l_L=-1$, while also \cite{Alexander:2004wk},
\begin{equation}
 l_{\vec{k}}=  \left\{
\begin{array}{ll}
      +1 & \theta< \pi/2 \\
     -1 &  \theta> \pi/2 \\
\end{array} 
\right. ,
\end{equation}
where $\theta$ denotes the polar angle of the arbitrary vector $\vec{k}$. Moreover, by construction, the following relations hold:
\begin{align}
   & e^{(L,R)}_{ij}(\vec{k})=e^{(L,R)}_{ij}(-\vec{k})\label{prop1}\\
    &e^{(L)}_{ij}(\vec{k})\;e^{(L)\; ij}(\vec{k})=e^{(R)}_{ij}(\vec{k})\;e^{(R)\; ij}(\vec{k})=0\label{prop2}\\
   & e^{(L)}_{ij}(\vec{k})e^{(R)\; ij}(\vec{k})=2\, .
   \label{prop3}
\end{align} 
Expanding the gravitational perturbations in Fourier space and helicity basis, we obtain:
\begin{equation}
    h_{ij}(\eta,\vec{x})=\frac{1}{(2\pi)^{2/3}}\int d^3k \sum_{\lambda=L,R}e_{ij}^{(\lambda)}(\vec{k})h_{\lambda,\vec{k}}(\eta)e^{i\vec{k}\cdot\vec{x}} \, .
    \label{FourierTensor}
\end{equation}
We may now show in an analytical way the derivation of the Chern - Simons part of the action, starting from the relative interaction term:
\begin{equation}
     S_{CS}^{int} = -\frac{1}{2}\int d^4 x \ A \ b \ R_{\mu\nu\rho\sigma}^{\ \ \ \ \ *}R^{\nu\mu\rho\sigma} \ .
     \label{CS_part_action}
\end{equation}
Up to second order in $h_{ij}$, $R_{\mu\nu\rho\sigma}^{\ \ \ \ \ *}R^{\nu\mu\rho\sigma}$ is expressed as~\cite{Lyth:2005jf}:
\begin{equation}
    R_{\mu\nu\rho\sigma}^{\ \ \ \ \ *}R^{\nu\mu\rho\sigma} =    2 \hat{\epsilon}^{ijk}\left( \partial_l h^{\prime m}_{j} \partial_m \partial_i h_{k}^l - \partial_l h^{\prime}_{jm}\partial^l \partial_i h_{k}^m + h^{\prime \prime}_{jl}\partial_i h^{\prime 
 l}_{k} \right)\, ,
    \label{R*R_hija}
\end{equation} 
where the prime denotes differentiation with respect to the conformal time $\eta$. Keeping only the antisymmetric part with respect to the indices $i$ and $j$, performing partial differentiations and using the cyclic permutation symmetry of $\hat{\varepsilon}^{ijk}$, we obtain,
\begin{equation}
    R_{\mu\nu\rho\sigma}^{\ \ \ \ \ *}R^{\nu\mu\rho\sigma} =  \partial_\eta \left[  \hat{\varepsilon}^{ijk}\left(\partial_l h^m_j \partial_m\partial_i h^l_k -\partial_l h_{jm}\partial^l\partial_i h^m_k+h^\prime_{jl}\partial_i h^{\prime l}_k \right)   \right] +\partial_i\left[\hat{\varepsilon}^{ijk}\left(\partial_lh^{\prime m}_k\partial^l h_{jm}-\partial_l h^m_j\partial_mh^{\prime l}_k - h^\prime_{jl}h^{\prime\prime l}_k\right)\right]\, , 
    \label{R*R_hij}
\end{equation} 
that is, a total divergence, $R_{\mu\nu\rho\sigma}^{\ \ \ \ \ *}R^{\nu\mu\rho\sigma}=\partial_\mu J^\mu$, with $J^\mu=\sqrt{-g}\mathcal{J}^\mu$ and $\mathcal{J}^\mu$ the covariant current in \eqref{totalder}.
Thus, combining with the part of the Einstein-Hilbert action given in  \eqref{S_2_A_zero1} ({\it cf.} Appendix \ref{sec:AppB}), we obtain,
\begin{equation}
    S_{GW} = \frac{1}{8\kappa^2}\int d^4x \;\left[ \alpha^2(\eta) \left(h_{ij}^\prime h^{\prime\,ij} -\partial_k h_{ij}\partial^k h^{ij}\right) + 4\,A\, \kappa^2\,b^\prime  \hat{\varepsilon}^{ijk}\left(\partial_l h^m_j \partial_m\partial_i h^l_k -\partial_l h_{jm}\partial^l\partial_i h^m_k+h^\prime_{jl}\partial_i h^{\prime l}_k \right)   \right] + S_B\, ,
    \label{ActionTensor}
\end{equation}
where $S_B$ denotes boundary terms,
\begin{align}
    &S_B=S^{(1)}_B+S^{(2)}_B\\
    &S^{(1)}_B=A\int d^4x \,\partial_\eta\left[  b\,  \hat{\varepsilon}^{ijk}\left(\partial_l h^m_j \partial_m\partial_i h^l_k -\partial_l h_{jm}\partial^l\partial_i h^m_k+h^\prime_{jl}\partial_i h^{\prime l}_k \right)     \right]\\
    &S^{(2)}_B=A\int d^4x \,\partial_i\left[ b\, \hat{\varepsilon}^{ijk}\left(\partial_lh^{\prime m}_k\partial^l h_{jm}-\partial_l h^m_j\partial_mh^{\prime l}_k - h^\prime_{jl}h^{\prime\prime l}_k\right)   \right]\, . \label{relB}
\end{align}

Substituting \eqref{FourierTensor} into \eqref{ActionTensor}, we obtain the classical action of GW in Fourier space, 
\begin{equation}\label{GWact}
    S_{GW}=\frac{1}{4\kappa^2}\sum_{\lambda=L,R}\int d\eta \int d^3\vec{k} \,z^2  _{\lambda,\vec{k}}(\eta) \left( 
      \vert h_{\lambda,\vec{k}}^\prime\vert^2-k^2\vert h_{\lambda,\vec{k}}\vert^2
\right)
\end{equation}
where, 
\begin{equation}
    z_{\lambda,\vec{k}}^2(\eta)=\alpha^2(\eta)\left(1-l_\lambda l_{\vec{k}} L_{CS}(\eta)\right) \, 
\end{equation}
with
\begin{equation}
    L_{CS}(\eta)=k \xi\,\, , \,\, \xi=\frac{4A b^\prime\kappa^2}{\alpha^2}, \, [\xi]=M^{-1}\, .
\end{equation}
The above considerations are classical. Canonical quantization of GW perturbations in our context has been discussed in detail in \cite{Dorlis:2024yqw}, following and extending the analysis in \cite{Alexander:2004us,Lyth:2005jf}. Basically, by representing the tensor perturbation $h_{ij}$ in \eqref{FourierTensor} as:
\begin{align}
  h_{ij}(\eta, \vec x) = \kappa \sum_{\lambda=L,R} \int \frac{d^3k}{(2\pi)^{3/2}} e^{i\vec k \cdot \vec x} \, \frac{\psi_{\lambda, \vec k}(\eta)}{a(\eta)\, \sqrt{1 - \ell_\lambda\, \ell_{\vec k} \, k\, \xi }}\, \epsilon^\lambda_{ij}\,, \quad \xi=\frac{4 \, A\, b^\prime (\eta) \, \kappa^2}{a(\eta)^2}   \,,
\end{align}
where $a(\eta)$ is the scale factor of the 
expanding inflationary Universe, the quantization of $h_{ij}$ is achieved by considering the Fourier transforms of the
``scalar'' field operators~\cite{Lyth:2005jf,Dorlis:2024yqw} 
\begin{align}\label{defscal}
    &\widehat \phi (\eta, \vec x) \equiv \psi_L(\eta, \vec x) = \int \frac{d^3k}{(2\pi)^{3/2}}\, e^{i\vec k \cdot \vec x}\, \widehat{\widetilde \phi_{\vec k}}(\eta)\,,
    \nonumber \\
    &\widehat \phi^\star (\eta, \vec x) \equiv \psi_R(\eta, \vec x) = \int \frac{d^3k}{(2\pi)^{3/2}}\, e^{i\vec k \cdot \vec x}\, \widehat{\widetilde \phi^\star_{-\vec k}}(\eta)\,, \nonumber \\
    &\widehat{\widetilde{\phi}}_{\vec{k}}(\eta) =\widetilde{v}_{\vec{k}}\hat{\alpha}_{\vec{k}}^-+v^*_{-\vec{k}}\hat{b}_{-\vec{k}}^+\,, \nonumber\\
   &\widehat{\widetilde{\phi}}^*_{-\vec{k}}(\eta) =v_{\vec{k}}\hat{b}_{\vec{k}}^- +\widetilde{v}^*_{-\vec{k}}\hat{\alpha}_{-\vec{k}}^+\,, 
\end{align}
where we have introduced two sets of creation and annihilation operators, $\alpha^{\pm}_{\vec{k}}$ and $b^{\pm}_{\vec{k}}$, for which $(\alpha_{\vec{k}}^-)^\dagger=\alpha_{\vec{k}}^+$ and $(b_{\vec{k}}^-)^\dagger=b_{\vec{k}}^+$, obeying the commutation relations:
\begin{equation}
\left[\hat{\alpha}_{\vec{k}}^-\;,\;\hat{\alpha}_{\vec{k}^\prime}^+ \right]=\left[\hat{b}_{\vec{k}}^-\;,\;\hat{b}_{\vec{k}^\prime}^+\right]=\delta^{(3)}(\vec{k}-\vec{k}^\prime)\,.
\label{creationcommutation}
\end{equation}

At this point we turn our attention to the quantization issues of the anomaly term \eqref{R*R_hija}. In the helicity basis \eqref{helicityexpansion}, and choosing, without loss of generality, the $z - axis$ as the direction of GW propagation, the CS term has the following structure 
\cite{Dorlis:2024yqw,Alexander:2004us,Lyth:2005jf}:
\begin{equation}
R_{\mu\nu\rho\sigma}^{\ \ \ \ \ *}R^{\nu\mu\rho\sigma} =   
   4 i \left(\partial^{2}_z h_L\partial_z h_{R}^{\prime}+  h^{\prime\prime}_L\partial_z h^{\prime}_R -\partial^{2}_z h_R\partial_z h^{\prime}_L-  h^{\prime\prime}_R\partial_z h^{\prime}_L \right)\,.
    \label{RCSconformal}
\end{equation}
\color{black}
The above expression makes clear our earlier assertion that the gCS anomaly term is non-trivial only in the presence of {\it chiral} (\emph{i.e.} left-right asymmetric) GW perturbations.\footnote{\color{black} We remark for completion at this point that, in generic torsional Einstein-Cartan~\cite{Cartan:1923zea} or teleparallel theories~\cite{Maluf:2013gaa,Cai:2015emx,Bahamonde:2021gfp} 
(including those with Nieh-Yan invariants~\cite{Rao:2023nip}), there is a plethora of works addressing the observation of cosmological effects of torsion, \emph{e.g.} on dark matter, via chiral GW~\cite{Izaurieta:2020vyh,Elizalde:2022vvc}, some of which might be detectable at current or future interferometers~\cite{deAndrade:2024mwl}. Our considerations in this work are different from those works, as they pertain to the effects of chiral GW on torsion-free CS anomaly condensates that lead to inflation. 
In this latter respect, 
as remarked in \cite{Dorlis:2024yqw},  the StRVM inflationary phenomenology is similar to the one proposed in \cite{Lue:1998mq,Alexander:2004wk,Alexander:2016hxk}
on inflationary birefringence induced by the gCS terms. 
Nonetheless, in view of the torsion interpretation of the 
antisymmetric tensor field strength $\mathcal H_{\mu\nu\rho}$, that characterises the underlying microscopic string theory, 
some of the methods of detection of the torsional effects mentioned in those references on Einstein-Cartan torsion~\cite{Cartan:1923zea} might also be applicable to our StRVM case. This will not be a topic discussed further here, but we may come back to it in the future.\color{black}} 
\color{black}

At this stage we make some important remarks regarding the canonical quantization of the gCS
theory, and in particular the anomaly term \eqref{RCSconformal}.
 The issue is the known ordering ambiguities that occur in a quantum theory when we replace the classical quantities, such as  $h_{L(R)}(x)$ in the above equation, by quantum operators.\footnote{We remark at this point that there exist similar examples of such quantum-ordering ambiguities
 in the framework of the Wheeler-de-Witt equation for the wave-function of the Universe~\cite{DeWitt:1967yk,Dewitt:1968lxx,Kiefer:2013jqa}. For concreteness, we mention below two of them. One deals with a path-integral approach, in which these ambiguities can be related~\cite{Halliwell:1988wc} to the choice of path-integral measure, and can be fixed, for instance, by requiring invariance under field redefinitions of the three (spatial components of the) metric and the lapse function. The other~\cite{Muniz:2019ahh} studies the influence of the quantum ordering on the (non)existence of an initial 
 singularity and the dark energy in the Universe, pointing to different scenarios of physical interest, according to the way of fixing the ambiguities.
 In our four-spacetime dimensional CS gravity case, a way of fixing the pertinent ordering ambiguities will be suggested below, based on the necessity of having a finite duration of the inflationary era, as suggested in the classical limit by the dynamical-system approach to the linear-axion inflation~\cite{Dorlis:2024yqw}.}

In general, we may present various orderings of quantum operators, 
$\widehat \phi_i$,
as:
\begin{align}\label{qord}
{\rm O}_{w_{\rm P}}\Big(\Pi_{i=1}
^N \widehat{\phi}_i\Big) = \sum_{\rm P}\, w_{\rm P}\, \Big(\widehat{\phi}_{i_1} \dots \widehat{\phi}_{i_N}\Big)_{\rm P}\,, \quad \sum_{\rm P} w_{\rm P}=1\,,
\end{align}
where $w_{\rm P}$ denote the weights of the various permutations P of the operators appearing on the right-hand-side of \eqref{qord}. 

Using the Fourier decomposition for each mode separately, we can arrive at the following relation for the operator form of \eqref{RCSconformal}, upon applying a given quantum-ordering scheme \eqref{qord}: 
\begin{align}
\widehat{R_{\mu\nu\rho\sigma}^{\ \ \ \ \ *}R^{\nu\mu\rho\sigma}} = 4 \int \frac{d^{3}\vec{k}d^{3}\vec{k'}}{(2\pi^3)} \ e^{i(\vec{k}+\vec{k'}) \cdot \vec{x}}\, &\Big[ k^2 k' l_{\vec{k'}}\left({\rm O}_{w_{\rm P_1}}\Big(\hat{h}_{L,\vec{k}}\hat{h}^{\prime}_{R,\vec{k'}}\Big) - {\rm O}_{w_{\rm P_2}}\Big(\hat{h}_{R,\vec{k}}\hat{h}^{\prime}_{L,\vec{k'}} 
\Big)\right) \nonumber \\
&- k' l_{\vec{k'}} \left({\rm O}_{w_{\rm P_3}}\Big(\hat{h}^{\prime \prime}_{L,\vec{k}} \hat{h}^{\prime}_{R,\vec{k'}} \Big)
- {\rm O}_{w_{\rm P_4}}\Big(\hat{h}^{\prime \prime}_{R,\vec{k}}\hat{h}^{\prime}_{L,\vec{k'}} \Big) \right) \Big] \ ,
    \label{R*R_quantum}
\end{align} 
where, for the sake of generality, we assumed different ordering schemes (provided by the different weights $w_{{\rm P}_i}$, $i=1,\dots, 4$) for each product of operators appearing in 
\eqref{R*R_quantum}, with the weights satisfying 
$$\sum_{{{\rm P}_i}=1}^2 \, w_{{\rm P}_i}=1\,, \quad \forall~ \,\, i=1, \dots, 4\,.$$ 

The hermitian conjugate operator of \eqref{R*R_quantum} is then given by:
\begin{align}
    \widehat{( R_{\mu\nu\rho\sigma}^{\ \ \ \ \ *}R^{\nu\mu\rho\sigma})}^\dag =  4 \int \frac{d^{3}\vec{k}d^{3}\vec{k'}}{(2\pi^3)} \ e^{i(\vec{k}+\vec{k'}) \cdot \vec{x}}\Big[ k^2 k' l_{\vec{k'}} \, &\left({\rm O}_{w_{\rm P_1}}\Big( \hat{h}^{\prime}_{R,\vec{k'}} \hat{h}_{L,\vec{k}} \Big) - {\rm O}_{w_{\rm P_2}}\Big(
    \hat{h}^{\prime}_{L,\vec{k'}}\hat{h}_{R,\vec{k}} \Big)\right) 
    \nonumber \\
    & - k' l_{\vec{k'}} \left({\rm O}_{w_{\rm P_3}}\Big(\hat{h}^{\prime}_{R,\vec{k'}}\hat{h}^{\prime \prime}_{L,\vec{k}}\Big) - {\rm O}_{w_{\rm P_4}}\Big(\hat{h}^{\prime}_{L,\vec{k'}}\hat{h}^{\prime \prime}_{R,\vec{k}} \Big)  \right) \Big] \,,
    \label{R*R_dagger_quantum}
\end{align}
and a potential non - hermiticity would imply that:\footnote{ The situation may be thought of as being somewhat analogous to that encountered in the Dirac Lagrangian for spin-1/2 fermions, which are also non-classical objects. The naively defined Lagrangian density $\mathcal L_F = \overline \psi \Big(i\, \gamma^\mu \, \partial_\mu  - m \Big)\, \psi$ is non-hermitian, with the imaginary parts being associated with the total divergence of the corresponding Noether current: ${\mathcal L}_F - {\mathcal L}_F^\dagger = i \partial_\mu \Big(\overline \psi \, \gamma^\mu \, \psi \Big)$\,. One can maintain hermiticity upon defining a particular ordering by replacing 
$\partial_\mu$
by the operator $\frac{1}{2}\stackrel{\leftrightarrow}{\partial_\mu}$, with $\stackrel{\leftrightarrow}{\partial_\mu}$ defined as:
$A\stackrel{\leftrightarrow}{\partial_\mu} B \equiv  A \, \partial_\mu B - (\partial_\mu A)\, B$\,. In our case, the operator on the left-hand side of \eqref{IMRCS} is proportional to the four-divergence of the  gCS topological current, and the symmetric ordering \eqref{symord} may be thought of as corresponding to the 
specific ordering that renders the operators in \eqref{R*R_quantum} and \eqref{R*R_dagger_quantum} hermitian. 
}

\begin{equation}\label{IMRCS}
     \widehat{R_{\mu\nu\rho\sigma}^{\ \ \ \ \ *}R^{\nu\mu\rho\sigma}} - 
      \widehat{( R_{\mu\nu\rho\sigma}^{\ \ \ \ \ *}R^{\nu\mu\rho\sigma})}^\dagger  = - 2\, i \,{\rm Im}\widehat{R_{CS}} \neq 0 \ .
\end{equation}
It is straightforward to see, for instance, that in the case of the {\it symmetric ordering} for the product of two operators, in which all the weights are $w_{\rm P}=\frac{1}{2}$, {\it i.e.}:
\begin{align}\label{symord}
    {\rm O}_{{\rm w}_{\rm P}=\frac{1}{2}} \Big(\widehat \phi_1 \, \widehat \phi_2 \Big) = \frac{1}{2} \Big(\widehat \phi_1 \, \widehat \phi_2  + \widehat \phi_2 \, \widehat \phi_1 \Big)\,,
\end{align}
the imaginary parts of the gCS operator {\it vanish}:
\begin{align}\label{imcsfree}
{\rm Im}R_{CS}\,\Big|_{\,{\rm symm.~ordering}~w_{{\rm P}_i} =\frac{1}{2}}
=  0\,.
\end{align}
This scheme may be considered as providing a well-defined quantum operator in the sense discussed in passing in ref.~\cite{Lyth:2005jf}, where the imaginary parts of the naively defined composite gCS operator $R_{CS}$ have been ignored.  
On the other hand, for any other ordering scheme the imaginary parts will depend on (be proportional to) the weights $w_{{\rm P}_i}$, and, thus, be scheme dependent.

Nonetheless, the reader can see easily that the real part of the gCS condensate ${\rm Re} \widehat{R_{CS}}$ is  {\it quantum-operator-ordering-scheme independent}, that is, the value of this quantity is independent of the $w_{{\rm P}_i}$, $i=1, \dots, 4$\,,
\begin{align}\label{reCSnoord}
    \frac{d}{d w_{{\rm P}_i}} {\rm Re}R_{CS}= \frac{1}{2} \, \frac{d}{d w_{{\rm P}_i}} 
    \Big(\widehat{R_{\mu\nu\rho\sigma}^{\ \ \ \ \ *}R^{\nu\mu\rho\sigma}} +
      \widehat{( R_{\mu\nu\rho\sigma}^{\ \ \ \ \ *}R^{\nu\mu\rho\sigma})}^\dag \Big) =0\,, \quad \forall \quad i=1, \dots, 4 \,.
\end{align}

Taking into account that each ordering defines a \textit{fundamental quantum theory}, this implies that the classical theory corresponds to (infinitely many in our case) quantum theories. 
There is no contradiction here. The classical limit is  not fundamental, in contrast to the quantum version of it. In our case, we are only dealing with effective low-energy theories obtained from an UV completion of string theory, which is a candidate theory for quantum gravity. The latter still eludes us. Once this issue is resolved, then the UV completion will correspond to a specific quantum ordering of the various operators entering the quantum theory. 

At this juncture, we make the important remark 
that, in the way constructed, the GW action \eqref{GWact} can be free from such ordering ambiguities, even if the $\widehat{R_{CS}}$ operator is not, since the relevant parts can be absorbed in a boundary term \eqref{relB}. Nonetheless, this characterizes only the case of a homogeneous and isotropic (pseudo-)scalar, which appears in the cosmological StRVM. In any other case, there will be ordering ambiguities also in the GW action, an issue that will be treated in a future work. 

In an effort to potentially identify a ``physical'' quantum ordering in our case, we next remark that, in the presence of the imaginary parts of the condensate, the Minkowski-signature space-time GW path-integral partition function, $\langle 0 | 0\rangle$, which expresses the amplitude of the probability that the StRVM system remains in the initial (inflationary in our case) ground state, acquires a damping factor, which thus determines the lifetime of inflation. In the HF approximation, for instance, we obtain
formally:\footnote{Notice that our arguments below on estimating the life time of the inflationary vacuum are based on the study of the path-integral partition function of GW perturbations $\mathcal Z$ in the case of a single source of GW. This suffices, because, as we discussed above, the case of $\mathcal N$ sources can be studied by simply raising the partition function of a single source $\mathcal Z$ to the $N$-th power, \eqref{ZN}, $\mathcal Z_{\mathcal N~\rm sources} = \mathcal Z^{\mathcal N}$. Hence all the necessary information is encoded in a single-source GW path integral. 
  } 
\begin{align}\label{dumping}
    \langle 0 | 0 \rangle^{\rm HF}_{\rm Minkowski-signature} \sim \exp\Big(- b\, \langle  R_{CS} \rangle  \int_{\rm dS} d^4 x \sqrt{-g} \, \Big) \Big[ \dots \Big]\,,
\end{align}
where $b$ denotes the axion background value, about which we consider GW perturbations in the path integral,  $\int_{\rm dS} d^4x \sqrt{-g}$ denotes the (approximately de Sitter) four volume during inflation, and the terms $\Big[ \dots \Big]$ denote the GW-perturbation path-integral in the absence of gCS anomalous terms. The life time 
would be formally estimated by setting the dumping exponent in \eqref{dumping} to one.

However, to give meaning to the path-integral one needs to analytically continue to a Euclidean four-dimensional spacetime. In such a case, the imaginary parts of the gCS appear as phases in the integrand of the respective path integral. 
If we represent the Euclidean version of the left-hand side  in terms of the vacuum energy as 
\begin{align}\label{EV}
 \langle 0 | 0 \rangle_{\mathcal E} = \exp\Big(-V^{(\rm \mathcal E)}\, E\Big)\,,   
\end{align}
 with $V^{(\rm \mathcal E)}$ the Euclidean ($\mathcal E$) four volume, then from a fundamental theorem of the complex calculus we note that the presence of a phase in the path integral can only make it smaller than in the case in which the phase is absent, provided that the measure of the path-integration is positive definite:
\begin{align}\label{fz}
  \Big|\int \mathcal D z\, | f(z) |\, e^{i\alpha (z)} \,\Big| \le \int \mathcal D z \, |f(z)|\,, \quad  z \in \mathbb C\,.   
\end{align}

From \eqref{EV}, \eqref{fz},  we then conclude that the imaginary parts of the gCS condensate result in a higher vacuum energy than in the case where the condensate is absent.\footnote{The situation is somewhat analogous to the celebrated Vafa-Witten theorem on the impossibility of the spontaneous breaking of Parity
in vector-like theories~\cite{Vafa:1984xh,Vafa:1984xg}, which is based on similar energetics arguments. Nonetheless, in our case, the presence of imaginary parts of the gCS condensate is not interpreted as implying that the formation of the condensate is not possible, but rather that, once it is formed via GW, quantum effects destabilize it, thereby causing the RVM-inflationary vacuum to decay, with a life time determined by these imaginary parts.  }

Naively, this would prompt one to conclude that, in our case, the symmetric ordering \eqref{symord} is the one with physical meaning, as in the latter scheme the gCS is free from imaginary parts \eqref{imcsfree}, and thus the vacuum energy would be minimized as compared to all other quantum orderings. Moreover, in the absence of imaginary parts of the operator \eqref{R*R_quantum}, one would find it natural to identify the classical ($\hbar \to 0$) limit of this {\it hermitian} operator with a {\it real} function, as  $R_{CS}$ is ({\it cf.} \eqref{RCS}). 
This would in turn imply that, in the context of our effective low-energy gCS theory, the inflationary phase would be eternal, unless one has mechanisms of particle production during inflation (conventional or otherwise) in order to induce a graceful exit from it. 
It has been recently discussed~\cite{Akhmedov:2024axn}, however, that such  particle-pair production mechanisms might be  associated with the existence of imaginary parts of the gravitational effective action. This would prompt us to consider ordering schemes other than the symmetric one, which are characterized by such complex effective actions in our CS gravity case.

Hence, adopting the symmetric ordering might not be a physically correct procedure in our case. As already mentioned, from the point of view that the quantum theory is the fundamental one, we cannot exclude the possibility that in our stringy-quantum-gravity  case of the StRVM, non hermitian Hamiltonians, implied by the other quantum ordering schemes, as described above, play a physical r\^ole in the scenario of GW-induced condensate inflation advocated here and in \cite{bms,ms1,Mavromatos:2022xdo}. 
Under this interpretation, the StRVM vacuum is {\it destabilized} by quantum effects. This implies that the running vacuum will decay so that the gravitational anomaly will be eradicated. This is the r\^ole of the chiral matter fermions, generated at the end of the RVM inflation as a result of the metastable vacuum decay in the scenario of \cite{bms,ms1}, which generate their own gravitational anomalies that cancel the primordial ones. Lacking a well-established framework of quantum gravity, such a scenario cannot be excluded based solely on hermiticity arguments.

\color{black} At this stage, we cannot resist in referring the reader briefly to the field of the so-called PT-symmetric approach to quantum mechanics and field theory~\cite{Bender1,Bender2} according to which, under such a condition, or, more generally the invariance of a system under some antilinear symmetry that can include PT~\cite{Mannheim}, non-hermitian Hamiltonians may be characterized by real eigenvalues and play a physical r\^ole. In the context of the StRVM, for instance, such PT-symmetric approach may provide novel scenarios for the acceleration of the Universe~\cite{CSPT}. In our context here, of course, we do not examine in detail such PT symmetric scenarios, viewing the presence of imaginary parts in the effective action as signifying an open-ness of the system. In the string theory context of interest to us here, for example, these imaginary parts may be linked to 
the existence of infinite towers of the massive string states (with masses that are integer multiples of the string mass scale $M_s$) that become important when the energy of the graviton modes approach the cutoff $M_s$~\cite{Mavromatos:2024pho}. The full string theory, with all its states taken into account should, of course be unitary. The non-unitarity (in the sense of the existence of non-hermitian action) is a feature of the effective theory, truncated to the local degrees of freedom only, with energies below $M_s$.\footnote{\color{black} Representing the decaying vacuum of the StRVM as an open system is  a concept familiar from the theory of decaying particles with a finite width, which are represented as open quantum  mechanical Lindblad systems~\cite{Bertlmann:2006fn,Bernabeu:2012au}. There, the environment, which  the unstable particle interacts with, is provided by enlarging the original Hilbert space by states representing the decay products. In the StRVM, as we have mentioned, the environment consists of the entire spectrum of the massive string states.\color{black}}

A highly non-trivial question is the definition of the inner product of states in such non-hermitian gravitational theories. At present, in the context of our stringy quantum gravity, this is far from being understood. The environment of the infinite towers of massive string states, which opens up the system, 
makes the situation non-local from a low-energy point of view. Nonetheless, one may attempt to define such inner products by staying within the non-Hermitian effective local CS gravitational field theory, which depends, however, explicitly on the UV cutoff (the string mass scale $M_s$). 
An inspiration towards such a definition in our StRVM, might be the work of \cite{Alexander:2022ocp}, 
which attempts to define an inner product of gravitational states 
on the four-dimensional quantum gravity in the simplified case in which the Ashtekar connection is diagonal. In this case, the constraints of quantum gravity are solved by the so-called Chern-Simons-Kodama (CSK) state~\cite{Kodama:1990sc}. The latter lead to de-Sitter space (hence, inflation) as an allowed configuration in the semi-classical limit. The analysis of \cite{Alexander:2022ocp}
has as starting point the four-dimensional Holst Action, which extends the Einstein-Cartan general relativity by the Holst topological term~\cite{Holst:1995pc}. As we have remarked, however, in the introduction of this article, this term is not topological (not a total derivative) in the presence of torsion, as is the case of our StRVM (which contains a totally antisymmetric component of the $\mathcal H_{\mu\nu\rho}$ torsion, as we discussed above). In such a case, the Holst term needs to be replaced by the Nieh-Yan invariant, and in our anomalous CS theory by the combination of the Nieh-Yan invariant plus the gCS term \eqref{nieh3}. This is the first complication in defining an appropriate inner product among gravitational states. The second, and most important, one arises because of the presence of the imaginary parts of the gCS condensate itself, as we stated above. In such case, unlike the situation in \cite{Alexander:2022ocp}, reality conditions on the effective action cannot be imposed, but one may attempt to formulate the theory as a PT symmetric one, which generalises the Hermitian theories, as mentioned above. This lies beyond the scope of the current work, however. We hope to  be able to address this important issue in a future work.\color{black}

The detailed estimation of the lifetime of the inflationary StRVM vacuum therefore is a complicated and, in general, a microscopic-string-theory dependent issue, given that  unless the string-embedded grand unified gravitational field theory is specified, the dominant decay channel of the RVM model cannot be computed in detail. Nonetheless, within our low-energy approximation, studied in \cite{bms,ms1,ms2,Dorlis:2024yqw} and here, we can still provide a hopefully reliable analysis that could allow  for an estimate of the appropriate string scale required to match the model predictions with the current astro-cosmological phenomenology of inflation~\cite{Planck:2018jri,Planck:2018vyg,Martin:2013tda}.

To this end, the reader should notice that the classical inflationary model of \cite{Dorlis:2024yqw}, in which inflation is induced by the linear-axion potential, is also characterized by a finite lifetime, which under appropriate initial conditions can be of the phenomenological order ${\mathcal O}(50-60)$ e-foldings, requiring in our StRVM a string scale $M_s=\mathcal O(10^{-1})M_{\rm Pl}$.
If we insist, therefore, on matching these classical results of the dynamical-system approach of \cite{Dorlis:2024yqw} with those obtained in the context of a (non-hermitian) quantum-gravity framework, where the presence of imaginary parts in the gCS condensate will be interpreted as destabilization of the pertinent quantum vacuum, then we may seek quantum-ordering schemes in which such an agreement is achieved. 

The simplest of such schemes, {\it i.e.} the one in which the quantization of the composite operators \eqref{R*R_quantum} (and \eqref{R*R_dagger_quantum}) is obtained by a simple replacement of 
the quantities $h_{L(R)}$ by the respective operators, 
$\widehat h_{L(R)}$,
in the order they appear classically, {\it i.e.} we set formally the operation $\rm O_{w_{{\rm P}_i}} =1$ in the relevant expressions, \eqref{R*R_quantum}, 
\eqref{R*R_dagger_quantum}. As we shall demonstrate below, this case is the one required to match the results with the approach of \cite{Dorlis:2024yqw}, in a non-trivial way.

In that case, we obtain for the imaginary parts \eqref{IMRCS}:
\begin{equation}
\begin{aligned}
     \widehat{R_{\mu\nu\rho\sigma}^{\ \ \ \ \ *}R^{\nu\mu\rho\sigma}} - 
      \widehat{( R_{\mu\nu\rho\sigma}^{\ \ \ \ \ *}R^{\nu\mu\rho\sigma})}^\dag  &= 4 \int \frac{d^{3}\vec{k}d^{3}\vec{k'}}{(2\pi^3)} \ e^{i(\vec{k}+\vec{k'}) \cdot \vec{x}}\left\{ k^2 k' l_{\vec{k'}}\left( \left[\hat{h}_{L,\vec{k}},\hat{h}^{\prime}_{R,\vec{k'}}\right] + \left[\hat{h}^{\prime}_{L,\vec{k'}},\hat{h}_{R,\vec{k}}\right] \right) \right\} \\ 
      &- 
       4 \int \frac{d^{3}\vec{k}d^{3}\vec{k'}}{(2\pi^3)} \ e^{i(\vec{k}+\vec{k'}) \cdot \vec{x}}\left\{ k' l_{\vec{k'}} \left( \left[\hat{h}^{\prime \prime}_{L,\vec{k}},\hat{h}^{\prime}_{R,\vec{k'}}\right] + \left[\hat{h}^{\prime }_{L,\vec{k'}},\hat{h}^{\prime\prime}_{R,\vec{k}}\right] \right)  \right\} \ .
       \label{nonhermitian_Rcs}
\end{aligned}
\end{equation}

  Upon taking vevs with respect to GW metric perturbations in the inflationary (Bunch-Davies-type) vacuum $|0\rangle$, which, in the path-integral approach, is interpreted as the 
integral over GW perturbations \eqref{Oaver}
of the appropriate composite operator, the shift symmetry \eqref{shiftsym} breaks down due to the formation of the gCS condensate. The presence of non-hermiticity, as implied by \eqref{nonhermitian_Rcs}, leads to  {\it metastability} of the {\it quantum} inflationary vacuum, with a life time which is given by the inverse of the imaginary part of the gCS condensate, as a rough estimate. 

On cutting-off the momentum integrals in the UV at $\mu$ \eqref{muMs}, and using \eqref{defscal} and \eqref{creationcommutation},
we obtain from \eqref{nonhermitian_Rcs})~\cite{Mavromatos:2024pho}:
\begin{align}
 2 {\rm Im}\langle \widehat{R_{CS}}\rangle = {\rm Im}\langle \widehat{R_{\mu\nu\rho\sigma}\, \widetilde R^{\nu\mu\rho\sigma}} \rangle = 
 \frac{16  A \text{$\dot{b}$} \mu ^7}{7 M^{4}_{\rm Pl} \ \pi ^2} \left[1+\left(\frac{H_I}{\mu}\right)^2\left(\frac{21}{10}-6\left(\frac{A\mu\text{$\dot{b}$}}{M_{\rm Pl}^2}\right)^2\right)\right] \ , 
\end{align}
where $\mu \simeq M_s$ ({\it cf.} \eqref{muMs}). Hence, in view of \eqref{HI}, and $\eqref{cutoff}$,
the second term inside the square brackets 
on the right-hand side of the above equation is subdominant, and will not play a r\^ole in our estimates of the inflationary vacuum lifetime.

This can be achieved by examining the Hamiltonian of GW from a single source alone. 
To put it in other words,  
the imaginary parts of the condensate back-react on the effective Lagrangian of GW, and thus on the corresponding Hamiltonian $\mathcal H$, which in this way also acquires an imaginary part. The latter is estimated as~\cite{Mavromatos:2024pho}:
\begin{equation}
 {\rm Im}\left(\mathcal H\right) = \int d^3 x \ \sqrt{-g}\,  \frac{1}{2} A \ b \ {\rm Im} \left(\langle R_{\mu\nu\rho\sigma}\widetilde{R}^{\nu\mu\rho\sigma}\rangle \right) \approx V^{(3)}_{dS} \ \frac{8  b A^2  \text{$\dot{b}$}  \mu ^7}{7 M^{4}_{\rm Pl} \ \pi ^2}~
\label{imaginaryhamiltonian}
\end{equation}
where $\approx$ implies that we keep only leading order contributions in the small quantity $\kappa^2 \dot b$, and $V^{(3)}_{dS}$ denotes the de Sitter 3-volume of the inflationary spacetime. 

Some important remarks are due at this point. The reader should notice 
that the imaginary parts of the gCS condensate depend on the UV cutoff $\mu$ in the context of the low-energy theory. This in turn implies 
the open and dissipative nature of the system of graviton modes with momenta below $\mu$, which in the UV complete (full string) theory interact with the ``environment'' of the infinite towers of the massive string states, with energies above the cutoff $M_s$.
In this sense, our graviton system is somewhat analogous to the representation of a decaying particle system as an open quantum system, where the states of the decay products play the r\^ole of the relevant
``environment''~\cite{Caban:2005ue,Bertlmann:2006fn,Bernabeu:2012au}. 

The dissipative nature of our open-system, due to the introduction of a cut-off $\mu$ in our effective description, provides us with a naively defined estimate of the lifetime $\tau$ of the inflation vacuum (in natural units, where $\hbar=1$):
\begin{align}\label{lifetime}
\tau \sim \left(
{\rm Im} \mathcal{H} \right)^{-1}\,,
\end{align}
which is the Euclidean version of what one would obtain formally by setting the dumping exponent of the Mikowski expression \eqref{dumping} equal to unity. However, this is the proper formal treatment within our approach to estimate the duration of inflation.

We next note that the (Euclidean) four volume $V^{(3)}_{dS} T^E$ of the (inflationary) de-Siter spacetime, 
with radius $\Lambda$, is given by \cite{Fradkin:1983mq}
\begin{equation}
   V^{(3)}_{dS} T^E =  \frac{24 \pi^2}{M^{2}_{\rm Pl} \Lambda} \ , \quad \Lambda =  3 H^{2}_I~,
   \label{4volume}
\end{equation}
with $T^E$ corresponding to the Euclidean time, defining the appropriate duration of inflation, which phenomenologically is of order~\cite{Planck:2018jri,Planck:2018vyg,Martin:2013tda} 
\begin{align}\label{durinfl}
T^E\sim (50-60)H_I^{-1}\,,
\end{align}
where the numerical coefficient on the right-hand side of the above relation is the number of e-foldings~\cite{DiMarco:2024yzn}. 
Substituting \eqref{durinfl} to eq.\eqref{imaginaryhamiltonian}, we obtain:
\begin{equation}\label{ImHb}
    {\rm Im}\left(\mathcal H\right) = \frac{64 b A^2 \text{$\dot{b}$} \kappa ^6 \mu ^7}{ 7 H_I }\cdot\frac{1}{H_I T^E}~.
\end{equation}
The order of magnitude of the quantities entering \eqref{ImHb} can be taken from \cite{Dorlis:2024yqw}, by leaving the string-scale $M_s = \mu$ as a free parameter. Then, we can easily obtain that the (naively defined) lifetime $\tau$ of the RVM inflationary vacuum of the StRVM is determined by evaluating the quantity: 
\begin{equation}
    H_I\tau\sim \frac{7H_I^2M_{\rm Pl}^6}{64bA^2\dot{b}M_s^7}\left(H_I T^E\right) \sim 10^{-2}\left(\frac{M_{\rm Pl}}{M_s}\right)^{3} \cdot (H_I T^E)~.
\end{equation}
From our previous discussion, this is the decay rate of the unstable RVM ``vacuum'', whose energy is higher than that of the case without the gCS condensate. For such a decay rate to be consistent with the duration of inflation \eqref{durinfl}, the following upper limit for the string-effective-theory UV cut-off $\mu=M_s$ ({\it cf.} \eqref{muMs}) has to be satisfied:
\begin{equation}
\frac{M_s}{M_{\rm Pl}}\lesssim
0.215~~.
\end{equation}
This is quite consistent with the findings of the linear-axion-potential dynamical-system analysis of \cite{Dorlis:2024yqw}, which concentrated on the real part of the CS condensate. 
The above discussion leads to the conclusion that the r\^ole of KR axions (or other string-compactification-induced axions, which also couple to the gravitational CS term) in inducing RVM inflation is instrumental, since, in their absence, the CS term, being a total derivative, would not contribute to the dynamics and the pertinent condensate would be trivially vanishing.

\section{Periodic modulations of the Axion Potential and slow-roll inflationary parameters}\label{sec:slowrollparam}

In the context of the StRVM, the dynamics of the primordial Universe is described by \eqref{eq:Action},  without any gauge fields, that is $\mathbf F =0$   ({\it cf.} \eqref{eq:ActionAB}). In this case, as discussed above and in \cite{Dorlis:2024yqw}  , inflation is driven alone by the linear axion potential \eqref{effectivepot}.
The (small) slow-roll parameters for inflation due to a single (pseudo)scalar field $\phi$ with potential $V(\phi)$ are given approximately by (the prime denotes derivative with respect to $\phi$)~\cite{Leach:2002ar,Martin:2013tda}:
\begin{align}\label{slowroll}
\epsilon_1 &= \frac{M_{\rm Pl}^2}{2} \, \Big(\frac{V^\prime}{V} \Big)^2\,, \nonumber \\
 \epsilon_2 &= 2M_{\rm Pl}^2 \,  \Big(\frac{V^\prime}{V} \Big)^2 - 2 M_{\rm Pl}^2\, \frac{V^{\prime\prime}}{V} \equiv 4 \epsilon_1 - 2 \eta\,, \nonumber \\
 \epsilon_2 \, \epsilon_3 & \simeq 2M_{\rm Pl}^4 \, \Big[\frac{V^{\prime\prime\prime}\, V^\prime }{V^2} - 3 \frac{V^{\prime\prime}}{V} \,\Big(\frac{V^\prime}{V}\Big)^2 + 2 \Big(\frac{V^\prime}{V}\Big)^4 \Big]\, , \,\,\, {\rm etc.}\,,
 \end{align}  
The scalar spectral index is approximately given by (we ignore subdominant higher-order terms proportional to $\epsilon_1 \epsilon_2, \, \epsilon_2\epsilon_3$ that appear in the expression for $n_s$~\cite{Leach:2002ar,Martin:2013tda}):
\begin{align}\label{scalarspectral}
n_s \simeq 1 - 2\epsilon_1 - \epsilon_2\,,
\end{align}
while the ratio $r$ of tensor to the scalar perturbations is:
\begin{align}\label{rratio}
r \simeq 16 \, \epsilon_1 \,.
\end{align}
Planck-collaboration constraints on inflation~\cite{Planck:2018jri}, after taking into account lensing and Baryon-acoustic-oscillation (BAO) measurements, indicate that: 
\begin{align}\label{nsplanck}
n_s & = 0.9649 \pm 0.0042 \quad (\rm{68\%~C.L.,~Planck~TT,~TE,~EE~+~lowE~+~lensing})\,, \nonumber \\
n_s & = 0.9665 \pm 0.0038 \quad (\rm{68\%~C.L.,~Planck~TT,~TE,~EE~+~lowE~+~lensing~+~BAO})
\end{align}
respectively.  For the tensor-to-scalar ratio, on the other, the combined measurements yield the upper bound (at pivotal scale $k^\star = 0.002$):\footnote{This bound depends somewhat on the underlying cosmological model; the bounds on $r$ in \eqref{rplanck} refer to some important extensions of $\Lambda CDM$, as considered by Planck collaboration~\cite{Planck:2018jri}.}
\begin{align}\label{rplanck}
r_{002} & <  0.11  \quad (\rm{95\%~C.L.,~Planck~TT,~TE,~EE~+~lowEB~+~lensing})\,, \nonumber \\
r_{002} & <  0.056  \quad (\rm{95\%~C.L.,~Planck~TT,~TE,~EE~+~lowE~+~lensing~+~BK15})\,.
\end{align}
In general, for a few important extensions of the $\Lambda CDM$, the bounds on $r$, and also constraints of $r$ versus $n_s$ are given in 
\cite{Planck:2018jri}, where we refer the interested reader for further details. For our purposes here, we mention that the combined Planck and BK data yield the following bounds for
the slow-roll $\epsilon_i$, $i=1,2$ parameters \eqref{slowroll}, using 
Planck TT,TE,EE+lowE+lensing (+BK15) data~\cite{Planck:2018jri}:
\begin{align}\label{epsbounds}
\epsilon_1 &<  0.0097 \, \quad (0.0044) \, \quad  (95\% ~\rm C.L.)\,, \nonumber \\
\epsilon_2 &= 0.032^{+0.009}_{-0.008} \, \quad (0.035 \pm 0.008) \quad (68\% ~\rm C.L.)\,, \nonumber \\
\epsilon_3 &= 0.19^{+0.55}_{-0.53} \, \quad (0.12^{+0.36}_{-0.42}) \quad (95\% ~\rm C.L.)\,.
\end{align}

In the StRVM the $b$-axion potential is strictly linear~\cite{Dorlis:2024yqw}, which implies  
\begin{align}\label{strvmei}
\eta=0\,, \quad \epsilon_1 = \frac{1}{4} \epsilon_2 = \frac{1}{4} \epsilon_3  = 0.5 M_{\rm Pl}^2\, b(t)^{-2}\,.
\end{align}
During inflation, the axion $b$ of the StRVM varies linearly with the cosmic time ({\it cf.} \eqref{bdotInflation})~\cite{bms,ms1,ms2,Mavromatos:2022xdo,Dorlis:2024yqw} 
\begin{align}\label{blinear}
b(t) = b(0) + c_1 \, t \,, \quad c_1 \simeq \mathcal O(10^{-1})\, M_{\rm Pl}\, H_I\,,
\end{align}
where $H_I$ is the (approximately constant) Hubble rate during inflation, which is bounded by Planck-collaboration data~\cite{Planck:2018vyg,Planck:2018jri} according to \eqref{HI}. The constant $b(0)$ is the value of the axion field at the onset of the RVM inflation, and it turns out to be negative. The analysis of \cite{Dorlis:2024yqw} leads to the lower bound for the magnitude of this quantity (in order of magnitude):
\begin{align}\label{b0bound}
\frac{|b(0)|}{M_{\rm Pl}} \gtrsim \mathcal O(10)\,, \quad  b(0) < 0\,.
\end{align}
This implies that in order of magnitude the axion remains approximately constant during the entire duration of inflation, which lasts for~\cite{DiMarco:2024yzn} $H_I^{-1} {\mathcal N}_{\rm e}$, with ${\mathcal N}_{\rm e} = \mathcal O(50-60)$ the number of e-foldings of the Universe, according to the cosmological data~\cite{Planck:2018jri,Martin:2013tda}.

Thus, it is reasonable to estimate the order of magnitude of the (small) slow-roll parameters $\epsilon_i$, $i=1,2,\dots ,$ \eqref{slowroll}, by replacing $b$ by $b(0)$ in the corresponding expressions \eqref{strvmei},  \eqref{rratio} and \eqref{scalarspectral}. This leads to 
\begin{align}\label{strvme1values}
\epsilon_1 \lesssim 0.005  \,, \quad \epsilon_2 = \epsilon_3 \lesssim 0.02 \,, \quad r \lesssim 0.08\,, \quad n_s \simeq 1 - 6\epsilon_1 \gtrsim 0.97\,.
\end{align}
  As can be therefore seen from the above discussion, 
the phenomenology of the StRVM, with a gCS condensate-driven inflation~\cite{bms,ms1,Dorlis:2024yqw}, whose dynamics is described by the action \eqref{eq:ActionAB}, containing only fields from the massless bosonic ground state of the superstring, and a gCS-condensate-induced linear axion potential \eqref{effectivepot}, is    
largely consistent with the inflationary data \eqref{epsbounds}. However, 
although the rest of the slow-roll parameters lie within the data-preferred allowed regions, the scalar spectral index lies on the boundary of such regions.
We therefore need to improve the model so as to achieve better agreement with the data as far as (the central value of) $n_s$ is concerned. 

To this end, we shall consider the possibility of including periodic modulations on the axion potentials, which can be generated through non-perturbative effects. The hope is that by introducing $\eta \ne 0$ through such modulations, one 
can reduce the scalar spectral index $n_s$ accordingly, thus providing better match with the data. The issue is to justify microscopically the appropriate non-perturbative effects. As we shall see, in the context of string theory this implies that we may modify the StRVM cosmology by considering, in addition to the KR (string-model independent)  axion $b$, also string-model-dependent axions $a$, arising from compactification, along the lines of ~\cite{sasaki,Mavromatos:2022yql} .

\subsection{Target-space gauge group instantons and the KR axion Potential}\label{sec:gauginst}

Let us first start by discussing what non-perturbative effects we may consider in the context of the initial StRVM \eqref{eq:ActionAB}, with a gCS-condensate-induced linear axion potential \eqref{effectivepot}. The first thing that comes to mind are target-space instantons, associated with the non-Abelian gauge group that characterizes the four-dimensional effective action, after string compactification. In the initial StRVM, these gauge sector had been assumed not to characterize the dynamics of the 
primordial Universe. However, given that such gauge fields also characterize the massless string ground state of appropriate gauge supergravity low-energy theories obtained from phenomenologically realistic string models, we may take these gauge sectors into account, especially from the point of view of their non perturbative (instanton) contributions. Indeed, the spirit of \cite{bms} will be maintained if we integrate out such non-perturbative configurations, assuming that gauge fields with trivial instanton numbers do not exist as external lines in the field theory that describes the primordial Universe.   

According to standard theory~\cite{Eguchi:1980jx}, the following quantity 
\begin{align}\label{chern}
\frac{1}{16\pi^2} \int d^4 x \sqrt{-g}\, \mathbf F_{\mu\nu} \, \widetilde{\mathbf F}^{\mu\nu} = n \,, \quad n  \in \, \mathbb Z\,,
\end{align}
is an integer $n$ (Pontryagin index), whose value defines the various topological sectors defined by the (non-Abelian) gauge-group instantons. This implies that in the presence of instanton backgrounds the global shift symmetry \eqref{shiftsym} breaks down, and is replaced instead by periodicity of the axion backgrounds
\begin{align}\label{axionperiodic}
16\, \pi^2 A \, b(x) \, \to \, 16\, \pi^2 A \, b(x) + 2\pi \,.
\end{align}
Integrating out the instantons, leads to periodic modulations of the $b$ axion potential, which now assumes the form\footnote{\label{f15} Notice that with the + sign choice of the cosine term in the potential, there is no mass term generated for the axion $b$ during the inflationary phase. In StRVM such masses  could be generated at post-inflationary eras, e.g. during the QCD epoch, as a result of the anomalous coupling of the $b$ field to the Pontryagin term involving the gluon-field strengths~\cite{bms2}.}
\begin{align}\label{axionperiodpot}
V_{\rm eff}^{\rm periodic} (b) = \Lambda_1^4 \, {\rm cos}\Big(16\pi^2 A \, b(x)\Big) \equiv \Lambda_1^4 \, {\rm cos}\Big(\frac{b(x)}{f_b}\Big)\,,
\end{align}
where $\Lambda_1$ is the scale of the gauge-group instantons, and $f_b $ is the axion coupling. 
In the StRVM, $f_b$ is given by ({\it cf.} \eqref{Aval}):
\begin{align}\label{fbval}
f_b = 0.37\, \frac{M_s^2}{M_{\rm Pl}}\,.
\end{align}
The scale $\Lambda_1$ depends on the particular gauge model considered. For our purposes here it 
is treated as a phenomenological parameter of our low-energy string-inspired effective theory. The important feature is that the scale $\Lambda_1$  is suppressed by the exponential of the (Euclidean) one-instanton action, $\mathcal S_{\rm inst}$, which is known to be bounded from below by (see, for instance, \cite{Coleman:1978ae,Diakonov:2002fq,Tong:2005un}):
\begin{align}\label{instbound}
\mathcal S_{\rm inst} \gtrsim \frac{8\pi^2}{g^2_{\rm YM}} |n| \,,
\end{align}
where $g_{\rm YM}$ is the Non-Abelian gauge group (Yang-Mills) coupling, and, as already mentioned, $n$ labels the respective topological instanton sector. The equality is achieved only for (anti)self-dual gauge-field strengths  
$\mathbf F = \pm \widetilde{\mathbf F}$, where the plus (minus) sign corresponds to  $n > (<) \,0$. 

In the context of our string-inspired theory, it is natural to assume therefore the following form of $\Lambda_1$:
\begin{align}\label{L1}
\Lambda_1 = \xi \, M_s \, \exp\Big(-\mathcal S_{\rm inst}\Big) \lesssim \xi \, M_s \, \exp\Big(-\frac{8\pi^2}{g^2_{\rm YM}}\, |n|\Big) \ll M_s\,,
\end{align}
where $\xi >0$ is a numerical factor, that can in principle be calculated within a specific gauge theory. For our purposes we assume $\xi =\mathcal O(1)$.

Thus, the inclusion of instantons in the (3+1)-dimensional non-Abelian gauge sector of the model leads to the following effective potential for the KR axion $b$ field of the StRVM in the inflationary epoch, where the gCS condensate $ \left<R_{CS}\right>^{total}_{I}$ (\eqref{VEV_inflation_source}), in the presence of a number of GW sources, has been formed:
\begin{align}\label{periodicpot}
V_{\rm eff}(b) = b(x) \, \Lambda_{\rm cond}^3 + \Lambda_1^4 \, {\rm cos}\Big(\frac{b}{f_b}\Big)\,, \qquad \Lambda_{\rm cond}^3  \equiv  A\, \left<R_{CS}\right>^{total}_{I}\,.
\end{align}
We remark that the inflation is still driven by the linear potential given the smallness of the scale $\Lambda_1 \ll \Lambda_{\rm cond}$, due to its exponential suppression by the large (Euclidean) instanton action \eqref{L1}.\footnote{  The 
pertinent equations of motion are therefore affected only very mildly by the presence of the shift-symmetry breaking periodic modulations of the axion potential, and thus the main qualitative features of the StRVM inflation, discussed previously in the literature~\cite{bms,ms1,Mavromatos:2022xdo,Dorlis:2024yqw}, and reviewed here, including the non-dilution of the KR field at the end of inflation, remain intact. A detailed, quantitative, discussion on the r\^ole of the  potential \eqref{periodicpot}
within the dynamical system approach, in the spirit of \cite{Dorlis:2024yqw}, will be presented in a
future publication.
  } Nonetheless
there are small but significant modifications of the slow-roll parameters due to the non-perturbative effects. 
Indeed, for the axion potential \eqref{periodicpot}, we obtain from \eqref{slowroll} for the slow-roll inflationary parameters (as per our previous discussion, in the evaluation of these parameters we make the approximation that during inflation the order of magnitude of $b(x)$ is well-approximated by $b \simeq b(0)$):\footnote{We stress once again that in our model, the cosmic time dependence of the axion field $b$ is suppressed during the entire duration of inflation, due to the order of magnitude of the initial value $|b(0)|$ \eqref{b0bound}. Otherwise, the parameters $\epsilon_i, $ $i=1,2, \dots,$ would exhibit appreciable oscillatory behavior with time.}
\begin{align}\label{eiperiodic}
\epsilon_1 & \simeq \frac{M_{\rm Pl}^2}{2}\, \Big[\frac{\Lambda_{\rm cond}^3 - \frac{\Lambda_1^4}{f_b}\, {\rm sin}\Big(\frac{b}{f_b}\Big)}{b(0)\, \Lambda_{\rm cond}^3 + \Lambda_1^4\, {\rm cos}\Big(\frac{b(0)}{f_b}\Big)}\Big]^2 \simeq \frac{M_{\rm Pl}^2}{2\, b^2(0)} \, \Big[ 1 + \frac{2\, \Lambda_1^4\, {\rm sin}\Big(\frac{b(0)}{f_b}\Big)}{f_b|\Lambda_{\rm cond}^3|} - \frac{2\, \Lambda_1^4\, {\rm cos}\Big(\frac{b(0)}{f_b}\Big)}{|b(0)\,\Lambda_{\rm cond}^3|} + \dots \Big] \simeq 
\frac{M_{\rm Pl}^2}{2\, b^2(0)}  \nonumber \\
&({\rm since} \quad \Lambda_1 \ll \Lambda_{\rm cond}, \, f_b\,, M_{\rm Pl}, M_s \,\, ({\it cf.}\, \eqref{L1}))\,, \nonumber \\
\epsilon_2 & \simeq  2\frac{M_{\rm Pl}^2}{b^2(0)} + 2 M_{\rm Pl}^2 \, \frac{\Lambda_1^4}{f_b^2}\, \frac{{\rm cos}\Big(\frac{b(0)}{f_b}\Big)}{|b(0)\, \Lambda_{\rm cond}^3|} \simeq  2\frac{M_{\rm Pl}^2}{b^2(0)} + 2 \frac{M_{\rm Pl}^2}{f_b^2}\, \frac{\Lambda_1^4}{|b(0)\, \Lambda_{\rm cond}^3|}\,,
 \end{align}
where in the last equality of the second line we have made the order of magnitude assumption that ${\rm cos}\Big(\frac{b(0)}{f_b}\Big) = \mathcal O(1)$. The reader should also recall that  $\Lambda_{\rm cond} ^3 <0$, 
$b(0) \, \Lambda_{\rm cond}^3 > 0$, which is indicated by the presence of absolute values in the respective terms in the expressions for $\epsilon_{1,2}$ in  \eqref{eiperiodic}. This means that the presence of the non-perturbatively-induced periodic modulations in the $b$-axion potential \eqref{periodicpot} leads to an increase in the value of $\epsilon_2$, as compared to the case of the purely linear-axion potential \eqref{effectivepot}, and, therefore, to smaller values of the spectral index $n_s$,  in view of \eqref{scalarspectral}.

Saturating, for concreteness,  the bound \eqref{b0bound}, and using \eqref{fbval}, \eqref{VEV_inflation_source}, \eqref{nosourcesI} (with $\mathcal N_S=\mathcal O(1)$) and \eqref{L1} (taking $n=1$ for naturalness), we obtain from \eqref{eiperiodic}:
\begin{align}\label{eiperiodicval}
\epsilon_1 \simeq 0.005\, , \quad \epsilon_2 &\simeq 0.02 +  14.6 \, \Big(\frac{M_s}{M_{\rm Pl}}\Big)^{-4} \, \frac{\Lambda_1^4}{|b(0)\, \Lambda_{\rm cond}^3|} \simeq  0.02 + 1.9 \times 10^{7} \, \Big(\frac{\Lambda_1}{M_s}\Big)^4 \nonumber \\
& \simeq \,  0.02 + 1.9 \, \xi^4 \, \times 10^{7} \, e^{-4\, \mathcal S_{\rm inst}} \lesssim   0.02 + 1.9 \, \xi^4 \, \times 10^{7} \, e^{-\frac{32\pi^2}{g^2_{\rm YM}}}\,.
\end{align} 
To reduce the value of $n_s$, as compared to the bound \eqref{strvme1values}, at the third-significant-figure level, so as to get better consistency with the central value of $n_s$ as measured by Planck collaboration~\cite{Planck:2018jri}, we should require that the non-perturbative contributions to the $\epsilon_2$ are of order $\mathcal O(10^{-3})$, that is, we impose:
\begin{align}\label{nonpertns}
\Delta n_s^{\rm non-perturb.} = - 1.9 \, \xi^4 \, \times 10^{7} \, e^{-4\, \mathcal S_{\rm inst}} = - \mathcal O(10^{-3}) \,  ,
\end{align}
which can be achieved provided the gauge-sector (Euclidean) instanton action has the order of magnitude (assuming for concreteness $\xi = \mathcal O(1)$) :
\begin{align}\label{instactionval}
\mathcal S_{\rm inst} \sim 5.9 \,,
\end{align} 
which would satisfy the bound \eqref{instbound} if  $5.9 \gtrsim 8\pi^2/g_{\rm YM}^2 $, corresponding to a fine structure constant $\alpha_{\rm YM}$ of the underlying Yang-Mills gauge theory (renormalized at the energy scale of the instantons) 
\begin{align}\label{fsc}
\alpha_{\rm YM} \equiv \frac{g_{\rm YM}^2}{4\pi} \gtrsim \frac{2\pi}{5.9} \sim 1.06, 
\end{align}
which is a natural value for a strongly coupled gauge theory. We note that with $\xi=\mathcal O(1)$, and the restriction on the string scale \eqref{cutoff}, the energy scale of the gauge instantons \eqref{L1} is set to $\Lambda_1 \sim 9 \times 10^{-4} \, M_{\rm Pl}$, that is in the GUT scale, and just above (by less than two orders of magnitude) the inflationary scale $H_I$ bound  \eqref{HI} set by Planck measurements~\cite{Planck:2018jri}. These are energy scales characterizing the (end of the) stiff era in the StRVM. Due to the assumed dilute gas approximation one should not expect dramatic changes in the transition from the stiff to the RVM inflationary epoch in the current variant of the StRVM with axion potential \eqref{axionperiodpot}, although this needs to be checked by repeating the dynamical system analysis of \cite{Dorlis:2024yqw} for this potential. There is margin in these considerations when the precise value of $\xi$ is taken into account. For instance, in the presence of $\xi \ne 1$, the one-instanton action becomes $\mathcal S_{\rm inst} \simeq 5.9 - {\rm ln}\xi $. If we require that the energy scale of instantons be equal to that of inflation $H_I$, so that instantons arise at the onset of inflation, then this implies $\xi \sim 10^{-2}$, which then yields $\mathcal S_{\rm inst} \sim 10. 5$, leading to an $\alpha_{\rm YM} \sim 0.6$.

Hence, upon considering models with gauge sectors that have instantons satisfying \eqref{instactionval}, \eqref{fsc}, one may get quite good agreement of the slow-roll inflationary parameters of this variant of StRVM with Planck data.\footnote{Note that the condition 
\eqref{nonpertns} is self consistent with the approximations made in the computation of $\epsilon_i, \, i=1,2, \dots$, in \eqref{eiperiodic}, specifically the order of magnitude of $\epsilon_1$, to which instanton corrections are subleading, of $\mathcal O(10^{-6})$.} 
The reader should also bear in mind that another crucial feature for consistency with the inflationary phenomenology is the sign  of the periodic modulation potential \eqref{axionperiodpot}. As already mentioned in footnote \ref{f15}, the chosen $+$ sign guarantees the absence of mass generation for the axion $b$ field, but also implies that the slow-roll parameter $\epsilon_2$, \eqref{slowroll}, increases its value relative to the case of the exactly-linear-axion potential. This in turn implies the relative reduction of the scalar spectral index $n_s$, \eqref{scalarspectral}, which in this way can match the Planck data~\cite{Planck:2018jri,Planck:2018vyg}. The exactly opposite situation (that is, relative reduction of $\epsilon_2$ and increase of $n_s$) would characterize the choice of the $-$ sign in the periodic modulation potential, which, in addition to being phenomenologically problematic, would also be responsible for the undesirable feature of mass generation for the $b$ 
axion during the inflationary epoch.

\subsection{World-sheet instantons and axions from string compactification}\label{sec:wsinst}

In realistic string theories, one faces a multiaxion situation, given that, apart from the string-model independent axion $b$, one also has axions arising from compactification~\cite{svrcek}. The latter are known to be characterized by additional non-perturbative effects, called world-sheet instantons, which can wrap around compactified cycles. These instantons exhibit a much richer structure  
than the gauge instantons discussed above in the case of the KR $b$ field, which notably does not couple to world-sheet instantons. The world-sheet-instanton structure is particularly rich in intersecting D-brane compactifications~\cite{Blumenhagen:2005mu,Blumenhagen:2006xt,Blumenhagen:2009qh}, which our StRVM can be consistently embedded to. 

The corresponding axion potentials contain now multiaxion configurations, which are characterized by both linear axion and periodic contributions due to world-sheet instantons.  For instance, in the simplified case, where only one compactification axion $a(x)$ is dominant, the effective axion potential of both, the KR axion $b(x)$ and the compactification axion $a(x)$, 
is of the form~\cite{silver,sasaki,Mavromatos:2022yql}
 \begin{equation}\label{effpot}
 V(a, \, b)={\Lambda_3}^4\left( \pm 1+ f_a^{-1}\, \tilde \xi_1 \, a(x) \right)\, \cos({f_a}^{-1} a(x))+\frac{1}{f_{a}}\Big(f_b \, {\Lambda_{\rm cond}}^3 + \Lambda_2^4 \Big) \, a(x) + V_{\rm eff}(b)\,, 
 \end{equation}
 where $V_{\rm eff}(b)$ is given by \eqref{periodicpot}, and 
 the parameters $\tilde \xi_1, f_a$, and the world-sheet-instanton-induced scales $\Lambda_2, \Lambda_3 $ are all determined from the underlying microscopic string theory model. The $\pm$ sign inside the parenthesis of the first term in the right-hand side of \eqref{effpot} indicates whether a mass  for the compactification axion $a$ is generated ($-$) or not ($+$). This could affect the phenomenology of the post-inflationary epoch, in the 
 sense that in the massive-axion-$a$ case one may encounter an early matter-dominated era interpolating between the RVM inflation and the radiation epoch, which could lead to distinctive signatures in the profiles of GW during radiation.
  The parameter $f_a$ is the compactification-axion coupling determined from the specific string compactification of the  underlying microscopic string model under consideration. Both axions $b$ and $a$ couple to the gCS anomaly terms ({\it cf.} \eqref{eq:Action}), which leads to the linear axion terms in the effective potential \eqref{effpot}. The coefficient $\Lambda_{\rm cond}^3$ is defined in \eqref{periodicpot} and is proportional to the real part of the gCS condensate \eqref{vevRCSfinal}.

The scale $\Lambda_2$ is associated with specific compactification effects.
Indeed, in the type IIB string model of \cite{silver}, involving D5-branes, wrapped around a two cycle $\Sigma^{(2)}$ of size 
$\ell \sqrt{\alpha^\prime} = \ell \, M_s^{-1}$, when compactification (dimensionless) axions ${\mathcal B}(x)$ are turned on, 
 the corresponding ${\mathcal B}(x)$ potential assumes the generic form 
\begin{align}\label{L1silver}
V_{\rm D5}^{\mathcal B} \sim \frac{\epsilon}{g_s \, (2\pi)^5}\, \sqrt{\ell^4 + \mathcal {\mathcal B}(x)^2}\, M_s^4  \, ,
\end{align}
where $g_s$ is the string scale, and $\epsilon$ is a parameter, which depends on the warp factor of the underlying string/brane models under consideration~\cite{silver}. 
In simplified scenarios of compactification, where the compact six-dimensional manifold is characterized by a single size of the compactification radii, $L\sqrt{\alpha^\prime}$, such that the compactified  volume is $\mathcal V^{(6)} \sim  (L \, \sqrt{\alpha^\prime})^6$, 
the axions ${\mathcal B}(x)$ are related to a canonically normalized axion $a(x)$, of mass-dimension $+1$, as~\cite{silver}: 
\begin{align}\label{typeIIB}
a(x)^2 \sim \frac{L^2}{3g_s^2 \, (2\pi)^7}\, {\mathcal B}(x)^2\, M_s^2 
\sim \frac{1}{6\, L^4} \, {\mathcal B}(x)^2 M_{\rm Pl}^2 \,,
\end{align}
where 
we took into account that 
in type IIB string models, we restrict ourselves for concreteness here, $\alpha^\prime M_{\rm Pl}^2 \sim \frac{2\, \mathcal V^{(6)}}{(2\pi)^7\, g_s^2}$. For large compactification axions $\mathcal B \gg \ell^2 $, as required by the consistency of the inflationary axion monodromy scenario of \cite{silver} with the cosmological data~\cite{Planck:2018jri}, one obtains from \eqref{L1silver} a linear axion-$a$ potential term of the form \eqref{effpot}, corresponding to a scale 
\begin{align}\label{L2} 
\frac{\Lambda_2^4}{f_a}  \sim \frac{\epsilon}{L}\, \sqrt{\frac{3}{(2\pi)^3}}\, M_s^3\,.
\end{align}
For the specific type IIB brane-compactification models of \cite{silver}, 
the dimensionless parameter $\widetilde \xi $ has been assumed relatively small, so that the structures 
$a \, {\rm cos}(a/f_{a})$  in \eqref{effpot}  do not to affect significantly the slow-roll inflation driven by the linear potential.

In our work, we can extend the considerations beyond the type IIB compactification models of \cite{silver}, and treat the effective potential \eqref{effpot} 
as phenomenological, with the constraint though that the following hierarchy of scales is valid~\cite{Mavromatos:2022yql}:
 \begin{align}\label{constr}
 \Big(\frac{f_b}{f_a} + \frac{\Lambda^4_2}{f_a\, \Lambda^3_{\rm cond}} \Big)^{1/3}\, \Lambda_{\rm cond}  \, < \, \Lambda_3 \, < \, \Lambda_1 \ll \Lambda_{\rm cond} ~, 
 \end{align}
where we remind the reader that the scale $\Lambda_1$ denotes the target-space gauge instanton scale that induces periodic modulations of the KR potential ({\it cf.} \eqref{periodicpot}). Such a hierarchy 
ensures that the dominant effects in the axion potential come from the anomaly condensate, so that the spirit of the StRVM examined above, as regards the induced RVM inflation
and its slow-roll parameters, 
is maintained.

We reserve a detailed analysis of the effects of more general multiaxion configurations with potentials of the form \eqref{effpot}, in which one deviates from the scale hierarchy 
\eqref{constr}, on the inflationary era for a future publication. 
For example, we may consider multiaxion StRVM models in which 
there are no target-space gauge instantons, but only world-sheet instantons pertaining to compactification axions. In such a case, one may set $\Lambda_1=0$ and remove this scale from \eqref{constr}, keeping only the gCS and the world-sheet instanton scales $\Lambda_2, \Lambda_3$, in order to examine their hierarchies so that the phenomenologically correct slow-roll inflationary parameters are obtained. 

Before closing the section we remark, for completeness, that the existence of periodic modulations in 
the effective multiaxion potentials
leads~\cite{sasaki,Mavromatos:2022yql} to an enhancement of the densities of rotating primordial black holes (PBH) produced during inflation. In the context of the StRVM, such an enhancement of PBH production can be very significant, since it can affect the 
duration of the reheating phase after RVM inflation~\cite{Mavromatos:2022yql}. Indeed, as already mentioned, under some circumstances and for some regions of the parameter space of the models, there is 
the possibility of existence of an intermediate matter-dominated phase between the RVM inflation and radiation eras~\cite{Carr:2018nkm}. This can enhance significantly the populations of PBHs, and, as a consequence, the  profiles of the GWs produced from the coalescence of such black holes. This would lead in turn to observable in principle modifications of the spectrum of the GWs during the early radiation era. In fact, as argued in \cite{Mavromatos:2022yql}, by looking at the details of such GW profiles via future interferometers, one can in principle distinguish the effects of the StRVM from generic string-inspired axion-monodromy inflationary models with linear axion potentials, such as those discussed in \cite{silver,sasaki}. Furthermore, the existence of early-matter-dominated eras, preceding radiation, in the StRVM might lead to additional constraints of the model, associated with the requirement of avoiding the overproduction of PBHs~\cite{Papanikolaou:2024rlq}.
In view of the above discussion in this section, we should also consider the target-space gauge-instanton contributions to the periodic modulations of the potential \eqref{periodicpot} and examine their effects on the aforementioned features. This is left for future work.

\section{Conclusions and Outlook}\label{sec:concl}

In this work, we have 
elaborated further in the properties of the so-called Stringy Running-Vacuum-Model (StRVM) of Cosmology. The fundamental fields of this approach belong to the massless gravitational multiplet of the underlying string theory, namely the graviton, dilaton (which in our approach is assumed constant), and the (3+1)-dimensional dual of the antisymmetric-tensor field strength, which is a pseudoscalar field, the so called KR axion.
The model entails a linear-in-cosmic time KR axion background, which makes a (remote) connection of this approach with the works of \cite{Antoniadis:1988aa,Antoniadis:1988vi,Antoniadis:1990uu} in stringy cosmologies, involving such configurations of stringy axions. The linear nature of the KR axion of the StRVM is linked to the formation of a condensate 
of the gravitational Chern-Simons (gCS) term in the effective action. The latter is induced by chiral primordial GW, which are produced by a sufficiently large number of sources.
The condensate leads to a linear KR-axion (monodromy) potential, driving an RVM-type inflationary period.

Specifically, within a weak (perturbative) quantum gravity path-integral approach of the chiral GW tensor perturbations about the expanding-universe spacetime FLRW background, we have shown first, rather rigorously, that the real part of the gCS condensate becomes proportional to the number of GW sources.  
We have proceeded, next, in evaluating 
the non-trivial imaginary parts of the gCS condensate.  Although in the symmetric quantum-operator ordering such parts vanish, nonetheless in the current article we have argued in favor of their physical significance, and therefore have evaluated them in a specific ordering scheme we adopted here, in order to match our current findings  with those of the classical dynamical-system analysis of \cite{Dorlis:2024yqw}. The latter also indicates a finite-life time of the linear-axion-potential-induced inflation, which in our weak quantum-gravity approach is linked to the non-trivial imaginary parts of the gCS condensate.  The presence of imaginary parts indicates metastability of the inflationary vacuum, which therefore implies that the StRVM de-Sitter phase survives swampland criteria, consistently with the embedding of the model into the frameworks of microscopic string theories. 
 A careful and detailed treatment of the spacetime boundary terms of the CS gravity model has also been given, with the aim of ensuring that the imaginary parts of the gCS condensate are not canceled by boundary terms. We have demonstrated that these imaginary parts are quantum effects, being proportional to appropriate commutators of chiral tensor operators.  We have  
also calculated the magnitude of 
the vacuum expectation value (with respect to the appropriate Bunch-Davies vacuum) of these commutators, and provided an estimate of the life time of the inflationary era.
On requiring agreement of our results 
with the inflationary phenomenology, we restricted the string scale in this StRVM framework to be of order of one tenth of the four-dimensional (reduced) Planck mass. This result is in agreement with the previous analysis by the authors using a dynamical-system approach to the gCS-condensate-induced StRVM inflation, which is essentially a linear-axion (monodromy) inflation.
 Notably, the real part of the gCS condensate, turns out to be independent of the quantum-operator-ordering scheme. 

An important feature of the StRVM, which we did not discuss here, but we only mention it for completeness, is that the model is characterized by unconventional Leptogenesis (and subsequent Baryogenesis) during the early radiation era~\cite{bms}, which is induced 
by the linear-in-cosmic-time background of the KR axion. Indeed, such a background remains undiluted during the RVM inflationary phase of the model, and is responsible for inducing the aforementioned generation of Lepton asymmetry during the post-inflationary early radiation era in models involving massive right-handed neutrinos (RHN), which can be accommodated within the underlying string theories. In the presence of the KR axion background there is an asymmetric rate between the decays of the RHN into standard-model particles and antiparticles, which leads to a lepton asymmetry, according to the mechanism advocated in Refs.~\cite{deCesare:2014dga,Bossingham:2017gtm,Bossingham:2018ivs,Mavromatos:2020dkh}. Such a mechanism lies in 
the context of the Lorentz- and CPT- Violating Standard Model Extension (SME)~\cite{kostel}, 
in an axial (constant) background field $B_\mu$, 
which the RHN Lagrangian in the (approximately) linear-in-cosmic-time KR-axion background resembles. As discussed in detail in \cite{bms}, after leptogenesis, the cosmic rate of the axion background $\dot b \equiv B_0(T)$ continues to drop with the temperature $T$ of the Universe, as the latter cools down. In this framework, the current-era value of $B_0(T_{\rm CMB})$, with $T_{\rm CMB} = 2.7$~K today's value of the Cosmic-Microwave-Background (CMB) temperature, lies comfortably below the current upper bounds of the respective parameters of Lorentz- and CPT Violation in the SME framework~ \cite{Kostelecky:2008ts}.

 There are several important issues that remain open in our approach. The first, and most important, is a precise calculation of the gCS condensate in a gauge invariant way, within a proper (non-perturbative) and gauge invariant framework of Quantum Gravity, which is still lacking.
What we have done in this and previous related works is to demonstrate the presence of non-trivial complex condensates of gCS terms in the context of StRVM, which have been computed only within weak GW perturbations, in a particular gauge-fixed Lagrangian.
The gauge invariance of the gCS condensate has not been demonstrated, as this relies on the existence of the still elusive UV complete theory of quantum gravity.
Although in the case of StRVM such an UV completion is provided by the underlying microscopic string theory,
the latter contains infinite towers of stringy massive states, whose handling is still in its infancy. The dependence of the gCS on the UV cutoff of the low-energy effective theory obtained from strings
(that is, the string scale) is a manifestation of ignoring these massive towers, which makes the low-energy theory an open system.
It has been conjectured~\cite{Mavromatos:2024pho} that the imaginary parts of the condensate, we found in our approximate, weak-quantum-graviton treatment, do indicate the open nature of the low-energy gravitational field theory, and therefore one hopes that the metastable nature of the StRVM inflation will survive a proper and complete string-theory treatment. 

Another open issue, of less fundamental nature though, concerns the precise r\^ole of compactification axions, which coexist in string theories with the string-model-independent KR axions.  Non-perturbative stringy effects, either due to target-space gauge instantons, associated with the KR axions, or world-sheet instantons, pertinent to the compactification axions, 
generate periodic modulations of the respective effective potential for the axions, whose scales 
need to be estimated precisely in phenomenologically realistic string models. This task is far from being complete, and in fact it is not easy, due to the complexity and plethora of the available string models. In the current work we have adopted a phenomenological approach, by means of which we estimated the pertinent non-perturbative energy scales of the instanton-induced (multi)axion potential modulations by requiring agreement with the inflationary phenomenology. For concreteness, we have done this in simplified situations where only one of compactification axions was dominant.
In realistic string theories, there may be more than a single compactification axion that are dominant, which will complicate the situation, especially from the point of view of quantifying the enhancement of the production of primordial black holes during the inflationary era in this framework, and the associated modifications of the profiles of GW at the early radiation era. Related issues, such as the potential existence of intermediate matter eras in prolonged reheating situations of the StRVM after inflation, still remain wide open, 
whose study is an urgent one, given that there are in principle observable consequences of such a situation in future GW interferometers~\cite{Mavromatos:2022yql,Papanikolaou:2024rlq}. In fact, in this respect we mention recent works, related to potential experimental/observational signatures of string, or string-ispired, theories~\cite{Basilakos:2024diz,Basilakos:2023jvp,Basilakos:2023xof}, in {\it e.g.} data from the NANOgrav-facility~\cite{NANOGrav:2023gor,NANOGrav:2023hfp} and beyond, which can be adopted to test StRVM as well.

Last, but not least in the catalogue of future research directions, is the potential r\^ole of StRVM in providing an alleviation of the current cosmological tensions (Hubble and growth-of-structure tensions)~\cite{Gomez-Valent:2023hov}, and more general, observable deviations from $\Lambda$CDM in modern eras, where an accelerated expansion of our Universe occurs, according to observations. It must be noted that the modern era of the StRVM is still not understood fully, especially as far as the microscopic origin of the current-era observed acceleration of the Universe is concerned~\cite{Planck:2018vyg}. As remarked in \cite{bms}, given that ordinary matter has been largely depleted today from the energy budget of the Universe, in favor of Dark energy, it is possible that GW perturbations resurface and dominate the current Universe (provided there is a sufficient number of appropriate sources available in the late Universe). 
This may lead again to the reappearance of a GW-induced condensate of the gCS (which had been canceled at the end of the RVM inflation, as we have discussed above, by the then generated chiral matter fermions). As discussed previously, in such a case, it is possible that the vacuum energy density contains a term of $\mathcal O(H^2)$, which in our case would emerge from the average term $\mathcal N\Big\langle \delta S^{(2)}[b,g,h]\Big|_{A=0}/\delta g_{\mu\nu} \Big\rangle $ in the graviton Euler Lagrange equation \eqref{HFstation} corresponding to the present epoch, in which $H=H_0$. If this turns out to be true, then a metastable vacuum energy of approximately de-Sitter type would also characterize the current era, in analogy to the RVM inflationary vacuum discussed in this work. Such a situation would share features with the generic RVM approach to cosmology, which predicts observable, in principle, deviations from the $\Lambda$CDM 
paradigm due to the dominant $H_0^2$ terms in the vacuum energy density of the present epoch~\cite{rvm1,rvm2,rvmqft1,rvmqft2,rvmqft3,rvmqft4,tsiapi,tsiapi2,Gomez-Valent:2024tdb}. This remains to be seen, and constitutes the subject of a forthcoming work. \\

\noindent {\it Affaires \`a suivre ...}

\acknowledgments

 Preliminary results of this work have been presented by N.E.M. in an invited talk in the international workshop ``Applications of Field Theory to hermitian and Non-Hermitian Systems'', 
September 10-13 2024, Science Gallery (King's College London), London, UK, (\texttt{https://indico.cern.ch/event/1360299/}). N.E.M. thanks the organisers W.-Y. Ai and Sarben Sarkar for the invitation and for providing a thought-stimulating and interesting meeting. 
The work of P.D. is supported by a graduate scholarship from the
National Technical University of Athens (Greece). The work of N.E.M. is supported in part by the UK Science and Technology
Facilities research Council (STFC) and UK Engineering and Physical Sciences Research Council (EPSRC) under
the research grants  ST/X000753/1 and EP/V002821/1, respectively. The work of S.-N.V. is supported
by the Hellenic Foundation for Research and Innovation (H.F.R.I. (EL.ID.EK.)) under the ``5th Call for H.F.R.I. Scholarships to PhD Candidates"
(Scholarship Number: 20572). 
NEM also acknowledges participation in the COST Association Actions CA21136 “Addressing observational tensions in cosmology with systematics and fundamental physics (CosmoVerse)” and CA23130 "Bridging high and low energies in search of quantum gravity (BridgeQG)".\\

\appendix 
\section{Well Defined Variational Principle in Gravitational Chern - Simons Theory}\label{sec:bound}

In this Appendix, we elaborate further on the variational principle of the Gravitational Chern - Simons theory under the assumption of Dirichlet boundary conditions. Specifically, in what follows we present analytically the bulk and boundary contributions in this theory, and the necessary introduction of specific boundary terms to the action in order to have a well defined variational principle. Although the final results may be familiar to General Relativists, we include details here for the benefit of the average reader, as some of the intermediate steps are non trivial.

The Chern-Simons Gravitational theory is given by the following action:
\begin{equation}
	S=\int_\mathcal{M} d^4x \left[\mathcal{L}_{EH}+\mathcal{L}_{matter}+\mathcal{L}_{int}\right]
\label{eq:ActionB}
\end{equation}
where
\begin{align}
   & \mathcal{L}_{EH}= \sqrt{-g}\frac{R}{2\kappa^2}\label{EH}\\
   &\mathcal{L}_{matter}= -\sqrt{-g}\left(\frac{1}{2}(\partial_\mu b)(\partial^\mu b)+V(b)\right)\label{matter}\\
   & \mathcal{L}_{int} = -\sqrt{-g}A\;b\;R_{CS}\label{interaction}
\end{align}
with $\kappa=M_{\rm Pl}^{-1}$ is the inverse of the reduced Planck mass and $R_{CS}$ is the Chern-Simons topological term, defined in \eqref{RCS}. Assuming the variation of the metric and the matter field in the following way (weak-field approximation),\footnote{Here $h_{\mu\nu}$ is dimensionless, and the perturbative expansion is in powers of the weak field $h_{\mu\nu}$. This is to be contrasted with a coupling-constant($\kappa$) expansion in which one considers $g_{\mu\nu} \to g_{\mu\nu} + \kappa \, \widetilde h_{\mu\nu}$, where $\tilde h_{\mu\nu}$ has dimensions of mass.}
\begin{align}
    & g_{\mu\nu}\rightarrow g_{\mu\nu}+h_{\mu\nu}\\
    & b\rightarrow b+\delta b
\end{align}
we find the following first - order terms of the Lagrangians.\footnote{To derive this and the following expressions, we used the "xpert" package of Mathematica.   Our results are checked, and found consistent, with classic analyses in the relevant literature~\cite{tHooft:1974toh,Christensen:1979iy}  .} 
\begin{align}
     &\delta\mathcal{L}_{EH}^{(1)}= \sqrt{-g}\left[  -\frac{1}{2\kappa^2}\left(R_{\alpha\beta}-\frac{1}{2}g_{\alpha\beta}R\right)h^{\alpha\beta} \right]+\sqrt{-g}\left[ \frac{1}{2\kappa^2}\nabla_\beta\left(\nabla_\alpha h^{\alpha\beta}-\nabla^\beta h^\alpha_\alpha\right)     \right]\label{EHvariation}\\
     &\delta\mathcal{L}_{matter}^{(1)}= \sqrt{-g}
\left[\square b - V^\prime(b)\right]\delta b  +\sqrt{-g} \frac{1}{2} \left[  \nabla_\alpha b \nabla_\beta b -g_{\alpha\beta}\left(\frac{1}{2} \nabla_\mu b \nabla^\mu b +V(b)   \right)   \right] h^{\alpha\beta}- \sqrt{-g}\nabla_\mu \left(\delta b\nabla^\mu b\right)\label{mattervariation}\\
&\delta\mathcal{L}_{int}^{(1)}=\sqrt{-g}\left( -A R_{CS}\right)\delta b + \sqrt{-g}\left[ -A \;b\; R_\alpha\!^{\mu\nu\rho}\left( \widetilde{R}_{\beta\mu\nu\rho}-\widetilde{R}_{\nu\rho\beta\mu}  \right)   \right] h^{\alpha\beta} +\sqrt{-g}\left[ -A \;b \;\epsilon_{\beta\mu\rho\sigma}\;R_{\alpha\nu}\!^{\rho\sigma} \nabla^\nu\nabla^\mu h^{\alpha\beta}    \right]\label{intvariation1}
\end{align}

We can easily check from the above expressions that the first - order variation with respect to the matter field yield the equation of motion for the matter field in the bulk, 
\begin{equation}
    \square \  b= V^\prime(b) + A R_{CS}
    \label{mattereom}
\end{equation}
while giving rise also to a surface-boundary contribution to the action, which by virtue of Gauss’ theorem reads,
\begin{equation}
    \delta S_{matter/boundary} = - \int_{\partial \mathcal{M}} dS^\mu \delta b \;\nabla_\mu b
\end{equation}
with $\partial \mathcal{M}$ denoting the boundary of the spacetime manifold $\mathcal{M}$. Such a term trivially vanishes under the assumption of  Dirichlet boundary conditions on $\partial\mathcal{M}$, 
\begin{equation}
    \delta b\vert_{\partial\mathcal{M}}=0
\end{equation}

In Eq.~\eqref{EHvariation} the first term gives the Einstein-Hilbert Lagrangian contribution to the equations of motion, through the Einstein tensor, $G_{\mu\nu}=R_{\alpha\beta}-\frac{1}{2}g_{\alpha\beta}R $, as well as a boundary contribution, 
\begin{equation}
    \delta S_{EH/boundary}= \frac{1}{2\kappa^2} \int_{\partial \mathcal{M}} dS^\mu \left(  \nabla^\alpha h_{\alpha\mu}-\nabla_\mu h     \right)
    \label{EH_boundary_contribution}
\end{equation}
where we use the notation \eqref{traceh}, 
$h \equiv h^\alpha_\alpha$, for the trace of the metric perturbation. Such a boundary contribution is non-vanishing under Dirichlet boundary conditions\footnote{Notice that covariant derivatives are reduced to partial derivatives upon the application of Dirichlet boundary conditions.},
\begin{equation}
h_{\mu\nu}\vert_{\partial\mathcal{M}}=0,\;\;\gamma^{\rho\sigma}\partial_\rho h_{\mu\nu}\vert_{\partial\mathcal{M}}=0,\,\gamma^{\rho_{1}\sigma_{1}}\gamma^{\rho_{2}\sigma_{2}}\dots\gamma^{\rho_n\sigma_{n}}\partial_{\rho_1}\partial_{\rho_2}\dots \partial_{\rho_n} h_{\mu\nu}\vert_{\partial\mathcal{M}}=0, \quad\, \forall \, n\in\mathbb{N} \ .
\label{DirichletBoundaryConditions}
\end{equation}
The conditions \eqref{DirichletBoundaryConditions} imply that the metric perturbation has to be constant and vanishing on the boundary $\partial\mathcal{M}$. On the other hand, 
the normal components of the derivatives of the metric perturbations are in principle non-vanishing, thus contributing non-trivially to \eqref{EH_boundary_contribution}. In order to cancel the boundary contribution of \eqref{EH_boundary_contribution}, the Gibbons-Hawking-York (GHY) term is introduced \cite{Gibbons:1976ue}, 
\begin{equation}
     S_{GHY} = \frac{1}{\kappa^2}\int_{\partial\mathcal{M}}\epsilon \;d^3y \sqrt{\vert\gamma\vert} K \ .
    \label{eq:GHYterm}
\end{equation}
where $K=g^{\alpha\beta}K_{\alpha\beta}$, with $K_{\alpha\beta}$ the extrinsic curvature,
\begin{equation}
K_{\alpha\beta}=\gamma^\rho_\alpha\gamma^\lambda_\beta\nabla_\rho n_\lambda
\label{eq:extrinsiccurvature_definition}
\end{equation}
and we have defined the induced metric on the spacelike or timelike boundary,  $\partial\mathcal M$, as:
\begin{equation}
    \gamma_{\alpha\beta}=g_{\alpha\beta}-\epsilon n_\alpha n_\beta
    \label{eq:inducedmetric_definition}
\end{equation}
with $n^\alpha$ the normal on $\partial\mathcal{M}$, satisfying the normalization condition $n^\alpha n_\alpha=\epsilon=\pm1$.  In order to invert \eqref{eq:extrinsiccurvature_definition}, so as to express $\nabla_\alpha n_\beta$ in terms of $K_{\alpha\beta}$, we make a decomposition of $\nabla_\alpha n_\beta$ into parallel and normal parts to the boundary $\partial\mathcal{M}$,\footnote{Suppose that $T^\mu$ is a vector. Then, we can decompose,
\begin{equation*}
    T^\mu= \gamma^\mu_\rho T^\rho + \epsilon n^\mu n_\rho T^\rho= T_{\parallel}^\mu+T_{\perp}^\mu
\end{equation*}
where the first term denotes the parallel, while the second one the normal, component of the vector on the boundary $\partial\mathcal{M}$.
Such a decomposition  is generalized to higher-rank tensorial quantities. 
For example, if we have a second rank 
contravariant tensor $T^{\alpha\beta}$ we obtain: 
\begin{equation*}
    T^{\mu\nu}= n^\mu n^\nu T^{\vec{n}\vec{n}}+\epsilon n^\mu \gamma^\nu_\rho T^{\vec{n}\rho}+\epsilon n^\nu \gamma^\mu_\rho T^{\rho\vec{n}}+\gamma^\mu_\rho \gamma^\nu_\sigma T^{\rho\sigma}
\end{equation*}
and so on, where we introduced the
notation~\cite{Grumiller:2008ie}, 
\begin{align}\label{normnot}
T^{\vec{n}}=n_\mu T^{\mu}\,,
\end{align}
that is, 
when the symbol $\vec n$ is in the place of an index in a tensorial quantity, then that index is contracted with the normal vector. We can easily see that, when an index of a tensor is contracted with the normal vector, only its normal component survives, for example, 
\begin{equation*}
    n_\mu T^{\mu\nu}= \epsilon n^{\nu}T^{\vec{n}\vec{n}} +\gamma^\nu_\rho T^{\vec{n}\rho}\,. 
\end{equation*}} from which we obtain, 
\begin{equation}
    \nabla_\alpha n_\beta=K_{\alpha\beta}+\epsilon n_\alpha\gamma^\rho_\beta\nabla_{\vec{n}} n_\rho\, ,
    \label{nablatoextrinsic}
\end{equation}
in which we see that the component induced by the normal vector has only tangential contribution, a fact that is implied by $\nabla_\mu\left(n_\alpha n^\alpha\right)=0$. Furthermore, the extrinsic curvature is the tangential derivative of the normal vector,
\begin{equation}
    \gamma^\alpha_\sigma\nabla_\alpha n_\beta=K_{\sigma\beta}
\label{eq:tangentialderivativeofnormalvector}
\end{equation}

\par 

The GHY term is necessary in order to cancel the variation of the boundary contribution \eqref{EH_boundary_contribution}, leading in this way to a well defined variational principle, in the sense that the field equations are indeed stationary points of the action, $\delta S=0$. We mention at this point that  such boundary terms do not affect the equations of motion in the bulk. However, they have a crucial impact, since they give the only non-zero contribution to the on-shell action and in this sense, they yield a non-trivial partition function, from which we extract the physical/thermodynamic properties of the system, as in the case of black hole thermodynamics \cite{Gibbons:1976ue}. 
In what follows we derive explicitly the variation of GHY term, under the imposition of Dirichlet boundary conditions \eqref{DirichletBoundaryConditions}. This 
will lay the groundwork for the analysis of the gCS interaction term. 

To this end, we first observe that, upon variation of  \eqref{eq:GHYterm} with respect to the metric, $g_{\mu\nu}$, we obtain:
\begin{equation}
    \delta S_{GHY} = \frac{1}{\kappa^2}\int_{\partial\mathcal{M}}\epsilon \;d^3y \ \delta \left(\sqrt{\vert\gamma\vert} K \right)
\end{equation}
The calculations are simplified considerably by  noting that the variation of the determinant of the induced metric to any order $n$ in the variation, $\delta^n \left(\sqrt{\abs{\gamma}}\right)$, is proportional to the perturbation, $h_{\mu\nu}$, and, since it is an overall multiplicative factor to the expression, we can ignore it from the start, under the imposition of Dirichlet boundary conditions \eqref{DirichletBoundaryConditions}. This implies:
\begin{equation}
    \delta^n S_{GHY} \simeq \frac{1}{\kappa^2}\int_{\partial\mathcal{M}}\epsilon \;d^3y \ \sqrt{\vert\gamma\vert} \ \delta^n K \, , \quad \forall n\in \mathbb{N}
    \label{variation_GHY_dirichlet}
\end{equation}
where, form now on, we follow the notation of \cite{Grumiller:2008ie}, using the symbol $``\simeq"$  to denote equality under the application of the Dirichlet boundary conditions \eqref{DirichletBoundaryConditions}. The term that is important in the above variation is the one of the extrinsic curvature scalar, $\delta K$, given by:
\begin{equation}
    \delta K = - h^{\alpha \beta }K_{\alpha\beta}+g^{\alpha \beta}\delta K_{\alpha\beta} 
    \label{general_variation_extrinsic_scalar}
\end{equation}
where, $\delta g^{\alpha \beta} \equiv -h^{\alpha \beta}$. From the definition of the extrinsic curvature \eqref{eq:extrinsiccurvature_definition}, we obtain,
\begin{equation}
\begin{aligned}
    \delta K_{\alpha\beta}=&\, h^\mu_\nu\left( \gamma^\nu_\alpha\gamma^\rho_\mu \gamma^\lambda_\beta + \gamma^\rho_\alpha\gamma_\mu^\lambda\gamma^\nu_\beta   \right)\nabla_\rho n_\lambda+ \gamma^\rho_\alpha\gamma^\lambda_\beta\delta\left(\nabla_\rho n_\lambda\right)\\
    =&\, h^\mu_\nu\left( \gamma^\nu_\alpha\gamma^\rho_\mu \gamma^\lambda_\beta + \gamma^\rho_\alpha\gamma_\mu^\lambda\gamma^\nu_\beta   \right)\nabla_\rho n_\lambda  +\gamma^\rho_\alpha\gamma^\lambda_\beta \left( \nabla_\rho\delta n_\lambda -
 n_\nu \delta\Gamma^\nu_{\rho\lambda}\right)\\
 \simeq&\,\gamma^\rho_\alpha\gamma^\lambda_\beta \left( \nabla_\rho\delta n_\lambda -
 n_\nu \delta\Gamma^\nu_{\rho\lambda}\right)
    \end{aligned}
\label{variation_of_extrinsic_curvature}
\end{equation}
and then,
\begin{equation}
   \delta K \simeq \gamma^{\rho\lambda} \left( \nabla_\rho\delta n_\lambda -
 n_\nu \delta\Gamma^\nu_{\rho\lambda}\right)\, .
   \label{variation_extrinsic_scalar}
\end{equation}
The variation of the normal vector is given by \cite{Grumillerbook}:
\begin{equation}
     \delta n_{\alpha} =\frac{\epsilon}{2}n_{\alpha}n^{\lambda}n^{\nu}h_{\lambda\nu}\equiv\frac{\epsilon}{2}n_{\alpha}h_{\vec{n}\vec{n}}\, ,
     \label{variation_normal_vector}\,,
\end{equation}
with the notation \eqref{normnot}. 
Thus,
\begin{equation}
    \nabla_{\rho} \delta n_{\lambda} = \frac{\epsilon}{2}\nabla_{\rho}\left(n_{\lambda} h_{\vec{n}\vec{n}}\right) \simeq \frac{\epsilon}{2} n_{\lambda} \nabla_{\rho} h_{\vec{n}\vec{n}} \, ,
    \label{nabla_delta_normal}
\end{equation}
and then, 
\begin{equation}
\gamma^{\rho\lambda}\nabla_\rho\delta n_\lambda\simeq 0\,.
\end{equation}
From the variation of the affine connection, 
\begin{equation}
    \delta \Gamma^\nu_{\rho\lambda}=\frac{1}{2}\left( \nabla_\rho h^\nu_\lambda+\nabla_\lambda h^\nu_\rho -\nabla^\nu h_{\rho\lambda}   \right)
    \label{affinevariation}
\end{equation}
one can easily see that
\begin{equation}
\gamma^{\rho\lambda}\delta\Gamma^\nu_{\rho\lambda}\simeq-\frac{1}{2}\gamma^{\rho\lambda}\nabla^\nu h_{\rho\lambda}\, ,
    \end{equation}
where we neglected terms that yield tangential derivatives on the boundary. Hence, we obtain,
\begin{equation}
    \delta K\simeq\frac{1}{2}n_\nu\gamma^{\rho\lambda}\nabla^\nu h_{\rho\lambda}=\frac{1}{2}\left( 
 n_\nu g^{\rho\lambda}-\epsilon n_\nu n^\rho n^\lambda   \right)\nabla^\nu h_{\rho\lambda}\simeq-\frac{1}{2}n^\nu\left( 
 \nabla^\rho h_{\rho\nu}-\nabla_\nu h \right)\, ,
\end{equation}
which we arrived at by first expanding $\gamma^{\rho\lambda}=g^{\rho\lambda} - \epsilon n^\rho n^\lambda$, and then replacing $\epsilon n_\nu n^\rho=\delta^\rho_\nu-\gamma^\rho_\nu$.
Thus, we finally conclude that:
\begin{equation}
    \delta S_{GHY} \simeq  - \frac{1}{2\kappa^2}\int_{\partial\mathcal{M}} \;dS^{\mu} \left( \nabla^{\alpha}h_{\alpha \mu} -   \nabla_{\mu} h  \right) 
\end{equation}
which gives us back the term that exactly cancels the variation of the Einstein-Hilbert boundary term  \eqref{EH_boundary_contribution}.

Next we turn our attention to Eq.~\eqref{intvariation1}, that is  the contribution of the gCS term. As we have already mentioned, the contribution to the matter-field equation of motion is straightforwardly obtained. We focus on the gravitational equations of motion. Our aim is to obtain the remaining boundary contribution,
in order to see, following \cite{Gibbons:1976ue}, which boundary term is required for the CS part of the action to cancel any boundary contribution, again aiming at establishing a well-defined variational principle under the  Dirichlet boundary conditions \cite{Grumiller:2008ie}.  

To this end, we first observe that 
Eq.~\eqref{intvariation1} can be equivalently written as:
\begin{equation}
\begin{aligned}
    \delta\mathcal{L}_{int}^{(1)}= &\sqrt{-g}\left(-AR_{CS}\right)\delta b +\sqrt{-g}\left\{ -A b R_\alpha\!^{\mu\nu\rho}\left( \widetilde{R}_{\beta\mu\nu\rho}-\widetilde{R}_{\nu\rho\beta\mu}   \right) -2A \nabla^\mu\left[ \nabla^\nu\left(b \widetilde{R}_{\alpha\nu\beta\mu}\right)  \right] 
  \right\}h^{\alpha\beta} +\\
  & + \sqrt{-g}\left[  \nabla^\nu\left( -2Ab\widetilde{R}_{\alpha\nu\beta\mu}\nabla^\mu h^{\alpha\beta}  \right)  +2A \nabla^\mu\left[\nabla^\nu\left(b\widetilde{R}_{\alpha\nu\beta\mu} \right) h^{\alpha\beta}\right] \right].
  \end{aligned}
  \label{intvariation2}
\end{equation}
The terms in the curly brackets contribute to the gravitational equations of motion in the bulk, while the terms in square brackets are the boundary contributions. We concentrate on the bulk-part first. Expanding the covariant derivatives, we obtain:
\begin{equation}
    \sqrt{-g}\left\{\dots\right\}h^{\alpha\beta} = \sqrt{-g}\left\{-A b R_\alpha\!^{\mu\nu\rho}\left( \widetilde{R}_{\beta\mu\nu\rho}-\widetilde{R}_{\nu\rho\beta\mu}   \right) 
  -2A\left( \nabla^\mu(\nabla^\nu b) \widetilde{R}_{\alpha\nu\beta\mu} +\nabla^\mu b \nabla^\nu \widetilde{R}_{\alpha\nu\beta\mu}+b\nabla^\mu\nabla^\nu\widetilde{R}_{\alpha\nu\beta\mu}      \right)        \right\}h^{\alpha\beta}
  \label{bulkCScontribution}
\end{equation}
where we omitted the term $\nabla^\nu b \nabla^\mu\widetilde{R}_{\alpha\nu\beta\mu}=0$, due to the second Bianchi identity (see for example \cite{Poisson:2009pwt}). 

We next focus on the last term in the curly brackets of \eqref{bulkCScontribution}, in order to show that it cancels out the first one. Again, due to the second Bianchi identity, we have:
\begin{equation}
\begin{aligned}
-2Abh^{\alpha\beta}\nabla^\mu\nabla^\nu\widetilde{R}_{\alpha\nu\beta\mu}=&-2A\;b\;h_{\alpha\beta}\left[\nabla_\mu,\nabla_\nu \right]\widetilde{R}^{\alpha\nu\beta\mu} \\
=&-2A\;b\;h_{\alpha\beta}\left( 
 R^{\alpha}\!_{\lambda\mu\nu} \widetilde{R}^{
\lambda\nu\beta\mu}-R_{\lambda\mu}\widetilde{R}^{\alpha\lambda\beta\mu}+R^\beta\!_{\lambda\mu\nu}\widetilde{R}^{\alpha\nu\lambda\mu}+R_{\lambda\nu}\widetilde{R}^{\alpha\nu\beta\lambda}  \right)\\
=&-2A\;b\;h^{\alpha\beta}\left(   R_{\alpha}\!^{\lambda\mu\nu} \widetilde{R}_{\lambda\nu\beta\mu} +R_{\beta}\!^{\lambda\mu\nu}\widetilde{R}_{\alpha\nu\lambda\mu}      \right)
\end{aligned}
\end{equation}
where in the first line we used the fact that $\nabla_\nu\nabla_\mu\widetilde{R}^{\alpha\nu\beta\mu}=0$, and in passing from the second to the third line we used the symmetry of the Ricci tensor, $R_{\mu\nu}=R_{\nu\mu}$. Using the fact that the perturbation is symmetric, $h_{\alpha \beta}=h_{\beta\alpha}$, we obtain,
\begin{equation}
-2Abh^{\alpha\beta}\nabla^\mu\nabla^\nu\widetilde{R}_{\alpha\nu\beta\mu}= -2A \;b\; h^{\alpha\beta} R_{\alpha}\!^{\lambda\mu\nu} \left(   \widetilde{R}_{\lambda\nu\beta\mu} + \widetilde{R}_{\beta\nu\lambda\mu} \right) =
-2A \;b\; h^{\alpha\beta} R_{\alpha}\!^{\mu\nu\rho} \left( \widetilde{R}_{\beta\rho\mu\nu} +\widetilde{R}_{\mu\rho\beta\nu}          \right) \,,
\end{equation}
where the last equality has been obtained by 
an appropriate re-arrangement of the dummy indices,

Now using the anti-symmetry of the Riemann tensor on its last two indices, we obtain,
\begin{equation}\label{a38eq}
-2A\;b\;h^{\alpha\beta}\nabla^\mu\nabla^\nu\widetilde{R}_{\alpha\nu\beta\mu}= -2A \;b\; h^{\alpha\beta} R_{\alpha}\!^{\mu\nu\rho}\left( \widetilde{R}_{\beta\rho\mu\nu} -\widetilde{R}_{\mu\nu\beta\rho}          \right) \ .
\end{equation}
We then observe that the term inside the parenthesis on the right-hand side of 
\eqref{a38eq} is anti-symmetric under the interchange of $\mu\leftrightarrow \nu$, which means that only the anti-symmetric part of the Riemann tensor in these respective indices contributes to the contraction. Thus, we have:
\begin{equation}\label{a39eq}
-2A\;b\;h^{\alpha\beta}\nabla^\mu\nabla^\nu\widetilde{R}_{\alpha\nu\beta\mu}= -A \;b\; h^{\alpha\beta} \left(R_{\alpha}\!^{\mu\nu\rho}-R_{\alpha}\!^{\nu\mu\rho}\right)\left( \widetilde{R}_{\beta\rho\mu\nu} -\widetilde{R}_{\mu\nu\beta\rho}          \right) .
\end{equation}
Using the first Bianchi identity, then, the term inside the first parenthesis on the right-hand side of \eqref{a39eq} can be written as, 
\begin{equation}
    R_{\alpha}\!^{\mu\nu\rho}-R_{\alpha}\!^{\nu\mu\rho} = -R_{\alpha}\!^{\rho\mu\nu}-R_{\alpha}\!^{\nu\rho\mu}-R_{\alpha}\!^{\nu\mu\rho}=-R_{\alpha}\!^{\rho\mu\nu}, 
\end{equation}
implying that \eqref{a39eq} 
can be expressed in the form:
\begin{equation}
-2A\;b\;h^{\alpha\beta}\nabla^\mu\nabla^\nu\widetilde{R}_{\alpha\nu\beta\mu}= A \;b\; h^{\alpha\beta}R_{\alpha}\!^{\rho\mu\nu} \left( \widetilde{R}_{\beta\rho\mu\nu} -\widetilde{R}_{\mu\nu\beta\rho}          \right) \, .
\end{equation}
We can easily see that such a term cancels the other contribution in \eqref{bulkCScontribution}, which is proportional to $b$. What remains is
\begin{equation}
    \sqrt{-g}\left\{\dots\right\}h^{\alpha\beta} = \sqrt{-g}\left\{
  -2A\left[ \nabla^\mu(\nabla^\nu b) \widetilde{R}_{\alpha\nu\beta\mu} +\nabla^\mu b \nabla^\nu \widetilde{R}_{\alpha\nu\beta\mu}      \right]        \right\}h^{\alpha\beta}\, .
\end{equation}
Given that covariant derivatives commute when acting on scalar quantities, and 
taking into account the contraction with the metric perturbation, which is symmetric in its indices, we finally obtain,
\begin{equation}
    \sqrt{-g}\left\{\dots\right\}h^{\alpha\beta} = \sqrt{-g} \left\{ -A \nabla^\nu \left[  (\nabla^\mu b) \left( \widetilde{R}_{\alpha\nu\beta\mu}+\widetilde{R}_{\beta\nu\alpha\mu}  \right)   \right]       \right\}h^{\alpha\beta}  = \sqrt{-g} \left(2A C_{\alpha\beta}      h^{\alpha\beta}\right)
    \, ,
\end{equation}
where $C_{\alpha\beta}$ is the Cotton tensor~\cite{Jackiw}, defined as :
\begin{equation}
C_{\alpha\beta}= -\frac{1}{2}\nabla^\nu \left[  (\nabla^\mu b) \left( \widetilde{R}_{\alpha\nu\beta\mu}+\widetilde{R}_{\beta\nu\alpha\mu}  \right)   \right] \, . 
\end{equation}

Hence, combining with \eqref{EHvariation} and \eqref{mattervariation}, we arrive at the gravitational field equations of the CS gravity~\cite{Jackiw}, 
\begin{equation}
    G_{\mu\nu}= \kappa^2 T^{(b)}_{\mu\nu}+ 4 A \kappa^2 C_{\mu\nu}\, ,
\end{equation}
where
\begin{equation}
T^{(b)}_{\mu\nu}=\nabla_\alpha b \nabla_\beta b -g_{\alpha\beta}\left(\frac{1}{2} \nabla_\mu b \nabla^\mu b +V(b)   \right)  \, ,
\label{matterenergymomentum}
\end{equation}
is the energy-momentum tensor of the $b$-axion field.

We proceed next by considering the boundary contribution coming from the gCS interaction. From \eqref{intvariation2}, in the second line, we see that the second term of the boundary contribution trivially vanishes upon assuming Dirichlet boundary conditions \eqref{DirichletBoundaryConditions}. As such, the only non-trivial contribution comes from the first term, given by, 
\begin{equation}
    \delta S^{(1)}_{int/boun dary}= 2A \int_{\partial\mathcal{M}}dS^\nu\;\; b \;\widetilde{R}_{\alpha\nu\mu\beta}\nabla^\mu h^{\alpha\beta}\, .
    \label{int/boundary1}
\end{equation}
 We shall now proceed in the same way as we did for the boundary contribution in the Einstein-Hilbert action. The analogous term in order to cancel the boundary contribution for the CS-coupling interaction given by \eqref{int/boundary1} is given by \cite{Grumiller:2008ie}:
\begin{equation}
    S^{\mathcal{B}}_{CS}= -8 A \int d^3 y \sqrt{\abs{\gamma}} \ b \ \mathcal{CS}(K)\, ,
\label{CS_counterterm_boundary}
\end{equation}
where 
\begin{equation}
   \mathcal{CS}(K) = \frac{1}{2}n_\alpha \varepsilon^{\alpha \beta \gamma \delta} K^{\rho}_{\beta}\nabla_{\gamma} K_{\delta \rho}\, .
   \label{CS_boundary_counterterm}
\end{equation}
Denoting with Latin indices the tangential components, the above quantity is written as,
\begin{equation}
    CS(K)=\frac{1}{2}\varepsilon^{ijk}K^l_i\mathcal{D}_j K_{kl}\, ,
    \label{CS(K)tangentialindices}
\end{equation}
where $\mathcal{D}_j$ is the covariant derivative on the boundary $\partial M$ compatible with the induced metric, $\gamma_{ij}$, where $\varepsilon^{ijk}=n_\alpha\varepsilon^{\alpha\beta\gamma\delta}$. Using the antisymmetry of $\varepsilon^{ijk}$ and the Gauss-Codazzi equation,
\begin{equation}
    R_{l\vec{n}jk}=\mathcal{D}_j K_{kl}-\mathcal{D}_k K_{jl}\, ,
\end{equation}
equation \eqref{CS(K)tangentialindices} is written as,
\begin{equation}
    CS(K)=\frac{1}{4}\varepsilon^{ijk}K^l_iR_{l\vec{n}jk}\, ,
\end{equation}
or, in covariant form as, 
\begin{equation}
    CS(K)=\frac{1}{4}n_\alpha n^\lambda \varepsilon^{\alpha\beta\gamma\delta}K_{\rho\beta} R^\rho_{\,\lambda\gamma\delta}=\frac{1}{2}n^\alpha n^\lambda K^{\rho\beta}\widetilde{R}_{\rho\lambda\alpha\beta}\, .
    \label{CS(K)withRiemann}
\end{equation}
The variation of \eqref{CS_counterterm_boundary} receives non-trivial contributions 
only from the variation of  \eqref{CS_boundary_counterterm}, since the variation of the induced metric trivially vanishes under the imposition of Dirichlet boundary conditions \eqref{DirichletBoundaryConditions}, 
\begin{equation}
    \delta S^{\mathcal{B}}_{CS}\simeq -8A\int d^3y\sqrt{\vert\gamma\vert} \ b \ \delta \  CS(K)\, .
\end{equation}
The variation of \eqref{CS(K)withRiemann} reads, 
\begin{equation}
    \delta CS(K) \simeq \frac{1}{2}n^\alpha n^\lambda \left( \widetilde{R}_{\rho \lambda\alpha\beta}\delta K^{\rho\beta} +K_{\rho}^{\beta}\delta\widetilde{R}^\rho_{\ \lambda\alpha\beta}   \right)\, ,
    \label{CS_variation_1}
\end{equation}
in which we omitted  the variations of the normal vectors due to \eqref{variation_normal_vector}. In order to determine the first term, we use $\delta K^{\rho\beta}\simeq g^{\alpha\rho}g^{\beta\lambda}\delta K_{\alpha\lambda}$, as well as \eqref{variation_of_extrinsic_curvature}, \eqref{nabla_delta_normal} and \eqref{affinevariation}.  We find, 
\begin{equation}
    \delta K^{\rho\beta}\simeq \frac{1}{2}\gamma^{\rho\mu}\gamma^{\beta\sigma}\nabla_{\vec{n}} h_{\mu\sigma}\, .
\end{equation}
For the variation of the dual of the Riemann tensor we obtain, 
\begin{equation}
\delta\widetilde{R}^{\rho}_{\ \lambda\alpha\beta}\simeq \varepsilon^{\mu\nu}_{\ \ \  \alpha\beta}\nabla_\mu\delta\Gamma^\rho_{\,\lambda\nu}\,
\label{dual_riemann_variation}
\end{equation}
in which we used the variation of the Riemann tensor, 
\begin{equation}
    \delta R^{\rho}_{\ \lambda\alpha\beta}=\nabla_\mu\delta\Gamma^\rho_{\lambda\nu}-\nabla_\nu\delta\Gamma^\rho_{\lambda\mu}\, .
    \label{RiemannTensorVariation}
\end{equation}
Then, according to \eqref{affinevariation}, we obtain,
\begin{equation}
\delta\widetilde{R}^{\rho}_{\ \lambda\alpha\beta}\simeq    \frac{1}{2}\varepsilon^{\mu\nu}_{\ \ \ \alpha\beta}\left( \nabla_\mu\nabla_\lambda h^\rho_\nu +\nabla_\mu\nabla_\nu h^{\rho}_\lambda-\nabla_\mu\nabla^\rho h_{\lambda\nu}
  \right)\, .
\end{equation}
Substitution of the above to \eqref{CS_variation_1}, yields,
\begin{equation}\label{a59}
\begin{aligned}
    \delta CS(K) \simeq&\,\frac{1}{4}n^{\alpha}n^{\lambda}\left[\widetilde{R}_{\rho\lambda\alpha\beta}\gamma^{\rho\mu}\gamma^{\beta\sigma}\nabla_{\vec{n}}h_{\mu\sigma}+K^\beta_\rho \ \varepsilon^{\mu\nu}_{\ \ \ \alpha\beta}\left( \nabla_\mu\nabla_\lambda h^\rho_\nu +\nabla_\mu\nabla_\nu h^{\rho}_\lambda-\nabla_\mu\nabla^\rho h_{\lambda\nu}
  \right)  \right]\\
  \simeq &\,\frac{1}{4}n_\alpha  \widetilde{R}^{\mu\lambda\alpha\sigma}n_\lambda n^\delta\nabla_{\delta} h_{\mu\sigma}+\frac{1}{4}n^\alpha n^\lambda K^\beta_\rho \ \varepsilon^{\mu\nu}_{\ \ \ \alpha\beta}\nabla_\mu\nabla_\lambda h^\rho_\nu\\
  \simeq&\, \frac{\epsilon}{4}n_\alpha  \widetilde{R}^{\mu\lambda\alpha\sigma}\nabla_\lambda h_{\mu\sigma}+\frac{\epsilon}{4}K_{\beta\rho} \ \varepsilon^{\mu\nu\lambda\beta}\nabla_\mu\nabla_\lambda h^\rho_\nu\\
  \simeq&\, \frac{\epsilon}{4}n_\alpha  \widetilde{R}^{\mu\lambda\alpha\sigma}\nabla_\lambda h_{\mu\sigma}\,.
  \end{aligned}
\end{equation}
In passing from the first to the second 
and then the third line
in the above equation we used the decomposition of the induced metric \eqref{eq:inducedmetric_definition} and the fact the $R_{\vec{n}\vec{n}\mu\nu}=R_{\mu\nu\vec{n}\vec{n}}=0$, due to the anti-symmetry of the Riemann tensor, and we have ignored the double tangential derivatives of the metric perturbations, as these terms vanish upon using the Dirichlet boundary conditions \eqref{DirichletBoundaryConditions}. The last line of \eqref{a59} is obtained as a result of the total anti-symmetry of the Levi-Civita tensor, since only the commutator of the covariant derivatives contributes, which is proportional to the metric perturbation, and as such, vanishes due to the Dirichlet Boundary conditions \eqref{DirichletBoundaryConditions}. The final result is then given by, 
\begin{equation}
    \delta S^{\mathcal B}_{CS}= -2A \int \epsilon \,d^3y\sqrt{\vert\gamma\vert}\,n_\alpha\,b\,  \widetilde{R}^{\mu\lambda\alpha\sigma}\nabla_\lambda h_{\mu\sigma}\, ,
\end{equation}
which exactly cancels \eqref{int/boundary1}. \par Summarizing, the complete gravitational action, with a finite boundary $\partial\mathcal M$, reads:
\begin{equation}
    S=\int_\mathcal{M} d^4x\sqrt{-g}\left[ \frac{R}{2\kappa^2}-\frac{1}{2}\partial_\mu b \ \partial^\mu b 
+V(b) -A \;b \;R_{CS}       \right] +\frac{1}{\kappa^2}\int_{\partial\mathcal{M}} d^3y\sqrt{\vert\gamma\vert}\left(\,\epsilon\,  K- \; 8\;A\;\kappa^2\;b\;\mathcal{CS}(K)\right)\, ,
\label{CSACtioneithBoundary}
\end{equation}
with $\mathcal{CS}(K)$ given by \eqref{CS_boundary_counterterm}. The action \eqref{CSACtioneithBoundary} admits now a well defined variational principle under the assumption of Dirichlet boundary conditions \cite{Grumiller:2008ie}.

\section{Graviton propagation at second order in perturbations and two scalar modes in FLRW spacetime}\label{sec:AppB}
  In this Appendix, we expand the Lagrangian  corresponding to the action  \eqref{CSACtioneithBoundary} to second order in the metric perturbations, $h_{\mu\nu}$, with the aim of obtaining the graviton Lagrangian, and the associated propagator. First we shall deal with a general background, and then we shall apply our findings to the specific case of a FLRW background, where we shall show that the formalism becomes equivalent to the propagation of two scalar modes.
   
To this end, we first consider the Einstein Hilbert term \eqref{EH} expanded to second order in $h_{\mu\nu}$,
\begin{equation}
\begin{aligned}
    \delta\mathcal{L}_{EH}^{(2)}= \frac{1}{2\kappa^2}\sqrt{-g}&\left[ \frac{1}{4}\left( 
  \frac{1}{2}h^2 -h_{\alpha\beta}h^{\alpha\beta}     \right) R +\left(  h_{\alpha}^{\gamma}h^{\alpha\beta}-\frac{1}{2}h\,h^{\beta\gamma}      \right)   R_{\beta\gamma} -\frac{1}{4}\nabla_\beta h\nabla^\beta h -\frac{1}{2}h\nabla_\gamma\nabla^\gamma h \right.\\
  &+h^{\alpha\beta}\nabla_\alpha\nabla_\beta h +\nabla^\beta h \nabla_\gamma h_{\beta}^\gamma+\frac{1}{2}h\nabla_\gamma\nabla_\beta h^{\beta\gamma} -h^{\alpha\beta}\nabla_\beta\nabla_\gamma h_\alpha^\gamma-\nabla_\alpha h^{\alpha\beta}\nabla_\gamma h_\beta^\gamma- h^{\alpha\beta}\nabla_\gamma \nabla_\beta h_\alpha^\gamma\\
&\left.+h^{\alpha\beta}\nabla_\gamma\nabla^\gamma h_{\alpha\beta}-\frac{1}{2}\nabla_\beta h_{\alpha\gamma}\nabla^\gamma h^{\alpha\beta}+\frac{3}{4}\nabla_\gamma h_{\alpha\beta}\nabla^\gamma h^{\alpha\beta}
      \right] \ ,
\end{aligned}
\label{eq:gravitonaction1}
\end{equation}
where 
\begin{align}\label{traceh}
h=h_\mu^\mu = h_{\mu\nu}\, g^{\mu\nu}
\end{align}
is the trace of the graviton perturbation with respect to the background metric.

Next, we perform partial integrations in the terms that contain second order derivatives of the perturbation. Such terms can be cast in the following form
\begin{equation}
\delta\mathcal{L}_{EH,b}^{(2)}=\frac{1}{2\kappa^2}\nabla_\beta\left[ 
     -\frac{1}{2}h\nabla^\beta h +h^{\alpha\beta}\nabla_\alpha h +\frac{1}{2} h\nabla_\alpha h^{\alpha\beta}-h^{\alpha\beta} \nabla_\gamma h^\gamma_\alpha -h^{\alpha\gamma}\nabla_\gamma h^\beta_\alpha  +h^{\alpha\gamma}\nabla^\beta h_{\alpha\gamma}    \right]  \simeq 0
     \label{EH_Boundary_second}
\end{equation}
which vanishes upon the imposition of the Dirichlet boundary conditions \eqref{DirichletBoundaryConditions}, since every contribution turns out to be  proportional to the perturbation, $h_{\mu\nu}$. 

In this approach, therefore, one has to check whether there is any non-trivial boundary contribution from the GHY term \eqref{eq:GHYterm} at second order in the graviton perturbation,    $h_{\mu\nu}$. We first consider the  second order variation of GHY term with respect to the metric tensor, $g_{\mu\nu}$, and then impose the Dirichlet boundary conditions \eqref{DirichletBoundaryConditions}. To proceed, we remind the reader that, any variation, and consequently the second order variation, of the determinant of the induced metric, $\gamma_{\mu\nu}$, will vanish upon considering Dirichlet boundary conditions, yielding for the second variation of the GHY term \eqref{eq:GHYterm} the following form:
\begin{equation}
    \delta^2 S_{GHY} \simeq \frac{1}{\kappa^2} \int d^3 y \sqrt{\abs{\gamma}} \ \epsilon  \ \delta^2 K
\label{second_variation_GHY_dirichlet}
\end{equation}
The task is to define the second variation of the extrinsic curvature scalar, i.e. vary \eqref{variation_extrinsic_scalar}. Then, we can determine if there is any non-trivial boundary contribution to second order in the perturbation $h_{\mu\nu}$. Note also that $\delta h_{\mu\nu}=0$, and $\delta h^{\mu\nu}=\delta(g^{\alpha \mu}g^{\beta \mu}h_{\alpha \beta}) \simeq 0$ (under the use of Dirichlet boundary conditions \eqref{DirichletBoundaryConditions}). Then, using \eqref{general_variation_extrinsic_scalar}, \eqref{variation_of_extrinsic_curvature}, we obtain:
\begin{equation}
    \delta^2 K \simeq g^{\alpha\beta}\delta^2 K_{\alpha \beta} \simeq \gamma^{\rho}_{\alpha}\gamma^{\lambda}_{\beta}\left[\delta\left(\nabla_\rho \delta n_\lambda\right) - n_\nu \delta^2 \Gamma^{\nu}_{\rho \lambda} \right]\ . 
    \label{second_variation_extrinsic_scalar}
\end{equation}
For the first term on the right-hand side we have, 
\begin{equation}
    \delta\left( \nabla_\rho \delta n_\lambda   \right) = \nabla_\rho \delta^2 n_\lambda -\delta\Gamma^\sigma_{\alpha\beta}\delta n_\sigma\simeq \nabla_\rho \delta^2 n_\lambda\, .
    \label{delta_nabla_delta_n}
\end{equation}
where we used the fact that the variation of the normal vector $\delta n_\sigma \sim h_{\mu\nu}$. In view of \eqref{variation_normal_vector}, one can easily obtain that:
\begin{equation}
    \delta^2 n_\lambda = \delta \left(\frac{\epsilon}{2}n_\lambda n^\alpha n^\nu h_{\lambda \nu}\right) = \frac{3\epsilon}{4}n_\lambda h_{\vec{n}\vec{n}}h_{\vec{n}\vec{n}} \, .
    \label{second_variation_normal}
\end{equation}
Then, we have,
\begin{equation}
   \nabla_\rho\delta^2 n_\lambda 
\simeq 0\, ,
\end{equation}
and we see that the first term on the right-hand side of \eqref{second_variation_extrinsic_scalar} vanishes, 
\begin{equation}
     \delta\left( \nabla_\alpha\delta n_\beta   \right)\simeq 0\, .
    \label{first_term_second_GHY}
\end{equation}
For the remaining term, by using the variation of the Christoffel symbol \eqref{affinevariation}, we easily obtain,
\begin{equation}
\delta^2 \Gamma^\nu_{\rho\lambda}   \simeq0\, .
\label{second_term_second_GHY}
\end{equation}
Collecting all the results, we therefore find that \eqref{second_variation_GHY_dirichlet} trivially vanishes, 
\begin{equation}
    \delta^2 S_{GHY}\simeq 0 \, ,
    \label{GHY_second_order_ZERO}
\end{equation}
implying that there is no boundary contribution to the action of the metric perturbations under Dirichlet boundary conditions \eqref{DirichletBoundaryConditions}.
Thus, we are left only with a bulk contribution coming from the remaining terms of \eqref{eq:gravitonaction1}, which reads,
\begin{equation}
\begin{aligned}
    \delta\mathcal{L}^{(2)}_{EH}= \frac{1}{2\kappa^2}\sqrt{-g}&\left[   \frac{1}{4}\left( 
  \frac{1}{2}h^2 -h_{\alpha\beta}h^{\alpha\beta}     \right) R +\left(  h_{\alpha}^{\gamma}h^{\alpha\beta}-\frac{1}{2}h\,h^{\beta\gamma}      \right)   R_{\beta\gamma}  +\frac{1}{4}\nabla_\beta h\nabla^\beta h -\frac{1}{2}\nabla_\beta h \nabla_\alpha h^{\alpha\beta}\right.\\
  &\left.+\frac{1}{2}\nabla_\beta h_{\alpha\gamma}\nabla^\gamma h^{\alpha\beta}-\frac{1}{4}\nabla_\gamma h_{\alpha\beta}\nabla^\gamma h^{\alpha\beta}   \right],
  \end{aligned}
  \label{GravitonLgrangian1}
\end{equation}
up to boundary terms, which, as shown above ({\it cf.} \eqref{EH_Boundary_second}), vanish under Dirichlet boundary conditions \eqref{DirichletBoundaryConditions}. 

The presence of the Ricci tensor and the scalar curvature in \eqref{GravitonLgrangian1}
indicates that the above Lagrangian can be further simplified through the equations of motion of the background. For example, for vacuum propagation, i.e. $R_{\mu\nu}=0$, the above Lagrangian reduces to 
\begin{equation}
\begin{aligned}
    \delta\mathcal{L}^{(2)}_{EH,vacuum}= \frac{1}{2\kappa^2}\sqrt{-g}&\left[\frac{1}{4}\nabla_\beta h\nabla^\beta h -\frac{1}{2}\nabla_\beta h \nabla_\alpha h^{\alpha\beta}+\frac{1}{2}\nabla_\beta h_{\alpha\gamma}\nabla^\gamma h^{\alpha\beta}-\frac{1}{4}\nabla_\gamma h_{\alpha\beta}\nabla^\gamma h^{\alpha\beta}   \right].
  \end{aligned}
\end{equation}
In order to take into account the equations of motion of the background properly, it is convenient to express the Lagrangian \eqref{GravitonLgrangian1} in terms of the Einstein tensor, $G_{\mu\nu}$. To this end, first we write, 
\begin{equation}
    \frac{1}{2}\nabla_\beta h_{\alpha\gamma}\nabla^\gamma h^{\alpha\beta}= -\frac{1}{2}R^\alpha_{\,\lambda\beta\gamma}h_\alpha^\gamma h^{\lambda\beta}-\frac{1}{2}R_{\lambda\gamma}h_\alpha^\gamma h^{\alpha\lambda}+\frac{1}{2}\nabla^\gamma h_{\alpha\gamma}\nabla_\beta h^{\alpha\beta},
\end{equation}
up to irrelevant boundary terms, which vanish due to Dirichlet boundary conditions \eqref{DirichletBoundaryConditions}. The curvature terms appeared through the commutator of the covariant derivatives acting on the metric perturbation, $[\nabla_\beta,\nabla_\gamma]h^{\alpha\beta}$. Secondly, we replace $R_{\beta\gamma}=G_{\beta\gamma}+\frac{1}{2}g_{\beta\gamma}R$, to get, 
\begin{equation}
    \frac{1}{8}h^2R-\frac{1}{2}h\,h^{\beta\gamma}R_{\beta\gamma}= -\frac{1}{8}h^2\,R-\frac{1}{2}h\,h^{\beta\gamma}G_{\beta\gamma}.
\end{equation}
Combining all of the above, we obtain, 
\begin{equation}
\begin{aligned}
    \delta\mathcal{L}^{(2)}_{EH}=& \frac{1}{4\kappa^2}\sqrt{-g}G_{\mu\nu}h_\alpha^\mu h^{\alpha\nu}-\frac{1}{4\kappa^2}\sqrt{-g}\,h\,h^{\alpha\beta}G_{\alpha\beta}\\
    &+\frac{1}{2\kappa^2}\sqrt{-g}\left[ -\frac{1}{8}h^2\, R-\frac{1}{2}R^\alpha_{\,\lambda\beta\gamma}h^\gamma_\alpha  h^{\lambda\beta}+\frac{1}{2}\nabla^\gamma h_{\alpha\gamma}\nabla_\beta h^{\alpha\beta}+\frac{1}{4}\nabla_\beta h \nabla^\beta h-\frac{1}{2}\nabla_\beta h\nabla_\alpha h^{\alpha\beta}-\frac{1}{4}\nabla_\gamma h_{\alpha\beta}\nabla^\gamma h^{\alpha\beta}\right] \ .
    \end{aligned}
    \label{GravitonLgrangian2}
\end{equation}

Now we consider the contribution coming from the matter Lagrangian \eqref{matter}, from which we get,
\begin{equation}\label{dl2}
\delta\mathcal{L}_{matter}^{(2)}= \sqrt{-g}\left[ \frac{1}{4}h\,h_{\alpha\beta}\nabla^\alpha b\nabla^\beta b+\frac{1}{4}\left(h_{\alpha}^\mu h^{\alpha\nu}-\frac{1}{2}h h^{\mu\nu}\right)g_{\mu\nu}\left( V(b)+\frac{1}{2}\nabla_\alpha b\nabla^\alpha b   \right)-\frac{1}{2}h_{\alpha\beta}h_{\mu}^{\beta} \nabla^\alpha b\nabla^\mu b\right].
\end{equation}
and, when combined with \eqref{GravitonLgrangian1}, yields the result of \cite{tHooft:1974toh}.\footnote{The reader's attention is called to the different conventions between our work and that of \cite{tHooft:1974toh}, especially the definition of the Ricci tensor.}  In the same spirit as previously, we have to express the right-hand side of \eqref{dl2}
in terms of the matter energy-momentum tensor \eqref{matterenergymomentum} of the $b$ field. This will help us in identifying the background equations of motion  and  the remaining couplings of the metric perturbations with the matter $b$ field. By doing that, we obtain,

\begin{equation}
\delta\mathcal{L}_{matter}^{(2)}=-\frac{1}{4}\sqrt{-g}\,T^{(b)}_{\mu\nu} \,h^\mu_\alpha\,h^{\alpha\nu}+ \frac{1}{4}\sqrt{-g}h\,h^{\alpha\beta}\,T^{(b)}_{\alpha\beta}+\sqrt{-g}\left[\frac{1}{8}h^2\left(V(b)-\frac{1}{2}\nabla_\mu b\nabla^\mu b\right)-\frac{1}{4}h^\mu_{\alpha}h^{\alpha\nu}\nabla_\nu b \nabla_\mu b\right].
\label{mattercontributiontogravitonlagrangian}
\end{equation}

Thus, it is easy to see that the background equations of motion appear when combining \eqref{mattercontributiontogravitonlagrangian} and \eqref{GravitonLgrangian2}, while the remaining part gives the Lagrangian when the background equations of motion are taken into account.
Combining these two results, we obtain the action of the metric perturbations in the presence of the matter field,
\begin{equation}
\begin{aligned}
S^{(2)}[b,g,h]\Big\vert_{A=0}= \frac{1}{2\kappa^2}\int d^4x \sqrt{-g}&\left[-\frac{1}{8}h^2\, R-\frac{1}{2}R^\alpha_{\,\lambda\beta\gamma}h^\gamma_\alpha  h^{\lambda\beta}+\frac{1}{2}\nabla^\gamma h_{\alpha\gamma}\nabla_\beta h^{\alpha\beta}+\frac{1}{4}\nabla_\beta h \nabla^\beta h-\frac{1}{2}\nabla_\beta h\nabla_\alpha h^{\alpha\beta}\right.\\
&\left.-\frac{1}{4}\nabla_\gamma h_{\alpha\beta}\nabla^\gamma h^{\alpha\beta}+  \frac{\kappa^2}{4}h^2\left(V(b)-\frac{1}{2}\nabla_\mu b\nabla^\mu b\right)-\frac{\kappa^2}{2}h^\mu_{\alpha}h^{\alpha\nu}\nabla_\nu b \nabla_\mu b\right]\, ,
\end{aligned}
\label{GravandMatterWithoutGaugeFixing}
\end{equation}
where the subscript $A=0$ denotes the absence of the CS coupling. Diffeomorphism invariance of the theory implies the gauge transformation for the metric perturbations, 
\begin{equation}
    h_{\alpha\beta}\rightarrow h_{\alpha\beta}-\nabla_\alpha\xi_\beta-\nabla_\beta\xi_\alpha\, .
\end{equation}
Due to this symmetry, we can always choose a gauge in which the redefined field,
\begin{equation}
    \widetilde{h}_{\mu\nu}= h_{\mu\nu}-\frac{1}{2}g_{\mu\nu}h\, ,
\end{equation}
satisfies the traversability condition, $\nabla_\mu \widetilde{h}^{\mu\nu}=0$, i.e.
\begin{equation}
    \nabla_\mu h^{\mu\nu}=\frac{1}{2}\nabla_\nu h\, .
    \label{DeDonderGauge}
\end{equation}
However, there is a residual freedom in choosing the gauge, provided by the $\xi$ vectors satisfying,
\begin{equation}
    \square \xi_\nu +R_{\lambda\nu}\xi^\lambda=0\, ,
    \label{FinalGaugeFReedom}
\end{equation}
which, when imposed, leaves the two independent degrees of freedom of the graviton. With \eqref{FinalGaugeFReedom} there are four independent conditions that can be further applied in order to fix the gauge and identify the two independent degrees of freedom. Such a choice is the traceless condition, $h=0$, together with the vanishing of the temporal components of the metric perturbation, $h_{0\nu}=0$ \footnote{In order to clarify that the conditions  $h=0$ and $h_{0\nu}=0$ consist of $4$ independent conditions, observe that these two conditions are linearly dependent. When this dependence is taken into account there are $4$ independent conditions, which read; $g^{ij}h_{ij}=0$ and $h_{0i}=0,\,\forall i=1,2,3$.  }. In the case of a minimally coupled scalar field background the Tracelless - Transverse  (TT)  conditions 
\begin{align}\label{TTgauge}
h=0,\quad \nabla^\mu h_{\mu\nu}=0\,, 
\end{align}
 can in general be imposed for the graviton \cite{Fanizza:2021ngq}. When such a gauge is imposed, the only dependence on the (pseudo-) scalar field, $b$, comes from the last term in \eqref{GravandMatterWithoutGaugeFixing}. However, the last condition of $h_{0\nu}=0$ together with a homogeneous (pseudo-) scalar field, i.e. $b=b(t)$, as is the case in the FLRW background, eliminates such a direct dependence on $b(t)$. Hence, when the two remaining independent modes of the graviton are taken into account, Eq.~\eqref{GravandMatterWithoutGaugeFixing} reduces to:
\begin{equation}
     S^{(2)}[b,g,h]\Big\vert_{A=0}=S^{(2)}[g,h]\Big\vert_{A=0}= \frac{1}{2\kappa^2}\int d^4x \sqrt{-g}\left[-\frac{1}{4}\nabla_\gamma h_{\alpha\beta}\nabla^\gamma h^{\alpha\beta}- \frac{1}{2}R^\alpha_{\,\lambda\beta\gamma}h^\gamma_\alpha h^{\lambda\beta}
 \right] \ , 
 \label{S_2_A_zero}
\end{equation}
which is valid for FLRW background with a homogeneous (pseudo-) scalar field, $b=b(t)$. According to equation \eqref{FLRWmetricPerturbed}, for the tensorial perturbations of the FLRW spacetime, \eqref{S_2_A_zero} takes the following form, 
\begin{equation}
   S^{(2)}[b,g,h]\Big\vert_{A=0}= \frac{1}{4\kappa^2}\int d^4x \;\alpha^2(\eta) \left( \frac{1}{2}h_{ij}^\prime h^{\prime\,ij} -\frac{1}{2}\partial_k h_{ij}\partial^k h^{ij}   \right) \, .
   \label{S_2_A_zero1}
\end{equation}
Expanding $h_{ij}$ to the linear polarization basis \eqref{plus_cross_polarizations}, in order to express \eqref{S_2_A_zero} with respect to the two independent degrees of freedom, we obtain, 
\begin{equation}
   \left( \frac{1}{2}h_{ij}^\prime h^{\prime\,ij} -\frac{1}{2}\partial_k h_{ij}\partial^k h^{ij}\right) =  \sum_{\lambda=+,\times}h_{\lambda}^\prime(x) h_{\lambda}^\prime(x) -\partial_k h_{\lambda}(x)\partial^k h_{\lambda}(x)  \, .
\end{equation}
Thus, performing \cite{Dodelson:2003ft} the redefinition
\begin{equation}
h_{\lambda}(x)=\Psi_{\lambda}(x)/a(\eta)
\label{redefinition}
\end{equation}
we obtain, 
\begin{equation}
   S^{(2)}[b,g,h]\Big\vert_{A=0}= \frac{1}{4\kappa^2}\sum_{\lambda=+,\times}\int d^4x \, \left[\Psi_{\lambda}^\prime(x)\Psi_{\lambda}^\prime(x) -\partial_k \Psi_{\lambda}(x)\partial^k \Psi_{\lambda}(x)- \frac{\alpha^{\prime\prime}}{\alpha}\Psi_{\lambda}(x)\Psi_{\lambda}(x)\right] \, ,
   \label{S_2_A_zero2}
\end{equation}
up to an irrelevant boundary term, which vanishes upon assuming Dirichlet boundary conditions \eqref{DirichletBoundaryConditions}. The above action \eqref{S_2_A_zero2} is equivalent to that of two independent scalar fields, $\Psi_\lambda$, $\lambda=+,\times$, in a flat background, with a time-dependent quadratic potential, encoding the effect of the FLRW curved background. The introduction of the gCS coupling introduces derivative couplings between the cross ($\times$) and plus ($+$) polarizations. The modes  decouple when transformed to the helicity basis, and the formalism is again equivalent to that of two independent scalar modes, albeit with an anisotropic frequency in the time-dependent quadratic potential, signifying the parity-violating nature of the interaction of the homogeneous axionic field with $R_{CS}$.  \par  

We may now turn our focus on the term that contains the interaction with the $R_{CS}$ \eqref{interaction}. Since we are interested in the case of the FLRW background, and the derivation of the action for the GW is shown in detail in the main text, we will focus here only on the boundary terms that arose in the derivation, for which, we aim to prove that they are vanishing, under the assumption of Dirichlet boundary conditions \eqref{DirichletBoundaryConditions}. The latter is more convenient to be dealt with in a covariant form, in contrast to the formalism of Fourier expansions obtained in the main text, and this is the goal of the following considerations. In the Einstein-Hilbert Lagrangian, combined with the GHY term, in order to have a well defined first order variational principle, we showed explicitly that there is no boundary contribution at second order. In the case of the axion coupling with the $R_{CS}$, we showed that, by adding an appropriate boundary term, given in \eqref{CS_counterterm_boundary} and \eqref{CS_boundary_counterterm}, we can indeed have a well defined variational problem in  first order. We have to examine, though, what happens when we move to the second order of the perturbation, $h_{\mu\nu}$, and see if in that case, there is a non-trivial behavior regarding the boundary terms. By non-trivial behavior we are actually referring to the possible existence of terms that survive after imposing the Dirichlet boundary conditions \eqref{DirichletBoundaryConditions}, {\it i.e.} normal derivatives of the metric perturbation. For the variation of the boundary term of the gCS interaction given by \eqref{CS_counterterm_boundary}, \eqref{CS_boundary_counterterm} and \eqref{CS(K)withRiemann}, we obtain:
\begin{equation}
    \delta^2 S^{\mathcal{B}}_{CS} \simeq -8A \int d^3 y \sqrt{\abs{\gamma}} \ b  \ \delta^2 \mathcal{CS}(K)
 \  .
  \label{delta_square_action_Bound_CS}
\end{equation}
The second variation of $\mathcal{CS}(K)$ can be expanded in the following way: 
\begin{equation}
    \delta^2 \mathcal{CS}(K) \simeq \frac{1}{2}n^\alpha n^\lambda \left\{ 2 \delta\left(\widetilde{R}_{\rho \lambda\alpha\beta}\right)\delta K^{\rho\beta} +\delta^2 \left(K_{\rho}^{\beta}\right)\widetilde{R}^\rho_{\ \lambda\alpha\beta} + K_{\rho}^{\beta}\delta^2 \left(\widetilde{R}^\rho_{\ \lambda\alpha\beta} \right) \right\}
    \label{delta_2_CS(K)_with_riemann}
\end{equation}
where we used \eqref{CS_variation_1}.
Let us examine first the term $\delta^2 K^{\rho}_\beta$ on the right-hand side of \eqref{delta_2_CS(K)_with_riemann}; Taking into 
account that the first variation of $K_{\beta\rho}$  is given by \eqref{variation_of_extrinsic_curvature}, we obtain:
\begin{equation}
    \begin{aligned}
    \delta^2 K^{\rho}_\beta \simeq g^{\alpha \rho} \delta^2 K_{\alpha\beta} \simeq \gamma^{\rho\sigma}\gamma^\lambda_\beta\delta^2\left(\nabla_\sigma n_\lambda\right)\simeq 0\, ,
\end{aligned}
\label{delta_square_K_up_down}
\end{equation}
where we used \eqref{variation_normal_vector}, \eqref{first_term_second_GHY} and  \eqref{second_term_second_GHY}.
For the second variation of the dual-Riemann tensor and the last term of \eqref{delta_2_CS(K)_with_riemann}, we have that:
\begin{equation}
    \delta^2 \left(\widetilde{R}^\rho_{\ \lambda\alpha\beta}\right) \simeq \varepsilon^{\mu\nu}_{\ \ \alpha \beta} \ \delta \left(\nabla_\mu \delta \Gamma^{\rho}_{\lambda \nu}\right)
    \label{delta_2_dual_1}
\end{equation}
We can write the term on the right-hand side of \eqref{delta_2_dual_1} as:
\begin{equation}
     \delta \left(\nabla_\mu \delta \Gamma^{\rho}_{\lambda \nu}\right) = \nabla_\mu \delta^2 \Gamma^{\rho}_{\lambda \nu} + \delta \Gamma^{\rho}_{\mu\sigma}\delta \Gamma^{\sigma}_{\lambda\nu} - \delta \Gamma^{\sigma}_{\mu\lambda}\delta \Gamma^{\rho}_{\sigma\nu} - \delta \Gamma^{\sigma}_{\mu\nu}\delta \Gamma^{\rho}_{\lambda\sigma}
     \label{delta_nabla_delta_christoffel}
\end{equation}
The second variation of the action term \eqref{delta_square_action_Bound_CS} will thus contain terms of the form:
\begin{equation}\label{b34}
\begin{aligned}
     \delta^2 S^{\mathcal{B}}_{CS} \ni & \int d^3 y \sqrt{\abs{\gamma}} \ b  \ n^\alpha n^\lambda K^{\beta}_{\rho} \ \delta^2 \left(\widetilde{R}^\rho_{\ \lambda\alpha\beta}\right) \simeq  \int d^3 y \sqrt{\abs{\gamma}} \ b  \ n^\alpha n^\lambda K^{\beta}_{\rho}  \ \varepsilon^{\mu\nu}_{\ \ \alpha \beta} \ \delta \left(\nabla_\mu \delta \Gamma^{\rho}_{\lambda \nu}\right) \\ 
     =&  \int d^3 y \sqrt{\abs{\gamma}} \ b  \ n^\alpha n^\lambda K^{\beta}_{\rho}  \ \varepsilon^{\mu\nu}_{\ \ \alpha \beta} \ \left[\nabla_\mu \delta^2 \Gamma^{\rho}_{\lambda \nu} + \delta \Gamma^{\rho}_{\mu\sigma}\delta \Gamma^{\sigma}_{\lambda\nu} - \delta \Gamma^{\sigma}_{\mu\lambda}\delta \Gamma^{\rho}_{\sigma\nu} - \delta \Gamma^{\sigma}_{\mu\nu}\delta \Gamma^{\rho}_{\lambda\sigma} \right] \\ 
     = & 
     \int d^3 y \sqrt{\abs{\gamma}} \ b  \ n^\alpha n^\lambda K^{\beta}_{\rho}  \ \varepsilon^{\mu\nu}_{\ \ \alpha \beta} \ \left[\nabla_\mu \delta^2 \Gamma^{\rho}_{\lambda \nu} + 2 \delta \Gamma^{\rho}_{\mu\sigma}\delta \Gamma^{\sigma}_{\lambda\nu} \right]
     \end{aligned}
\end{equation}
 where, we replaced \eqref{delta_nabla_delta_christoffel} in the above integrated quantity and simplified the expression, using the total antisymmetric nature of $\varepsilon^{\mu\nu\alpha\beta}$ and the symmetry of $\Gamma^{\sigma}_{\mu\nu}$ under the exchange $\mu \leftrightarrow \nu$.  Using the variation of the Christoffel symbols \eqref{affinevariation}, the second term on the right-hand side of \eqref{b34} can be written as, 
 \begin{equation}
 \begin{aligned}
2\,n^\alpha n^\lambda K^\beta_\rho\varepsilon^{\mu\nu}_{\ \ \alpha \beta}\delta\Gamma^\rho_{\mu\sigma}\delta\Gamma^\sigma_{\lambda\nu}\simeq&\, \frac{1}{2}n^\alpha n^\lambda K^\beta_\rho\varepsilon^{\mu\nu}_{\ \ \alpha \beta}\nabla_\sigma h^\rho_\mu\left( \nabla_\lambda h^\sigma_\nu-\nabla^\sigma h_{\nu\lambda}
 \right)\\
\simeq & \,\frac{1}{2}K^\beta_\rho \varepsilon^{\mu\nu}_{\ \ \alpha \beta}n^\alpha n^\lambda n^\gamma n^\delta \nabla_\gamma h^\rho_\mu\left( \nabla_\lambda h_{\delta\nu}-\nabla_\delta h_{\nu\lambda}
 \right)
 \simeq \,0\, ,
 \end{aligned}
 \end{equation}
in which we omitted tangential derivatives of the metric perturbation. The last line vanishes, since the term inside the parenthesis on the left-hand side of the last equation is anti-symmetric under the interchange  $\lambda\leftrightarrow\delta$ and is contracted with the symmetric tensor $n^\lambda n^\delta$. On expanding the second variation of the Christoffel symbol,
we obtain for the first term on the right-hand side (last equality) of \eqref{b34}:
\begin{equation}
\begin{aligned}
\delta^2\Gamma^\rho_{\lambda\nu}= 
&  -\frac{h^{\rho\sigma}}{2}\left(\nabla_\lambda h_{\delta\nu}+\nabla_\nu h_{\delta\lambda}-\nabla_\delta
  h_{\lambda\nu}  \right)\\
  &+\frac{1}{2}g^{\rho\delta}\left(-\delta\Gamma^\sigma_{\lambda\delta} h_{\sigma\nu}-\delta\Gamma^\sigma_{\lambda\nu}h_{\delta\sigma}-\delta\Gamma^\sigma_{\nu\delta}h_{\sigma\lambda}-\delta\Gamma^\sigma_{\nu\lambda}h_{\delta\sigma}-\delta\Gamma^\sigma_{\delta\lambda}h_{\sigma\nu}-\delta\Gamma^\sigma_{\delta\nu}h_{\lambda\sigma}\right)\, .
  \end{aligned}
\end{equation}
As such, when $n^\alpha \varepsilon^{\mu\nu}_{\ \ \alpha \beta}\delta\Gamma^\rho_{\mu\sigma}\nabla_\mu$ acts on $\delta^2\Gamma^\rho_{\lambda\nu}$ only terms proportional to the metric perturbation and terms proportional to the tangential derivative appear, implying in this way that these terms are also vanishing under Dirichlet boundary conditions \eqref{DirichletBoundaryConditions}. By replacing this remaining term inside the integral, only the tangential derivative of the perturbation appears, i.e. $n^{\alpha}\varepsilon^{\mu\nu}_{\ \ \alpha \beta} \nabla_\mu h^{\rho\sigma}$ which is vanishing under Dirichlet conditions, thus leading to:
\begin{equation}
    \int d^3 y \sqrt{\abs{\gamma}} \ b  \ n^\alpha n^\lambda K^{\beta}_{\rho} \ \delta^2 \left(\widetilde{R}^\rho_{\ \lambda\alpha\beta}\right) \simeq 0 \ .
\end{equation}
We are thus left only with the first term of 
the right-hand side of \eqref{delta_2_CS(K)_with_riemann}, which yields:
\begin{equation}
    \begin{aligned}
         \delta^2 S^{\mathcal{B}}_{CS} \ni & \int d^3 y \sqrt{\abs{\gamma}} \ b  \ n^\alpha n^\lambda \delta\left(K^{\beta}_{\rho}\right) \ \delta \left(\widetilde{R}^\rho_{\ \lambda\alpha\beta}\right) \\
         \simeq & \ 
         \frac{1}{2}\int d^3 y \sqrt{\abs{\gamma}} \ b  \ n^\alpha n^\lambda \varepsilon^{\mu\nu}_{\ \ \alpha \beta} \ \gamma^{\beta}_{\gamma} \gamma^{\delta}_{\rho} \ \nabla_{\vec{n}}h^{\gamma}_{\delta}  \ \nabla_\mu \delta \Gamma^{\rho}_{\lambda \nu}\\
         = & \  
         \frac{1}{2}\int d^3 y \sqrt{\abs{\gamma}} \ b  \ n^\alpha n^\lambda \varepsilon^{\mu\nu}_{\ \ \alpha \beta} \ \gamma^{\beta}_{\gamma} \gamma^{\delta}_{\rho} \ \nabla_{\vec{n}}h^{\gamma}_{\delta}  \ \left( \nabla_\mu\nabla_\lambda h^{\rho}_\nu +  \nabla_\mu\nabla_\nu h^{\rho}_\lambda -  \nabla_\mu\nabla^\rho h_{\lambda \nu} \right) \\ 
         \simeq & \  
         \frac{1}{2}\int d^3 y \sqrt{\abs{\gamma}} \ b  \ n^\alpha n^\lambda \varepsilon^{\mu\nu}_{\ \ \alpha \beta} \ \gamma^{\beta}_{\gamma} \gamma^{\delta}_{\rho} \ \nabla_{\vec{n}}h^{\gamma}_{\delta}  \  \nabla_\mu\nabla_\lambda h^{\rho}_\nu \\ 
         \simeq & \  
         \frac{1}{2}\int d^3 y \sqrt{\abs{\gamma}} \ b  \ \varepsilon^{\mu\nu\alpha\beta} \ \gamma_{\beta\gamma} \gamma^{\delta}_{\rho} \ \nabla_{\vec{n}}h^{\gamma}_{\delta}  \  \nabla_\mu\nabla_\alpha h^{\rho}_\nu \\
         \simeq & \ 0\, ,
    \end{aligned}
\end{equation}
where, in the second and third line, we substituted the expression for the respective variations, using \eqref{variation_of_extrinsic_curvature},\eqref{dual_riemann_variation} and \eqref{affinevariation}. In the fourth line we ignored second tangential derivatives of the perturbation since they vanish under Dirichlet conditions, and in the final line we used the decomposition \eqref{eq:inducedmetric_definition} of $n^{\alpha}n^\lambda = \epsilon g^{\alpha \lambda}- \epsilon \gamma^{\alpha \lambda}$, and from the contraction of $\nabla_\mu \nabla_\alpha h^{\rho}_{\nu}$ with $\varepsilon^{\mu\nu\alpha\beta}$ only the antisymmetric part survives, i.e. the commutator $\left[\nabla_\mu,\nabla_\alpha\right]h^{\rho}_{\nu} \sim h^{\rho}_{\nu}$, which vanishes under Dirichlet boundary conditions \eqref{DirichletBoundaryConditions}. With all these in mind, the result for the second variation of the boundary contribution for the CS-part of the action \eqref{interaction} vanishes,
\begin{equation}
     \delta^2 S^{\mathcal{B}}_{CS} \simeq 0 \ .
\end{equation}

  The above results, when restricted to the TT gauge \eqref{TTgauge} within the framework of an expanding-Universe FLRW background in the presence of chiral GW perturbations, 
have been used in section \ref{sec:reclass}, in order to develop our mean-field HF approach to the study of the dynamics underlying the gCS condensate. In this gauge, the 
quadratic action 
\eqref{GravandMatterWithoutGaugeFixing} becomes independent of the $b$-axion field, which has important consequences on the form of the dynamical equations \eqref{HFstation}, as discussed in the main text. This completes our discussion.

\bibliography{bibliographyDMVslowroll} 

\end{document}